\documentclass[12pt]{article}
 \pdfoutput=1
 \usepackage{graphicx}
  \usepackage{hyperref}
  \usepackage{epsfig}
  \usepackage{amsmath}
  \usepackage{amsthm}
  \usepackage{amssymb}
  \usepackage{color}
  \usepackage{mciteplus}
  \usepackage{epstopdf}
  \newlength{\abstractwidth}
  \usepackage{subfigure}

  \renewcommand{\title}[1]{\vbox{\center\bf{\Large{#1}}}\vspace{5mm}}
  \renewcommand{\author}[1]{\vbox{\center#1}\vspace{5mm}}
  \newcommand{\address}[1]{\vbox{\center\em#1}}
  \newcommand{\email}[1]{\vbox{\center\tt#1}\vspace{5mm}}

  \newcommand{\be}{\begin{equation}}
  \newcommand{\ee}{\end{equation}}
  \newcommand{\tr}{\text{tr}}
  
  \newcommand{\OO}{\mathcal{O}}

  \newcommand{\ket}[1]{| #1 \rangle }
  \newcommand{\bra}[1]{\langle #1 |}
  
  \newtheorem*{conj*}{Conjecture}

  \definecolor{darkgreen}{rgb}{0,.5,0}

\begin{document}

\begin{titlepage}
\rightline{MIT-CTP/4733}
\begin{center}
\hfill \\
\hfill \\
\vskip .5cm

\title{Chaos in quantum channels}

\author{Pavan Hosur,${}^{a}$ Xiao-Liang Qi,${}^{a}$ Daniel A. Roberts,${}^{b}$ and Beni Yoshida${}^{c,d}$}

\address{$^{a}$ Department of Physics, Stanford University \\ Stanford, California 94305, USA

\vspace{10pt}

$^{b}$ Center for Theoretical Physics {\it and} \\  Department of Physics, Massachusetts Institute of Technology \\
Cambridge, Massachusetts 02139, USA 

\vspace{10pt}

$^{c}$ Perimeter Institute for Theoretical Physics,\\ Waterloo, Ontario N2L 2Y5, Canada

\vspace{10pt}

$^{d}$ Walter Burke Institute for Theoretical Physics, California Institute of Technology, Pasadena, California 91125, USA
}

\email{pavanh@stanford.edu, xlqi@stanford.edu,\\drob@mit.edu, byoshida@perimeterinstitute.ca}

\end{center}

\begin{abstract}
We study chaos and scrambling in unitary channels by considering their entanglement properties as states. 
Using out-of-time-order correlation functions to diagnose chaos, we characterize the ability of a channel to process quantum information. 
We show that the generic decay of such correlators implies that any input subsystem 
must have near vanishing mutual information with almost all partitions of the output. 
Additionally, we propose the negativity of the tripartite information of the channel as a general diagnostic of scrambling. This measures the delocalization of information and is closely related to the decay of out-of-time-order correlators.
We back up our results with numerics in two non-integrable models and analytic results in a perfect tensor network model of chaotic time evolution. These results show that the butterfly effect in quantum systems implies the information-theoretic definition of scrambling.
\end{abstract}

\end{titlepage}

\tableofcontents
\baselineskip=17.63pt

\section{Introduction}

Quantum information is processed in quantum circuits, or more generally, quantum channels. A useful way to characterize fault-tolerance and computational power of such channels is by whether input information remains localized or is spread over many degrees of freedom. This delocalization of quantum information by a quantum channel over the entire system is known as scrambling \cite{Hayden:2007cs,Sekino:2008he,Lashkari:2011yi}. Scrambling implies that information about the input cannot be deduced by any local measurement of the output \cite{Page:1993df}. 

The information-theoretic phenomenon of scrambling is closely related to the notion of chaos in thermal systems. A vivid feature of quantum chaos is the butterfly effect: simple localized operators grow under time evolution to have large commutators with almost all other operators in the system.  Consider a pair of local Hermitian operators $W$ and $V$ supported on non-overlapping subsystems such that $[W,V]=0$. Under a chaotic time evolution with Hamiltonian $H$, a local operator $W(0)$ will evolve into a complicated operator $W(t) = e^{iHt} W e^{-iHt}$ which has an expansion as a sum of products of many local operators 
\be
W(t) = W + it\, [H, W] - \frac{t^2}{2!}\,[H,[H,W]] - \frac{i t^3}{3!}\,[H,[H,[H,W]]] + \dots. \label{BCH-expansion}
\ee
For a generic $H$ with local interactions, the $k$th-order nested commutator of $H$ with $W$ can lead to a product of as many as $k$ local operators that acts non-trivially on a large volume of the system \cite{Roberts:2014isa}. This implies that $[W(t),V]\not=0$ and will generically be a large operator of high weight.

The degree of non-commutativity between $W(t)$ and $V$ can be measured by their group commutator: $W(t)\, V\, W(t)\, V$. In fact, the generic decay of out-of-time-order (OTO) correlators of the form 
\be
\langle W(t)\, V \, W(t)\, V \rangle_\beta \equiv Z^{-1} \, \tr\, \{ e^{-\beta H}\, W(t)\, V \, W(t)\, V\, \} , \label{out-of-time-ordered-corr}
\ee
is a distinguishing feature of quantum chaos \cite{Shenker:2013pqa,Shenker:2013yza,Roberts:2014isa,Roberts:2014ifa,Shenker:2014cwa,Maldacena:2015waa}, 
where $\beta$ is the inverse temperature, and $Z=\tr\, e^{-\beta H}$. While this definition of chaos is very direct in terms of operators and observables, it should be possible to understand chaos solely as a property of the system itself.

In this paper, we will consider unitary quantum channels as states in order to characterize their ability to process quantum information in terms of entanglement. 
A unitary quantum channel is simply a unitary circuit where we allow the inputs to be mixed states. To use standard quantum-information measures, we introduce a mapping from a unitary quantum channel to a quantum pure state in a doubled Hilbert space. Using this map, we propose that the tripartite information of a subsystem of the input $A$ and a division of the output $CD$ into two subsystems $C$ and $D$
\be
I_3(A:C:D) = I(A:C) + I(A:D) - I(A:CD), \label{eq:def-of-scrambling}
\ee
provides a basic measure of scrambling in such channels. The negativity of the tripartite information is a natural measure of multipartite entanglement, and
in particular, channels that scramble will have near maximally negative tripartite information for all possible input/output subsystem combinations.

Next, we will use this definition of scrambling to make a direct connection between the OTO correlator diagnostic of chaos \eqref{out-of-time-ordered-corr} and the information-theoretic definition of scrambling \eqref{eq:def-of-scrambling}. It is intuitive that chaotic time evolution should correspond to scrambling: the growth of the operator $W(t)$ in time means that in order to recover or reconstruct the simple operator $W$ one will need to measure a growing fraction of the degrees of freedom of the system. By averaging OTO correlators of the form \eqref{out-of-time-ordered-corr} over a complete basis of operators on an input subsystem $A$ and output subsystem $D$, we can directly relate these correlators to the second R\'enyi entropy between subsystems of the input and output
\be
|\langle \OO_D(t)\,  \OO_A \,  \OO_D(t)\, \OO_A  \rangle|~\sim~2^{ -S_{AC}^{(2)} },
\ee 
where $\OO_A$ represents an operator in the input $A$, $\OO_D$ represents an operator in the output $D$, and $S_{AC}^{(2)}$ is the second R\'enyi entropy of the input/output system $AC$. With this result, we can show that the butterfly effect implies scrambling
\be
|\langle \OO_D(t)\,  \OO_A \,  \OO_D(t)\, \OO_A \rangle_{\beta=0}| =\epsilon ~\implies~I_3(A:C:D) \leq I_{3,\, \min}+2\log_2\frac{\epsilon}{\epsilon_{\rm min}}.
\ee
Here, $I_{3,\, \min}$ is the minimum of $I_3$, and $\epsilon_{\rm min}$ is the minimum of averaged OTO correlations. We will support all of these results with numerics in two non-integrable models, the one-dimensional Ising spin chain with parallel and transverse field, and a four-Majorana-fermion model introduced by Kitaev \cite{kitaev2} known to be a fast scrambler. Our results show that for chaotic systems, all of the information-theoretic quantities relevant to scrambling approach their Haar-scrambled values. 

Relatedly, we will comment on the relationship between the butterfly effect and the ballistic spreading of information in a channel. 
For systems with a notion of spatial locality, we will show that the information contained in an input subsystem $A$ will expand ballistically in the output  with a characteristic velocity $v_B$ (usually denoted the ``butterfly velocity'' \cite{Shenker:2013pqa,Roberts:2014isa}). This supports the idea that $v_B$ is the velocity of information in strongly-chaotic systems. We will also comment on the conceptual differences between the butterfly velocity and the entanglement or tsunami velocity $v_E$ of \cite{Hartman:2013qma,Liu:2013iza,Liu:2013qca}.

Finally, we will use our enhanced understanding of the relationship between scrambling, chaos, and entanglement in time to propose a solvable model of a unitary quantum channel that exhibits scrambling. Building on the work of \cite{Hartman:2013qma} and \cite{Pastawski:2015qua}, we discuss a perfect tensor model of a chaotic Hamiltonian time evolution. This can be thought of as a toy model for an Einstein-Rosen bridge that connects the two sides of the eternal black hole in AdS.

The plan of this paper is as follows. In \S\ref{sec:entanglement-in-time} we discuss unitary quantum channels and elaborate on the notion of entanglement in time. There, we consider the entanglement properties of such channels and introduce the tripartite information as a measure of scrambling. In \S\ref{sec:bounds}, we show that the decay of OTO correlation functions implies strong bounds on information-theoretic quantities, directly connecting chaos to scrambling. We provide evidence for our claims via numerical studies of qubit systems in \S\ref{sec:numerics} and with a perfect tensor model of chaotic time evolution in \S\ref{sec:perfect-tensor-model}. We conclude in \S\ref{sec:discussion} with a discussion of the relationship between chaos and computation. Some extended calculations, tangential discussions, and lengthy definitions are left to the Appendices.

\section{Unitary quantum channels} \label{sec:entanglement-in-time}

To study the scrambling properties of different unitary operators $U$ by using information-theoretic quantities, we will interpret them as states.
To be concrete, let us assume that the quantum system consists of $n$ qubits with a time independent Hamiltonian $H$. We will be interested in a particular unitary operator, the time evolution operator $U(t) = e^{-i H t}$. This will let us study a one parameter family of unitary operators indexed by $t$.

\begin{figure}
\begin{center}
\includegraphics[width=0.65\linewidth]{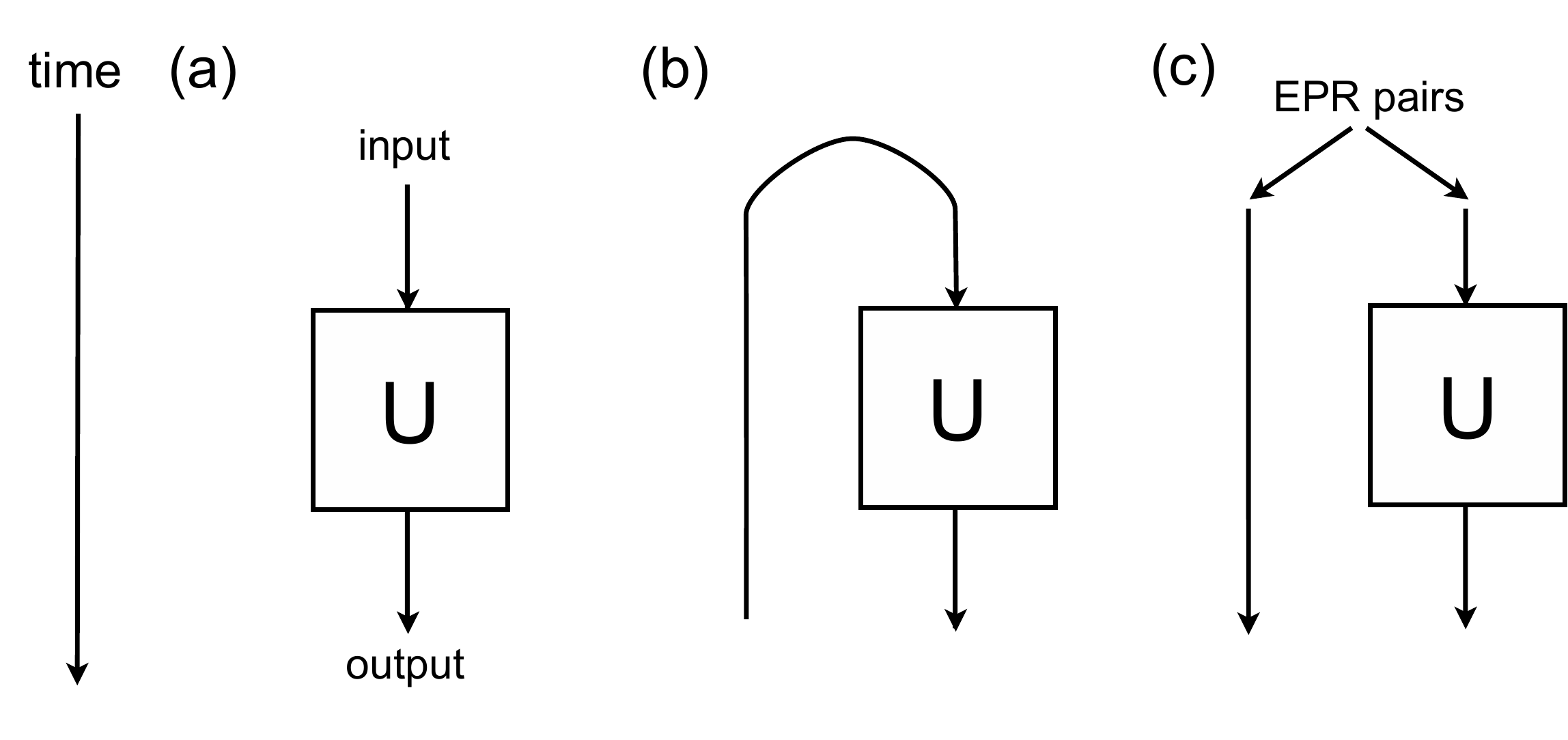}
\caption{Interpretations of a unitary channel: 
{\bf (a)} a unitary operator $U$ with input and output legs. 
{\bf (b)} state interpretation $|U\rangle$ of the unitary operator $U$. By bending the input legs down, we treat input/output equally.
{\bf(c)} the state interpretation is equivalent to the creation of a maximally entangled state followed by acting with $U$ on half the EPR pairs, which gives $|U\rangle$.
}\label{fig:operator-state-ref}
\end{center}
\end{figure}

A unitary operator $U(t)$ that acts $n$ qubits is described by a $2^n \times 2^n$ dimensional matrix 
\be
U(t) = \sum_{i,j=0}^{2^n-1} u_{ij} |i\rangle \langle j |.\label{eq:U-as-op}
\ee  
which we usually choose to think of in terms of a tensor with $n$ input and $n$ output legs, as shown in Fig.~\ref{fig:operator-state-ref}(a). However, it is also natural to map this to a $2n$-qubit state by treating the input and output legs on equal footing
\be
|U(t) \rangle = \frac{1}{2^{n/2}}\sum_{i,j=0}^{2^n-1} u_{ji} |i\rangle_{in}\otimes |j \rangle_{out}.\label{eq:U-as-state}
\ee
This is depicted graphically in Fig.~\ref{fig:operator-state-ref}(b). Clearly $|U(t) \rangle$ encodes all the coefficients ($u_{ij}$) necessary to represent the unitary operator $U(t)$.

When $t=0$, the unitary operator is simply the identity operator, and \eqref{eq:U-as-op} reduces to a state consisting of $n$ EPR pairs
\be
|I\rangle = \frac{1}{2^{n/2}}\sum_{i,j=0}^{2^n-1} |i\rangle_{in}\otimes |i \rangle_{out},\label{eq:U-as-I}
\ee
where each input leg is maximally entangled with each output leg, and there is no entanglement between different EPR pairs. For finite $t$, inserting a complete set of states into \eqref{eq:U-as-state} and using \eqref{eq:U-as-op}, we can rewrite it as
\be
|U(t)\rangle = \frac{1}{2^{n/2}}\sum_{i,j=0}^{2^n-1} |i\rangle_{in}\otimes U(t)|i \rangle_{out}=\mathbb{I}\otimes U(t)|I\rangle\label{circuit-state},
\ee
with $\mathbb{I}$ the identity operator acting on the incoming states.  This offers the following interpretation to the state $|U(t)\rangle$: a maximally entangled state is created between a reference system ``in'' and a system of interest ``out.'' Next, the operator $U$ acts on ``out,'' giving $|U(t)\rangle$. This perhaps offers a more physical interpretation of this operator-state mapping, as shown in Fig.~\ref{fig:operator-state-ref}(c). 

{\bf Note:} in this paper we will adopt the perspective of the mapping shown in Fig.~\ref{fig:operator-state-ref}(b) and use language that refers to the entanglement properties of channels (in time) between subsystems of inputs and outputs. Additionally, we will always draw our channels from the operator perspective as in Fig.~\ref{fig:operator-state-ref}(a).

It is natural to ask whether the choice of maximally entangled state $|I\rangle$ is artificial. Although different choices of initial state $|I\rangle$ can be made which define different mappings from $U$ to $|U(t)\rangle$, all our discussions remain insensitive to the choice as long as $|I\rangle$ is a direct product of EPR pairs. The two qubits in each pair are required to be the qubits at a given real-space position in the input and output systems, which are maximally entangled with each other. This choice guarantees that all quantum entanglement between different real-space locations in $|U(t)\rangle$ are created by the unitary evolution $U(t)$.

The operator-state mapping can be further generalized by considering a more generic statistical ensemble as the input state. 
Let $\{|\psi_{j}\rangle\}$ be a set of orthonormal states, and imagine that we input an initial state $|\psi_{j}\rangle$ with probability $p_{j}$ to a unitary quantum channel $U$. This means that the initial statistical ensemble is $\rho_{in}=\sum_{j}p_{j}|\psi_{j}\rangle\langle \psi_{j}|$. After time evolution, each input state evolves to $|\phi_{j}\rangle = U |\psi_{j}\rangle$, and the output statistical ensemble is given by $\rho_{out}=\sum_{j}p_{j}|\phi_{j}\rangle\langle \phi_{j}|$. The time evolution of a given input ensemble $\rho_{in}$ can be mapped to the following pure state 
\begin{align}
|\Psi\rangle = \sum_{j}\sqrt{p_{j}}\, |\psi_{j}\rangle_{in} \otimes |\phi_{j}\rangle_{out}=\mathbb{I}\otimes U(t)\sum_{j}\sqrt{p_{j}}\, |\psi_{j}\rangle_{in}\otimes|\psi_{j}\rangle_{out}. \label{eq:isomorphc-state}
\end{align}
The isomorphic state $|\Psi\rangle $ contains all the information required to characterize the properties of the channel. Namely, if one traces out the output system, then the reduced density matrix is the input state ($\rho_{in}=\tr_{out}\, |\Psi\rangle \langle \Psi|)$ while if one traces out the input system, then the reduced density matrix is the output state ($\rho_{out}=\tr_{in}\, |\Psi\rangle \langle \Psi|$). The state interpretation in \eqref{circuit-state} corresponds to the special case of a uniform input ensemble (i.e. $\rho_{in}= 2^{-n}\mathbb{I}$). In general, we will simply refer to the state given in \eqref{eq:isomorphc-state} as a unitary quantum channel.  In quantum information theory, such correspondence between quantum channels and quantum states is named as the channel-state duality.\footnote{In fact, the channel-state duality in quantum information theory extends to any quantum channels with decoherence as well as those with different sizes of the input and output Hilbert spaces~\cite{Nielsen_Chuang}.
}

A familiar example of a unitary channel is the thermofield double state 
\be
|TFD\rangle = \frac{1}{\sqrt{Z}}\sum_{i} e^{- \beta E_{i} /2 }|i\rangle \otimes |i\rangle,
\ee
where $E_i$ are eigenvalues of the Hamiltonian $H$ and $Z = \tr\, e^{-\beta H}$. 
Applying evolution for time $t$ to the right output system, one obtains the following time evolved state
\begin{align}
|TFD(t)\rangle =  \frac{1}{\sqrt{Z}}\sum_{i} e^{-\beta E_{i}/2} e^{-iE_{i}t}|i\rangle \otimes |i\rangle.\label{eq:time-evolved-TFD}
\end{align} 
One can interpret this state as a quantum channel $U=e^{-iHt}$ whose input is given by the thermal ensemble. Note this expression reduces to \eqref{circuit-state} for $\beta=0$.

\subsection{Entanglement in time}

\begin{figure}
\begin{center}
\includegraphics[width=0.45\linewidth]{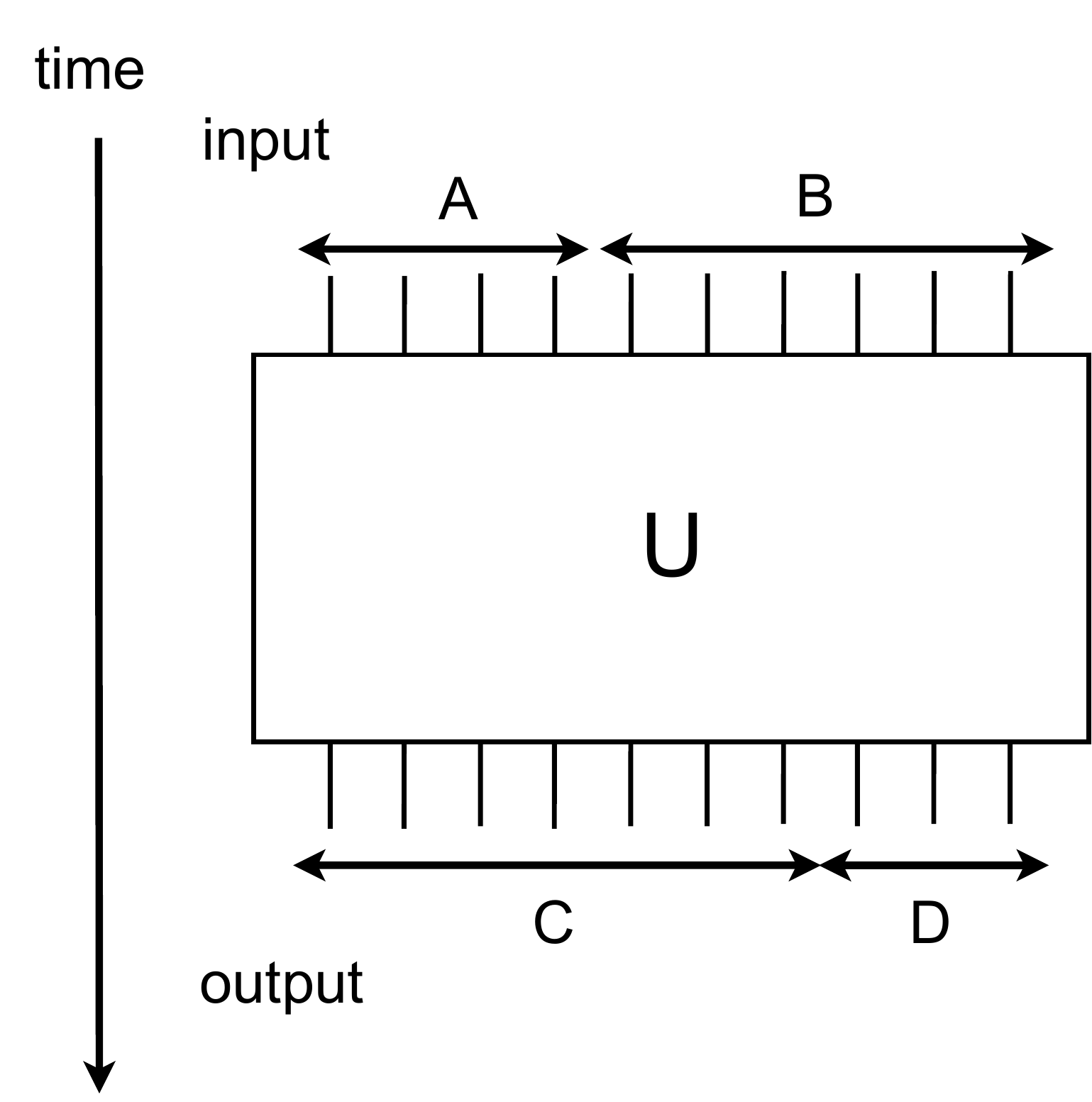}
\caption{Setup to study scrambling in a unitary quantum channel $U$. Even though we draw our channels with input and output legs, when we discuss entanglement we always mean of the state $|U\rangle$ given by the mapping to the doubled Hilbert space as in \eqref{eq:U-as-state}.
}\label{circuit}
\end{center}
\end{figure}

Since the state as defined in \eqref{eq:isomorphc-state} contains all the information concerning the inputs and dynamics of the channel, we would like to use it to establish a general measure 
for scrambling. 
We will do this by studying the entanglement properties of a unitary $U$ via the state $|\Psi\rangle$. Our setup is as follows. The input system is divided into two subsystems $A$ and $B$, and the output system is divided into two subsystems $C$ and $D$, as shown in Fig.~\ref{circuit}. The subsystems do not necessarily have to be of the same size (i.e. it is possible that $|A| \neq |B|$ or $|A| \neq |C|$), and at $t=0$ the input and output partitions do not necessarily need to overlap (i.e. it could be that $A\cap C = \emptyset$). Additionally, despite how it is drawn, there does not need to be any spatial organization to the partitions. For example, the subsystem $A$ could be an arbitrary subset of the input qubits.

With this state interpretation of the channel \eqref{eq:isomorphc-state}, we can form a density matrix $\rho =  \ket{\Psi}\bra{\Psi}$ to compute joint entropies of subsystems that include both input and output degrees of freedom. For example, the entanglement entropy $S_{AC}$ is given by
\be
S_{AC} = - \tr\, \{ \rho_{AC} \log_2 \rho_{AC} \},
\ee
where the notation $\rho_{AC}$ means the usual partial trace
\be
\rho_{AC} = \tr_{BD}\, \{ \rho \}.
\ee
Additionally, the mutual information between $A$ and $C$ is given by
\be
I(A:C) = S_A + S_C - S_{AC}, \label{def-of-mutual}
\ee
and we will sometimes compute R\'enyi entropies
\be
S^{(N)}_{AC} = \frac{1}{1-N} \log \tr\, \{ \rho_{AC}^N \}. \label{def-of-renyi}
\ee

Finally, let us note that, for a uniform input ensemble (or $\beta=0$), 
\be
S_A = a, \qquad S_B = b, \qquad S_C =c, \qquad S_D = d, \label{single-sub-system-EE}
\ee
\be
S_{AB} = S_{CD} = n
\ee
where $a,b,c,d$ are the numbers of qubits on $A,B,C,D$ respectively. These relations are true because the inputs are always maximally entangled with the outputs. Therefore, any subsystem that is only a partition of the inputs or only a partition of the outputs (including non-partitions such as $AB$ and $CD$) is maximally mixed. Even if we consider the more general channel \eqref{eq:isomorphc-state}, any subsystem that does not involve both input and output systems still has an entropy that is time-independent. Therefore the scrambling effect only appears in the entropy of regions on both sides, and (thus) the mutual information terms such as $I(A:C)$ and $I(A:D)$.  For this reason, we will primarily be interested in the mutual information between region $A$ and different partitions of the outputs. When region $A$ is taken to be small, such as a single lattice site, the mutual information of $A$ with part of the output system tracks how the information about local operators in $A$ spreads under time evolution.

\subsection{Scrambling}\label{sec:tripartite}
Scrambling is usually considered as a property of a state. In \cite{Sekino:2008he}, a reference state evolved with a random unitary sampled from the Haar ensemble is called ``Haar-scrambled.''  A much weaker notion of scrambling of a state (which \cite{Sekino:2008he} calls ``Page scrambling,'' or usually just ``scrambling'') is given by a state that has the property that any arbitrary subsystem of up to half the state's degrees of freedom are nearly maximally mixed. Said another way, a state is scrambled if information about the state cannot be learned from reasonably local measurements. Naturally (and proven in \cite{Page:1993df}), Haar scrambling implies Page scrambling.

We are interested in extending the notion of scrambling to unitary quantum channels of the form \eqref{eq:isomorphc-state}. Let us try to understand the properties of scrambling channels by considering entanglement across the channel. The identity channel is just a collection of EPR pairs connecting the input to the output. An example of a channel that does not scramble is the ``swap'' channel, where the arrangement of the EPR pairs are simply swapped amongst the degrees of freedom in the output system. In that case, localized quantum information in the input system remains localized in the output system, though residing in a different particular location. Instead, for a channel to scramble it necessarily must convert the EPR pairs into a more complicated arrangement of multipartite entanglement between the input and output systems. Such local indistinguishability of output quantum states corresponding to simple orthogonal input states in quantum channels enables secure storage of quantum information: a realization of quantum error-correcting codes.

Let us try to formalize this idea using our setup in Fig.~\ref{circuit}. If our channel is a strong scrambler, then local disturbances to initial states cannot be detected by local measurements on output states. This implies that measurements on a local region $C$ cannot reveal much information on local disturbances applied to $A$. Therefore the mutual information $I(A:C)$ must be small. From a similar reasoning, one also expects $I(A:D)$ to be small when the channel is a good scrambler since $D$ is a local region too. On the other hand, the mutual information $I(A:CD)$ quantifies the total amount of information one can learn about $A$ by measuring the output $CD$ jointly. Since we are interested in the amount of information concerning $A$ which is hidden non-locally over $C$ and $D$, a natural measure of scrambling would be
\begin{align}
I(A:CD) - I(A:C) - I(A:D).
\end{align}
This quantity accounts the amount of information about $A$ that are non-locally hidden over $C$ and $D$ such that any local measurements, exclusively performed on $C$ or $D$, cannot know. If the channel scrambles, we expect that this quantity is large. It is well known that the above quantity is equal to minus the tripartite information\footnote{In the condensed matter community, the tripartite information is referred to as the topological entanglement entropy, which measures the total quantum dimension in a $(2+1)$-dimensional TQFT \cite{Kitaev06, Levin06}.}
\begin{align}
I_3(A:C:D)  &= S_{A} + S_{C} + S_{D} - S_{AC} - S_{AD} - S_{CD} + S_{ACD}\nonumber\\
&\equiv I(A:C) + I(A:D) - I(A:CD)
\end{align}
The tripartite information $I_3(A:C:D)$ must be negative and have a large magnitude for systems that scramble. We propose that this is a simple diagnostic of scrambling for unitary channels.

Scrambling in unitary channels is closely related to other notions of scrambling of states. For example, if the input to the channel is fixed to be a direct product state, then tripartite scrambling implies that subsystems of the output state will be near maximally mixed. Thus, scrambling in terms of the tripartite information implies ``Page scrambling'' of the output state.\footnote{In fact, this operator notion of scrambling is stronger than ``Page scrambling''  since the latter only refers to a single state. The operator scrambling implies that the information about a local operator in the input system cannot be recovered from a subsystem $C$ of the output system (as long as it is not very close to the entire output system) even if $C$ is bigger than half of the system. 
} In Appendix~\ref{appendix-haar}, we analyze Haar-random channels and show that Haar scrambling also implies that the tripartite information of the channel is very negative.

Importantly, it should be noted that the tripartite information is a measure of four-party entanglement, not three-party entanglement. Namely, consider a state with three-party entanglement only, $|\Psi\rangle = |\psi_{A}\rangle \otimes |\psi_{BCD}\rangle$ where $A$ and $BCD$ are not entangled. Then one always has $I_3(A:B:C)=0$.  
$I_3$ for other choices of regions also vanish, because for pure states the tripartite information is symmetric in any partitions into four regions $A,B,C,D$
\begin{align}
I_3(A:B:C)=I_3(A:B:D)=I_3(A:C:D)=I_3(B:C:D).\label{eq:symmetry-of-I3}
\end{align}
Thus, in channels $I_3$ is really a measure of four-party entanglement between the input system and the output system. In this paper, we will often choose to write the tripartite information as $I_3(A:C:D)$ in order to emphasize a particular decomposition. However, for unitary channels the arguments are unnecessary due to the symmetry \eqref{eq:symmetry-of-I3}.

\begin{figure}[htb!]
\centering
\includegraphics[width=0.85\linewidth]{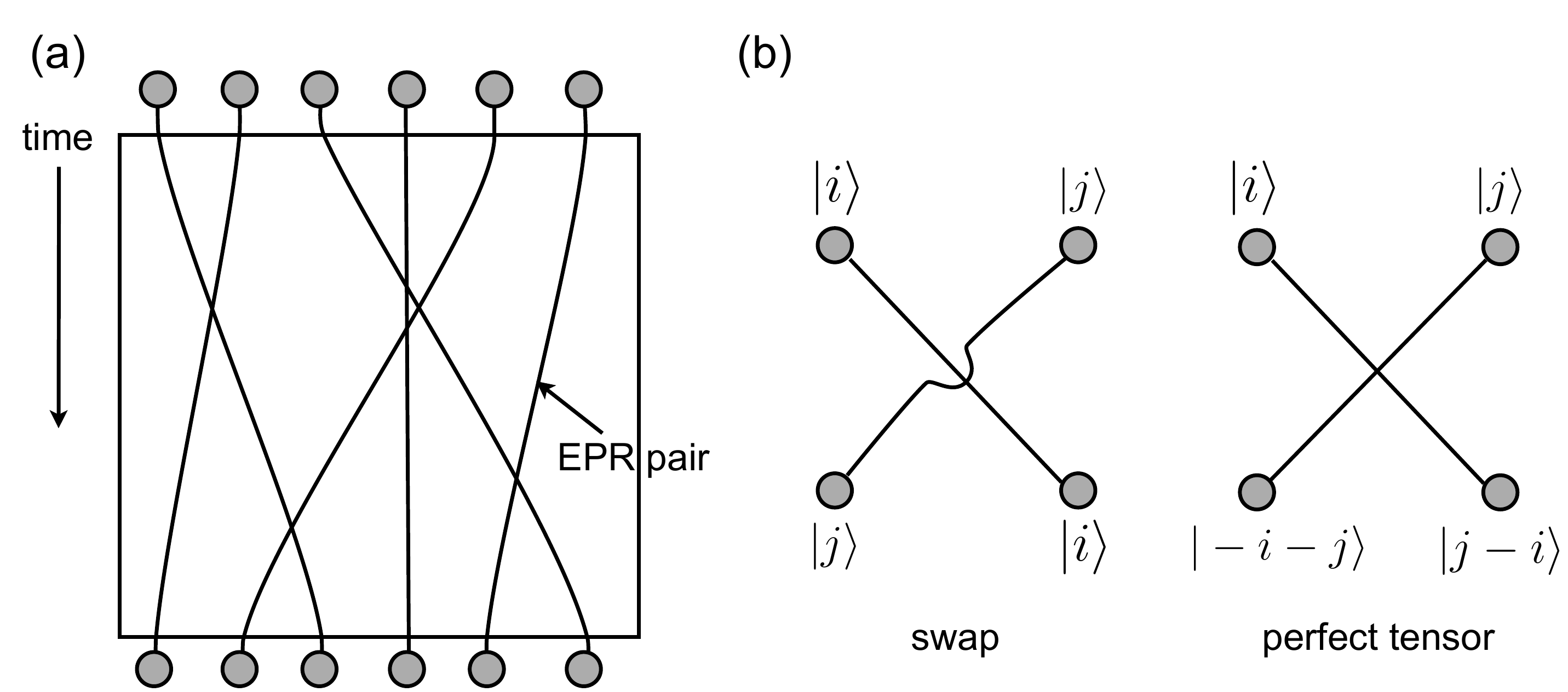}
\caption{
{\bf (a)} Permutations of qubits. The isometric pure state $|\Psi\rangle$ consists of EPR pairs between input and output qubits. 
{\bf (b)} A swap gate and a unitary corresponding to a perfect tensor. Note the similarity to the Feynman diagrams of a $2\to2$ scattering process for a free theory and interacting theory, respectively.
} 
\label{fig_swap}
\end{figure}

Finally, note that the condition $I_3(A:C:D) \le I(A:D)$ is known as \emph{strong subadditivity} and must always holds among entropies. On the other hand, $I_3(A:C:D) \le 0$ is often referred to as \emph{monogamy of mutual information} and doesn't necessarily hold for arbitrary states.\footnote{e.g. $I_3(A:C:D) = 1$ for the four-qubit GHZ state $\frac{1}{\sqrt{2}}(|0000\rangle + |1111\rangle)$.} However, for holographic systems the tripartite information must always be negative \cite{Hayden:2011ag}. This result is usually only discussed for holographic states but it also applies to holographic channels (such the eternal black hole in AdS) \cite{Gharibyan:2013aha}. 
It is natural to suggest that the negative $I_3$ is related to the fact that such holographic systems are strongly-chaotic and fast-scramblers \cite{Shenker:2013pqa}. (See also \cite{Rangamani:2015qwa} for a study of monogamy and other properties of entanglement in qubit systems.)

\subsection{Examples}
  
Here, we present a few examples of using the tripartite information of a channel to measure scrambling.

\subsubsection*{Swap channel}

Let us revisit the example discussed at the beginning of this section. Consider a system of $n$ qubits and assume that the unitary operator is the identity operator: $U=I$. The channel description is given by \eqref{eq:U-as-I}. This is a collection of EPR pairs connecting input qubits and output qubits. Since the state consists only of two-party entanglement, the tripartite information is zero. Similarly, consider a time-evolution that consists only of permutations of qubits, a ``swap'' channel. Namely, let us assume that $j$th qubit goes to $a_{j}$th qubit where $1\leq j,a_{j}\leq n $. Then, the isomorphic state consists of EPR pairs between $j$th input qubit and $a_{j}$th output qubit, and the tripartite information is zero (Fig.~\ref{fig_swap}(a)). Permutations of qubits can be thought of as a classical scrambling. Two initial nearby classical states may become far apart after permutations of qubits, yet it is still possible to distinguish two initial states by some local measurement on the output states. 

\subsubsection*{Perfect tensor}

Next, let us look at an example where the tripartite information is maximally negative. Consider a system of two qutrits (spins with three internal states $|0\rangle, |1\rangle, |2\rangle$). We consider the following unitary evolution
\begin{align}
|i\rangle \otimes |j\rangle \rightarrow |-i-j\rangle \otimes |j-i\rangle
\end{align}
where addition is defined modulo $3$. It can be directly verified that single qutrit Pauli operators transform in the following way\footnote{Pauli operators for $p$-dimensional qudits are defined by $X\ket{j}=\ket{j+1}$ and $Z\ket{j}=\omega^{j}\ket{j}$ with $\omega=e^{i\frac{2\pi}{p}}$ for all $j$ where addition is modulo $p$.}
\begin{align}
Z \otimes I \rightarrow Z \otimes Z  \qquad
X \otimes I \rightarrow X^{\dagger} \otimes  X^{\dagger}  \qquad
I \otimes Z \rightarrow Z \otimes Z^{\dagger}   \qquad
I \otimes X \rightarrow X^{\dagger} \otimes X.
\end{align}
In this unitary evolution, all local operators evolve to two-body operators. As such, information concerning local disturbances to initial states cannot be detected by any single qutrit measurements on output states. We can represent this unitary evolution as the following pure state
\begin{align}
|\Psi\rangle= \frac{1}{3}\sum_{i,j=0}^{2}(|i\rangle \otimes |j\rangle)_{AB} \otimes (|-i-j\rangle \otimes |j-i\rangle)_{CD} \label{eq:qutrit}
\end{align}
where addition is modulo $3$.
It is known that this pure state is maximally entangled in any bipartition. Namely, one has $S_{A}=S_{B}=S_{C}=1$, $S_{AB}=S_{BC}=S_{CA}=2$, $S_{ABC}=1$, and $I_{3} = -2$, where entropy for qutrits is measured in units of $\log 3$. This is a so-called perfect state. In general, for any pure state $ABCD$ one can show that $I_{3}\geq - 2\min(S_A,S_B,S_C,S_D)$. Therefore, this qutrit state has minimal value of $I_{3}$. 

Here, we note that the difference in the depiction of this qutrit perfect tensor and the swap gate ($|i\rangle \otimes |j\rangle \rightarrow |j\rangle \otimes |i\rangle$) resembles the difference in the Feynman diagrams of a $2\to2$ scattering process between a free theory and an interacting theory (see Fig.~\ref{fig_swap}(b)-(c)). We will comment on this in much greater detail in the context of conformal field theory in Appendix~\ref{sec:memory-effect}.

\subsubsection*{Black hole evaporation}
\begin{figure}[htb!]
\centering
\includegraphics[width=0.40\linewidth]{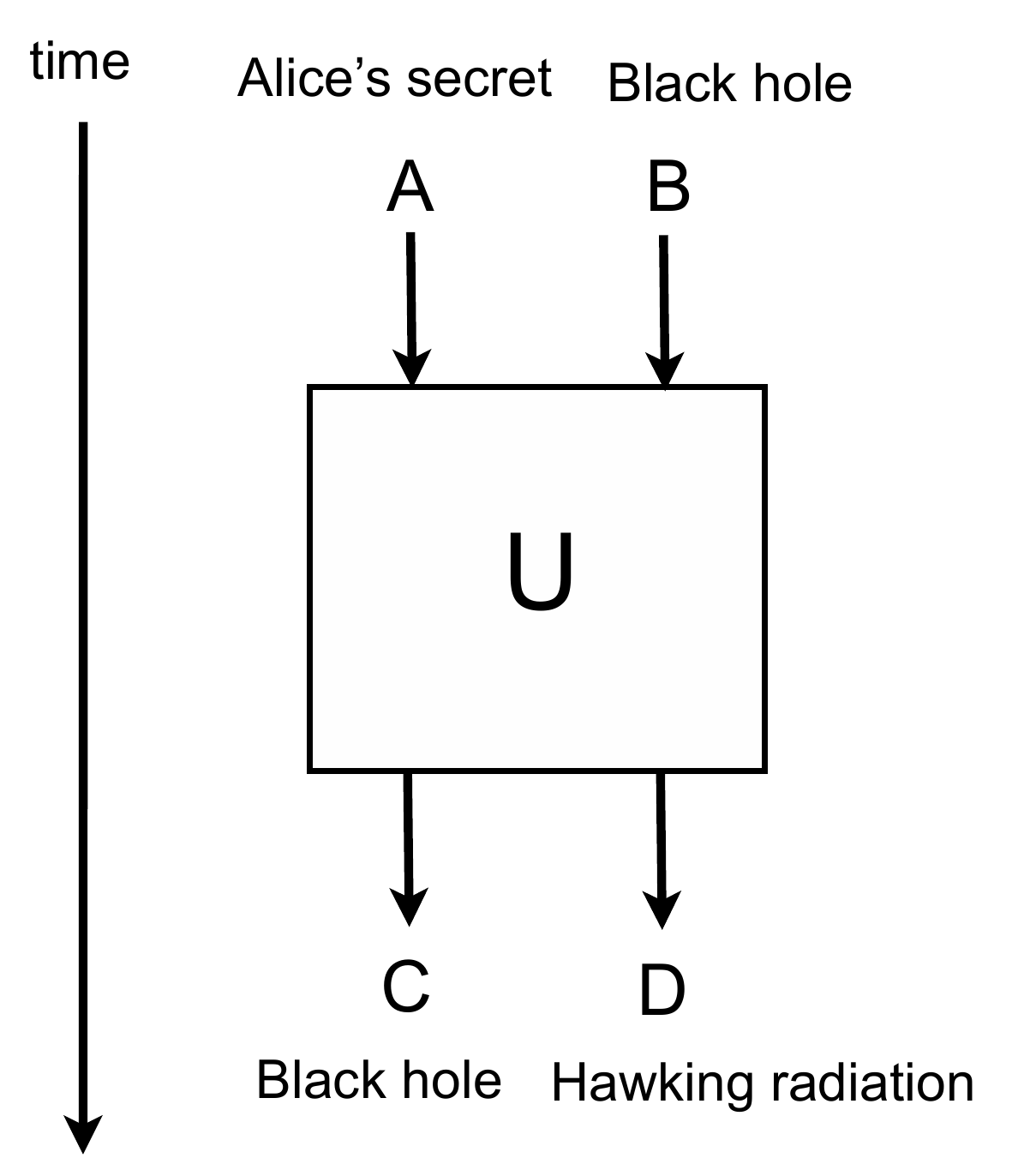}
\caption{The Hayden-Preskill thought experiment. 
} 
\label{Hayden-Preskill}
\end{figure}

Another interesting example is the thought experiment by Hayden and Preskill \cite{Hayden:2007cs} as shown in Fig.~\ref{Hayden-Preskill}. They considered the following scenario. Alice throws her secret ($A$), given in the form of some quantum state of $a=|A|$ qubits, into a black hole ($B$) with the hope that the black hole will scramble her secret so that no one can retrieve it without collecting all the Hawking radiations and decoding them. Bob tries to reconstruct a quantum state of Alice by collecting some portion of Hawking radiation ($D$) after a scrambling unitary evolution $U$ applied to the black hole, consisting both of Alice's secret $A$ and the original content of the black hole $B$. The remaining portion of the black hole after the Hawking radiation is denoted by $C$. So, as usual, this channel is split into four segments $A,B,C,D$ as shown in Fig.~\ref{Hayden-Preskill}. 

First, assume that Bob only knows the dynamics of the black hole (i.e. the operator $U$). In order for Bob to successfully reconstruct Alice's system $A$, the mutual information between Alice's secret and the Hawking radiation must be $I(A:D)\approx 2a$. (Recall that $I(A:D)/2$ is roughly the number of EPR pairs shared by $A$ and $D$.) However this is possible only when $c \approx 0$ since the channel is maximally entangled along any bipartition due to the assumption of $U$ being a scrambling unitary. Namely, $I(A:D)\approx 0$ as long as $D$ is smaller than $B$. Next, let us assume that Bob not only knows the dynamics $U$, but also knows the initial state of the black hole $B$. This is possible in principle if Bob has been observing the black hole since its formation. In this case, Bob has an access to both $B$ and $D$. Then for $d>a $, the mutual information between $A$ and $BD$ becomes nearly maximal; $I(A:BD)\approx 2a$ because $BD$ contains more than half of the entire qubits in channel $ABCD$. In this case, the tripartite information is given by
\be
I_{3}(A:B:D)= I(A:D) + I(A:B) - I(A:BD).
\ee
Since $I(A:D), \, I(A:B)\approx 0$, we find $I_{3}\approx - I(A:BD) \approx - 2a$ which implies that Bob can indeed learn about Alice's secret.\footnote{The firewall paradox of \cite{Almheiri:2012rt,*Braunstein:2009my,Almheiri:2013hfa} is related to the fact the scrambled channel with near maximally negative $I_3$ cannot allow bipartite entanglement $I(C:D)$ between the evaporating black hole $C$ and the recently evaporated Hawking radiation $D$. This is a consequence of the monogamy of entanglement, which is captured by the negativity of the tripartite information.
}

\subsubsection*{Holographic channels}
In the final example, we will consider thermal systems with Einstein gravity bulk duals. Under the holographic duality, the unitary quantum channel representing the CFT time evolution operator is geometrized as the black hole interior or Einstein-Rosen bridge \cite{Hartman:2013qma,Roberts:2014isa}. Such holographic states are already known to be fast scramblers \cite{Shenker:2013pqa}, so here we will simply confirm that holographic channels scramble in the sense of $I_3$.

\begin{figure}[ht]
\begin{center}
\includegraphics[scale=.55]{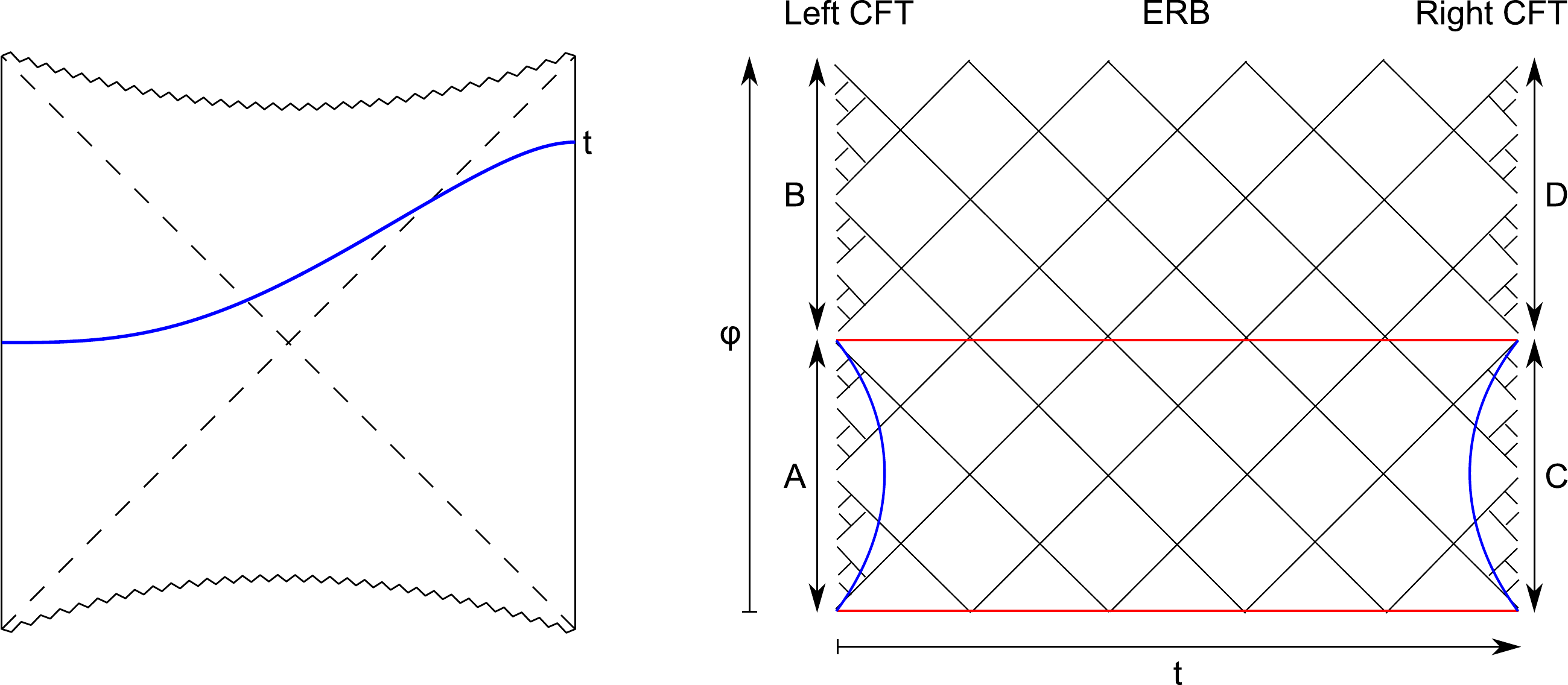}
\end{center}
\caption{The eternal AdS black hole interior is a geometric representation of the unitary quantum channel given by the time evolution operator of the dual CFT. {\bf Left:} Penrose diagram for the eternal AdS black hole geometry with a spacelike slice (blue) anchored on the  left boundary at the middle of the diagram anchored at time $t$ on the right boundary.  {\bf Right:} Geometric depiction of the spacelike slice through the Einstein-Rosen bride (ERB). The spatial coordinates on the boundary CFT are represented by $\varphi$. The renormalized length of the ERB is proportional to $t$. For small $t$, the RT surface used to compute the entanglement entropy $S_{AC}$ goes across the ERB (red). After a time proportional to the size of $A$ or $C$, the disconnected RT surface (blue) is preferred and the entanglement entropy is a sum of disjoint contributions ($S_{AC} = S_A + S_C$).
}\label{fig:eternal-black-hole}
\end{figure}

It's a trivial extension of the ideas in \cite{Hartman:2013qma} and \cite{Shenker:2013pqa} to calculate the tripartite information across the eternal AdS black hole (the holographic state dual to thermofield double state \cite{Maldacena:2001kr} of two entangled CFTs) so we will be brief. 
For simplicity, we will take $A$ to be aligned with $C$ across the Einstein-Rosen bridge, and $A, B, C, D$ to have the same size, as shown in Fig.~\ref{fig:eternal-black-hole}. For simplicity, we will consider time evolution only on the right boundary $U(t) = e^{-iHt}$. For any finite time, the mutual information $I(A:D)$ is always zero for any finite regions $A,D$. The Ryu-Takayanagi (RT) surfaces used to compute the entanglement entropy are always disconnected. The only interesting behavior is from $I(A:C)$, which was computed in this setup in \cite{Hartman:2013qma}. The initial RT surface extends across the Einstein-Rosen bridge, and $I(A:C)$ begins equal to the finite part of $S_A + S_C$, since the finite part of $S_{AC}$ is vanishing. Under time evolution, the finite part of $S_{AC}$ will increase linearly in time with a characteristic entanglement velocity $v_E$ \cite{Hartman:2013qma}, and $I(A:C)$ will decrease linearly to zero. Since $I(A:CD)=S_{BH}$ the Bekenstein-Hawking entropy of the black hole, after a time of at most $O(S_{BH}/2)$, we find
\be
I_3(A:C:D)  = -S_{BH},
\ee
which is its minimal possible value. 

\subsubsection*{Haar-random channels}

In Appendix~\ref{appendix-haar}, we analyze Haar-random unitaries. Using those results, we can bound the tripartite information in a Haar random channel
\be
I_{3,\,Haar} \le  -2\min(S_A,S_B,S_C,S_D)+1  + \dots .
\ee
The tripartite information of a random channel is near maximally negative plus one ``residual'' bit of information (independent of the overall system size). As we mentioned before, Haar scrambling implies tripartite scrambling.

\subsection{Chaotic channels vs. integrable channels}\label{sec:int-vs-chaos}

In this section, we will focus on the aspects of unitary quantum channels built from time-evolution operators that can be used to differentiate chaotic systems from integrable systems. A system, defined by a time independent Hamiltonian $H$, can be chaotic or integrable. A unitary operator $U(t)=e^{-iHt}$ can scramble. In \S\ref{sec:bounds}, we will learn that channels $U(t)$ that scramble for most values of $t$ must be built from chaotic systems. To be concrete, we will take integrable systems to be those that have a quantum recurrence time that is polynomial in their number of degrees of freedom $n$. On the other hand, chaotic systems will generally have recurrence times that are doubly exponential in $n$, $O(e^{e^n})$.

The important point about scrambling in channels is that all bipartite mutual informations between the input subsystems and output subsystems become small. As an extreme example of an integrable system, let's return to the swap channel (Fig.~\ref{fig_swap}(b)). As a reminder, for all $t$  this channel has $I(A:C) + I(A:D) = I(A:CD)$ or $I_3(A:C:D)=0$. Swap gates preserve bipartite entanglement and cannot create multipartite entangled output states. As such, an input localized at one lattice site can only be moved around to $n$ possible locations. The recurrence time must scale like $O(n)$. 

In order for the tripartite information of the channel to vanish for all times, the decrease of $I(A:C)$ must be exactly compensated by the increase of $I(A:D)$. If we take $C$ to contain $A$ at $t=0$, the initial values of mutual information are $I(A:C) = 2a$ and $I(A:D) = 0$. If $I_3$ vanishes for $t>0$, $I(A:C) $ must decrease in order for $I(A:D) >0$. If $I(A:C) $ returns to the initial value $2a$ at a later time (i.e. there is a recurrence), in the meantime we must have $I(A:D) = 0$. Therefore the signature of an integrable system is a sharp peak in $I(A:D) = 0$ (or equivalently, a dip in $I(A:C)$). The sharpness of the peak is determined by the relative sizes of the systems $A$ and $C$. In chaotic systems for times shorter than the Poincare recurrence time $O(e^{e^n})$, $I(A:D)$ and $I(A:C)$ will asymptote to the channel's Haar-random value (see Appendix~\ref{appendix-haar}).

As we will see numerically in \S\ref{sec:numerics}, for unequal divisions of the input ($a \ll b$) and output ($c \ll d$) the signature of an integrable system is a spike in $I(A:D)$ which occurs at the time signals from $A$ arrives at $D$. In such systems the tripartite information might become negative, but it will never become close to the Haar-scrambled value, and it will quickly return near zero. However, for equal divisions of input and output, the tripartite information of an integrable channel will tend to a constant much greater (i.e. less negative) than the Haar-scrambled value. This equal-sized subsystem configuration appears to be the most robust measure of scrambling.

This discussion of the spike in $I(A:D)$ highlights an important point: in integrable systems, such as the transverse Ising model, it is still possible to have early time exponential decay and ballistic growth of operators \cite{Roberts:2014isa} (see the top-middle panel of Fig.~\ref{fig:spin-chains}), and in integrable CFTs some (but not all!) OTO correlators can decay at late times \cite{Roberts:2014ifa}.  However, for these systems the growth of operators (or the decay of $I(A:C)$) must always be followed by a later decrease in size (or a recorrelation in the OTO correlator) as the system exhibits a recurrence.\footnote{Another relevant difference between chaotic and integrable systems is in terms of the expansion \eqref{BCH-expansion} for the time evolution of a simple operator. Due to ergodicity, such an expansion will have a number of terms exponential in the size of the system for chaotic dynamics. Integrable systems are not ergodic, so the expansion will only have a linear number of terms.} We will see this behavior explicitly in our numerics in \S\ref{sec:numerics}.

\section{Butterfly effect implies scrambling}\label{sec:bounds}

Now, we will show that the generic decay of OTO correlators of the form $\langle W(t) \, V \, W(t) \, V \rangle$ implies that the mutual information between any small subsystem in the inputs and any partition of the output should be small. We will provide an exact formula relating the operator average of OTO correlators in different size subsystems to the second R\'enyi entropy for a subsystem consisting of both inputs and outputs. Using tripartite information as our diagnostic of scrambling in a unitary channel, we will show the butterfly effect implies scrambling.

\subsection{Average over OTO correlators}

For simplicity of discussion, we consider a system consisting of qubits. We consider a complete basis of Hermitian operators $D_i$ in subsystem $D$, which satisfies the orthonormal condition
\be 
\tr \, \{D_iD_j\}=2^d\delta_{ij}. \label{eq_orthonormal}
\ee
Similarly, we define an orthonormal basis $A_i$ in subsystem $A$. 
 As a reminder, our state lives in a $2^n$-dimensional Hilbert space, and $AB$ and $CD$ are two different decompositions of that Hilbert space. Hilbert space $A$ is $2^a$-dimensional, and the operators $A_i$ act on $A$. A similar statement holds for $D$. Both $D$ and $A$ are setup as in Fig.~\ref{circuit}, and $S_D = d$, $S_A = a$. 
If $a = 1$, then one possible basis choice for $A_i$ the three Pauli operators $X, Y, Z,$ and the identity $I$. In general, there are $4^a$ independent operators in $A$. We can think of this as choosing one of the four operators $I,X, Y, Z$ at each site. If the Hilber space decompositions $A$ and $D$ do not overlap, then $[A_i, D_j]=0$ for all $i,j$. However,

As a measure of OTO correlation functions for generic operator choices, we consider the following quantity 
\be
\begin{split}
|\langle \OO_D(t)\,  \OO_A \,  \OO_D(t)\, \OO_A \rangle_\beta| :&= \frac{1}{4^{a+d}}\sum_{ij}  \langle D_i(t)\, A_j \, D_i(t)\, A_j \rangle_\beta \\
&= \frac{1}{4^{a+d}} \cdot \frac{1}{Z}\sum_{ij}  \tr\,  \{e^{-\beta H} D_i(t)\, A_j\, D_i(t)\, A_j \}, \label{corr-sum}
\end{split}
\ee
where the sums $i,j$ run from $1$ to $4^d,4^a$, respectively. Here, $\langle \cdot \rangle_{\beta}$ represents a thermal average, and $|\cdot|$ represents an operator average over the complete bases $D_{i},A_{j}$. Additionally, we will take the infinite temperature limit $\beta=0$  so we don't have to worry about the Euclidean evolution. For $t=0$, every correlator is unity $\langle D_i(0)\, A_j\, D_i(0)\, A_j \rangle_{\beta=0}=1$ 
due to the orthonormal condition and the fact that $A_j$ and $D_i(0)$ commute.

Under chaotic time evolution, $D_i(t)=e^{iHt} D_i e^{-iHt}$ will evolve into a high weight operator and cease to commute with $A_j$. (To see this, consider the BCH expansion for $D_i(t)$ as in \eqref{BCH-expansion}. As $t$ increases, the later terms with high weight will become important. These terms will no longer be confined to subspace $D$.  Under chaotic evolution the $D_i(t)$ will grow to reach $A$.)  This will lead to the decay of the OTO correlation functions $\langle D_i(t) \, A_j \,D_i(t) \, A_j \rangle_{\beta=0}$ for generic $i,j$ (as long as the $D_i$ or the $A_j$ are not the identity in which case $\langle D_i(t) \, A_j \,D_i(t) \, A_j \rangle_{\beta=0}=1$ for all $t$) and all such partitions $A,D$.

At early times, the average will be very close to unity. With chaotic time evolution, the butterfly effect will cause most of the correlation functions in the average to decay exponentially. Using standard techniques, one can relate \eqref{corr-sum} to  the second R\'enyi entropies of the time evolution operator considered as a state
\be
|\langle \OO_D(t)\,  \OO_A \,  \OO_D(t)\, \OO_A  \rangle_{\beta=0}|~=~2^{ n-a-d-S_{AC}^{(2)} }, \label{corr=renyi}
\ee
where $S_{AC}^{(2)}$ is the second R\'enyi entropy of $AC$ defined in \eqref{def-of-renyi}, $2^n$ is the dimension of the input or output Hilbert space, and the subsystems $A$ and $D$ have dimension $2^a$ and $2^d$, respectively. A proof of this result and its generalization to finite temperature is given in Appendix~\ref{appendix-proof}. 

At first glance, it is a little surprising that a R\'enyi entropy appears here: the entropy determines mutual information, which bound two-point functions, and chaos is distinctly measured by OTO four-point functions. However, the key point is that when $A$ and $D$ are small (so that they only contain approximately local operators), $B$ and $C$ contain highly nonlocal operators covering almost the entire input and output systems, respectively. As a result, $S_{AC}^{(2)}$ is sensitive to correlations between the few operators in $A$ and the complete (and nonlocal and high weight) set of operators in $C$. 

The OTO average \eqref{corr=renyi} is an operator-independent information-theoretic quantity that is constrained by chaotic time evolution. To understand its behavior, let's consider its maximum and minimum values. The average will be the largest at $t=0$, when all the correlators are unity. On the other hand, the average will be minimized when the R\'enyi is maximal: $\max S_{AC}^{(2)} = \min(a+c,\, d+b)$. 
Let us assume for the rest of this section that $a\le d$, so $\max S_{AC}^{(2)} = a+c$. This means the OTO average is bounded from below by $4^{-a}$. (It's worth mentioning that the Haar-scrambled value of the average is generally larger than this lower bound.\footnote{Since \eqref{corr-sum} includes $4^d + 4^a - 1$ terms where $A_j = I$ or $D_i = I$ (one for each term where $A_j = I$ or $D_i = I$, and minus one to prevent overcounting when they both are), if all the non-identity correlation functions decay the OTO average will be $4^{-a} + 4^{-d} - 4^{-a-d} > 4^{-a}$. Using the results from Appendix~\ref{appendix-haar}, we can show that this is larger value is exactly the Haar-scrambled value of the OTO average. To get lower value, some of the correlation functions need to cross zero so that the operators are negatively correlated.}) 
Therefore, we see the OTO average is bounded by
\be
4^{-a} \le |\langle \OO_D(t)\,  \OO_A \,  \OO_D(t)\, \OO_A \rangle_{\beta=0}|  \le 1.\label{corr=renyi-range}
\ee 

Now, we will recast this result in terms of mutual information in order to make a connection to our scrambling diagnostic. Let's assume that after a long time of chaotic time evolution the OTO average asymptote to a small positive constant $\epsilon$. This means that the entropy $S_{AC}$ is bounded:
\be
S_{AC} \ge S_{AC}^{(2)} = n - a -d - \log_2 \epsilon,  
\ee
where in the first part we used the fact that $S_R^{(i)} > S_R^{(i+1)}$ for R\'enyi entropies, and in the second part we used \eqref{corr=renyi}. In terms of mutual information, we have
\be
I(A:C) \le  2a +  \log_2 \epsilon, \label{mutual-information-main-point}
\ee
where here we have used the fact that $S_A$ and $S_C$ are always maximally mixed.

Eq.~\eqref{mutual-information-main-point} is one of our main results. At $t=0$, $I(A:C) = 2a$. The information about the input to the channel in $A$ is $2a$ bits and that information is entirely contained in the output subsystem $C$. Since there are $4^a$ linearly independent basis operators in $A$'s Hilbert space, we can interpret these $2a$ bits as the information about which of the $4^a$ operators was input into the channel. (For instance, if $a=1$, it takes two bits to index the operators $I, X, Y, Z$.) Under chaotic time evolution, $\epsilon \ll 1$, and the mutual information between $A$ and $C$ becomes small. The smallest possible value for $\epsilon$ is $2^{-2a}$, which occurs when $I(A:C) = 0$. In practice, there is always residual information between $AC$. Using the results from Appendix~\ref{appendix-haar}, we see that for Haar scrambling the mutual information can be bounded as
\be
I(A:C)_{Haar} \le \log_2(1 + 4^{a-d} - 4^{-d}),
\ee
which corresponds to all the non-identity terms in the OTO average decaying to zero. If the information-theoretic quantities constructed from the second R\'enyi approach their Haar-scrambled value, then all the nontrivial OTO correlators $\langle D_i(t) \, A_j \,D_i(t) \, A_j \rangle_{\beta=0}$ must approach zero.

Next, we note that \eqref{mutual-information-main-point} implies 
\be
I(A:C_\alpha)\le  2a +  \log_2 \epsilon,
\ee
for any partitioning of $C= C_{\alpha} \cup C_{\bar{\alpha} }$. This can be seen from subadditivity
\be
S_{AC_\alpha} + S_{AC_{\bar{\alpha}} } \ge S_{AC}, \label{subadditivity-dividing-system}
\ee
and the definition of mutual information. (This is also intuitive: any information contained about region $A$ in region $C$ must necessarily be more than or equal the information in a partition of $C$.) Therefore, we learn that in chaotic channels, local information in the input must get delocalized in the output (i.e. cannot be recovered in an output subsystem smaller in size than the total system $n$). Since the partitions $A$ and $C$ were completely arbitrary, we conclude that the decay of OTO correlators implies that all bipartite mutual informations are small. If the OTO correlator average is given by $\epsilon$ after a long time, then we have
\begin{eqnarray}
I(A:C),~I(A:D)&\leq &2a+\log_2 \epsilon\nonumber\\
I_3(A:C:D)&\leq & 2a+2\log_2\epsilon=-2a+2\log_2\frac{\epsilon}{\epsilon_{\rm min}}
\end{eqnarray}
When $\epsilon$ approaches the minimum value $\epsilon_{\rm min}=2^{-2a}$, $I_3$ approaches the most negative value $I_{3, \rm min}=-2a$. From this discussion, we conclude that the butterfly effect implies scrambling.

\subsection{Early-time behavior}\label{sec:rate-of-loss}

In this section, we will attempt to connect the universal early-time behavior of OTO correlators in strongly chaotic systems with the information-theoretic quantities we use to diagnose scrambling.

In strongly chaotic systems, all OTO correlation functions of operators with nontrivial time evolution will decay to zero. However, the behavior of the OTO correlation function $\langle W(t) \, V \, W(t) \, V \rangle_\beta$ as it asymptotes to zero is not universal. The approach will depend on the specific choices of operators $W, V$. For instance, in two-dimensional CFTs with large central charge and a sparse low-lying spectrum at late times the OTO correlator decays as 
\be
\frac{\langle W(t) \, V \, W(t) \, V \rangle_\beta}{\langle W W \rangle_\beta \, \langle V V \rangle_\beta}~\sim~e^{-4\pi h_v t / \beta},
\ee
where $h_v$ is the conformal weight of the $V$ operator, and it is assumed $1 \ll h_v \ll h_w$ \cite{Roberts:2014ifa}.

On the other hand, at early times the behavior of $\langle W(t) \, V \, W(t) \, V \rangle_\beta$ usually takes a certain form. The initial decay is known to fit the form
\be 
\frac{\langle W(t) \, V \, W(t) \, V \rangle_\beta}{\langle W W \rangle_\beta \, \langle V V \rangle_\beta}~\approx~1\,-\,\#e^{\lambda_L t} + \dots, \label{eq:early-time-OTO-corr-behavior}
\ee 
where in analogy to classically chaotic systems $\lambda_L$ has the interpretation of a Lyapunov exponent \cite{kitaev}.\footnote{However, the analogy is imprecise. In weakly coupled systems,  $\lambda_L$ has a semiclassical analog that does not map onto the classical Lyapunov exponent. Despite this, we will follow convention and refer to $\lambda_L$ as a Lyapunov exponent. We are grateful to Douglas Stanford for emphasizing this point. } In \cite{Maldacena:2015waa}, it was shown that quantum mechanics puts a bound on $\lambda_L$
\be
\lambda_L \le \frac{2\pi}{\beta}  \label{bound-on-chaos},
\ee
with saturation for strongly-interacting conformal field theories that have holographic descriptions in terms of Einstein gravity. This Lyapunov exponent is expected to be universal---independent of the choice of operators $W, V$---and as such any model that saturates the bound \eqref{bound-on-chaos} is expected to be a toy model of holography \cite{kitaev,kitaev2}. (In \S\ref{sec:numerics}, we will use numerics to explore the Majorana fermion model proposed in Ref. \cite{kitaev2} that in a certain limit is expected to have this property.)

Since the partitioning of the inputs into $A,B$ and outputs into $C,D$ was entirely arbitrary, let's first consider channels that operate on $0$-dimensional systems, e.g. fast scramblers in the sense of \cite{Sekino:2008he}, such as the Majorana fermion model of Kitaev \cite{kitaev2} (a simplification of the Sachdev-Ye model of $N$ $SU(M)$ spins \cite{sachdev1993gapless}, see also \cite{Sachdev:2015efa}) or a large $N$ strongly interacting CFT holographically dual to Einstein gravity near its Hawking-Page point.\footnote{We require this limit so we can think of the black hole as unit sized and not yet worry about the operator growth in spatial directions.} These systems still have low-weight $k$-local Hamiltonians wth $k \ll N$, but each degree of freedom interacts with every other degree of freedom. For these systems,  the OTO correlation functions decay as 
\be
\langle W(t) \, V \, W(t)\, V \rangle_\beta  = f_0 -  \frac{f_1}{N^2} e^{\lambda_L t} + O(N^{-4}), \label{corr-decay}
\ee
where $N^2=n$ is the total number degrees of freedom, and the constants $f_0$ and $f_1$ depend on properties of the $W,V$ operators (e.g. their CFT scaling dimensions) \cite{Maldacena:2015waa}. The decay of the correlator is delayed by the large number of degrees of freedom at time $t_* = \lambda_L^{-1} \log N^2$. This is usually referred to as the scrambling time \cite{Sekino:2008he}. Plugging \eqref{corr-decay} into \eqref{mutual-information-main-point}, we find that at early times the mutual information between $A$ and $C$ is bounded as
\be
I(A:C) \le 2a - \# e^{\lambda_L (t-t_*)} + \dots.\label{corr-decay-fast-scrambling}
\ee
Thus, the information between $A$ and $C$ must begin to decay by the scrambling time $t_*=\lambda_L^{-1} \log N^2$. This inequality would be an equality if we instead considered the mutual information constructed from the second R\'enyi $S_{AC}^{(2)}$.

Now, let's consider systems arranged on a spatial lattice but do not have a large number of degrees of freedom per site, e.g. spin chains. For these systems, the butterfly effect implies ballistic growth of operators in spatial directions \cite{Roberts:2014isa}. For local operators $W$ and $V$ separated by large distance $|x| \gg \beta$, in many known examples (such as holography and numerical investigations of one-dimensional spin chains) strong chaos implies that OTO correlation functions decay as
\be
\langle W(t) V W(t) V \rangle_\beta = f_1' -  f_2' \, e^{\lambda_L (t - |x|/v_B)} + O(e^{- 2|x| \lambda_L /v_B}),\label{corr-decay-butterfly-velocity}
\ee
with additional constants $f_1', f_2'$ that depend on the details of the operators.\footnote{As emphasized in \cite{Maldacena:2015waa}, the butterfly effect is relevant for systems with a large hierarchy of scales. For the $0$-dimensional systems we just considered, the hierarchy is provided by the parametric difference between the thermal time $\beta$ and the fast scrambling time $\beta \log N^2$. In the present case, the role of the large scale is instead being played by the large spatial separation $|x|$ between the operators.} 
In this case, the early-time decay of the correlator is suppressed by the large spatial separation between the operators. 
Under chaotic time evolution, the operator $W(t)$ will grow ballistically with some characteristic ``butterfly'' velocity we denote $v_B$.\footnote{The velocity $v_B$ can depend details of the theory that do not affect $\lambda_L$. For instance, it is modified in Gauss-Bonnet gravity \cite{Roberts:2014isa} and for certain Einstein gravity theories can even acquire a temperature dependence \cite{RobertsSwingleVBUpcoming}.
} Thus, $W(t)$ and $V$ will commute until a time $t > |x| / v_B$ when $V$ enters the ``butterfly'' light cone of $W$. Let's focus on a lattice of spins in $d$-spatial dimensions. We will pick our subsystem $A$ to be a ball of $a$ sites surrounding the origin with a radius $r_a$. We will pick $C$ to also be a ball surrounding the origin with a radius $r_c$ such that $r_c - r_a = |x|$. Then, after a scrambling time of $t_* = v_B t$, the mutual information between $A$ and $C$ must begin to decay 
\be
I(A:C) \le 2a - \# e^{\lambda_L (t -|x|/v_B)} + \dots.
\ee

Since we will study this quantity in \S\ref{sec:numerics}, we note that we can directly equate (rather than bound) the behavior of the second R\'enyi entropy to the Lyapunov behavior of the OTO correlators. Let's restrict to a one-dimensional spin chain, and take $A$ to be the first spin of the input and $D$ to be the last spin of the output. In that case, if we assume a form of the correlator \eqref{corr-decay-butterfly-velocity}, plug into \eqref{corr-sum} and compute the average, then we find
\be
S_{AC}^{(2)}(t) = S_{AC}^{(2)}(0) \, + \, \# e^{\lambda_L t} + \dots,\label{eq:lyapunov-second-renyi}
\ee
showing that at early times the R\'enyi entropy can grow exponentially with characteristic Lyapunov exponent $\lambda_L$.\footnote{This assumes that $\lambda_L$ is independent of the choice of operators $W$ and $V$ and that the ansatz \eqref{corr-decay-butterfly-velocity} is the correct form for the initial decay of the correlator. Both of these assumptions are not necessarily true for some spin systems. Additionally, if the constant in front of the exponential is not small (for example, in holographic systems), then the expansion will not be valid and one cannot see the exponential growth behavior. We thank Tarun Grover and Douglas Stanford for emphasizing these points.} We will roughly see this behavior in Fig.~\ref{fig:spin-chains}.

\subsection{Butterfly velocity vs. entanglement velocity}
There are two nontrivial velocities relevant to the growth of information-theoretic quantities in unitary quantum channels arranged on a lattice.

The butterfly velocity $v_B$ \cite{Shenker:2013pqa,Roberts:2014isa} is the speed at which the butterfly effect propagates. It is the speed at which operators grow under chaotic-dynamics. Such behavior is reminiscent of the Lieb-Robinson bound on the commutator of local operators separated in time for systems with local interactions \cite{Lieb:1972wy,Hastings:2005pr,hastings2010locality} and suggests identifying $v_B$ with the Lieb-Robinson velocity.
The butterfly velocity is often difficult to compute directly, but in holographic theories with Einstein gravity duals it is known to be \cite{Shenker:2013pqa}
\be
v_B = \sqrt{\frac{d}{2(d-1)}},  \qquad \mathrm{(Einstein~gravity)},\label{eq:holographic-vB}
\ee
 where $d$ is the spacetime dimension of the boundary CFT. 
 This value is modified in Gauss-Bonnet gravity \cite{Roberts:2014isa} and for certain theories can even acquire a temperature dependence \cite{RobertsSwingleVBUpcoming}.

The entanglement velocity (sometimes called the tsunami velocity) studied in \cite{Hartman:2013qma,Liu:2013iza,Liu:2013qca,Leichenauer:2015xra,Casini:2015zua} is often described as the rate at which entanglement spreads. It is rate of growth of entanglement entropy after a quench, and in holographic systems dual to Einstein gravity it can be computed directly \cite{Hartman:2013qma,Liu:2013iza,Liu:2013qca}
\be
v_E = \frac{\sqrt{d}(d-2)^{\frac{1}{2} - \frac{1}{d}}}{[2(d-1)]^{1-\frac{1}{d}}},  \qquad \mathrm{(Einstein~gravity)}, \label{eq:holographic-vE}
\ee
a different nontrivial function of the CFT spacetime dimension $d$.
In these theories, $v_E \le v_B$, and $v_B = v_E = 1$ for $d=1+1$.

\begin{figure}
\begin{center}
\includegraphics[width=.8\linewidth]{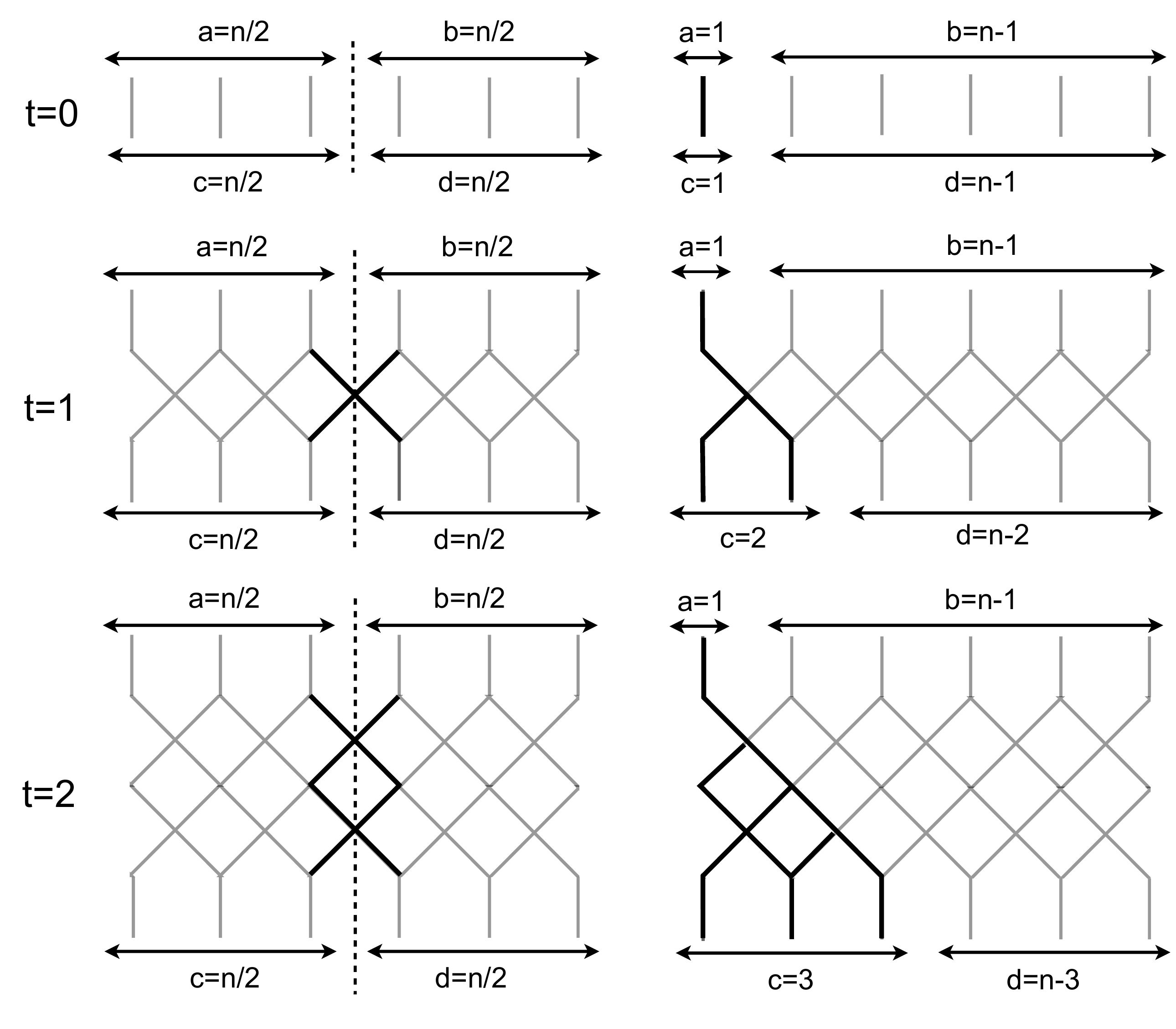}
\caption{A one-dimensional tensor network model of a channel with increasing vertical network depth $t$. This makes it clear that $v_E$ is related to vertical propagation and $v_B$ is related to horizontal propagation.
{\bf Left:} $I(A:C)$ decreases by $2$ with every increase in circuit depth. By measuring time in terms of the network depth, the entanglement velocity is trivial ($v_E=1$). 
{\bf Right:} the ``information'' or ``butterfly'' light cone of the input is controlled by the butterfly velocity ($v_B=1$). The information contained in this light cone is constant: $I(A:C(t)) =2a$.
}\label{fig:vb-and-ve}
\end{center}
\end{figure}

In the context of unitary quantum channels, these velocities have a very specific interpretation in terms of different mutual informations, see Fig~\ref{fig:vb-and-ve}. Consider a lattice of $n$ degrees of freedom and divide the input up such that $a=b=n/2$, the output such that $c=d=n/2$, and such that all subsystems are contiguous. If $A$ and $C$ are aligned such that at $t=0$, $I(A:C)=n$, then under time evolution (for chaotic \emph{and} integrable systems!) it is expected that for a long stretch of time that the mutual information will decrease linearly as
\be
I(A:C) = n -  v_E s t,\label{eq:information-flow-quench}
\ee 
until near when it saturates at $I(A:C) =0$. This is often referred to as a quench, which we discuss in depth in Appendix~\ref{sec:memory-effect} in the context of CFT. Here, $s$ is the ``entropy density,'' which converts $v_E$ from a spatial velocity (with units lattice-sites/time) to an information rate (bits/time). For spin systems, the state representation of the channel requires two qubits per addition to the spatial lattice (one to represent the input leg and one to represent the output leg), so $s=2$.

While this tells us about the accumulation of entanglement between $AC$ and $BD$, it tells us nothing about how information propagates in the spatial directions through the channel. To study the latter, we need a different configuration of subsystems. First, let's pick $A$ to be a small region $a=O(1)$ center at the origin. Then, instead of considering a fixed radius $C$, let us let pick subsystem $C$ to be a growing ball of radius $r_c = v_B t$ surrounding the origin in the output system. 
Then, the information between the input $A$ and output $C(t)$ will be constant
\be
I(A:C(t)) = 2a,
\ee
but the output is growing ballistically as a ball of radius $v_B t$. Thus, $v_B$ is the rate that information propagates spatially in unitary quantum channels.\footnote{A similar interpretation of the butterfly velocity was recently made by Stanford the context of holography. It was shown that the reconstruction of a particle falling into a black hole can occur for a region that grows ballistically with velocity $v_B$ \cite{scramblingEntanglementWedge}.
} For finite systems, eventually $C(t)$ will grow to encompass the entire output of the channel, and the information will stay delocalized until the recurrence time. Before then, any local measurement of any subsystem of the output will be insufficient to recover the information in $A$. The information is scrambled.

In Fig.~\ref{fig:vb-and-ve} , a tensor network model of a channel exhibits both of these velocities. The left side shows the linear decrease of $I(A:C)$ with increasing circuit depth in the vertical direction. Every increase in depth is accompanied by a decrease in $I(A:C)$ of $v_E s \Delta t$ bits, with $\Delta t$ the time interval corresponding to each step of the circuit. The bold tensors make it clear that this decrease is a simple consequence of the increase of circuit depth. In a gate or tensor model of a channel, $v_E$ is simply a conversion in units between the network's depth and conventional time. More generally, it is the velocity associated with the ``vertical'' direction of the channel. 

The right side of Fig.~\ref{fig:vb-and-ve} shows the growth of the region $C(t)$ required to capture all of the information from the input $A$. This has a natural interpretation of a light cone for information propagation and can be very different depending on the properties of the channel. This velocity is associated with the ``horizontal'' direction of the channel. In the simple tensor network picture and $1+1$-dimensional CFT, $v_E = v_B = 1$, but more generally $v_E \le v_B$ as in holographic Einstein gravity \eqref{eq:holographic-vB}-\eqref{eq:holographic-vE}.

\section{Numerics in qubit channels}\label{sec:numerics}

In this section, we will study chaos and scrambling in qubit channels. We will show explicitly that the tripartite information is a simple diagnostic of scrambling,  and we will verify our main result \eqref{corr=renyi} by directly showing that the butterfly effect implies scrambling. Finally, we will also comment on the relationship between the size of the subsystems and the expected behavior of the entropies in the channels.

We will study four different channels. First, we will directly compare two $1$-dimensional Ising spin chains, one integrable and one strongly chaotic, both of which can be expressed with the Hamiltonian
\be
H_{sc} = -\sum_i Z_i Z_{i+1} + gX_i + hZ_i.\label{eq:spin-chain-H}
\ee
Here, $i=1,\dots n$, and $X_i, Y_i, Z_i$ are the Pauli operators on the $i$th site. The integrable system we study is the transverse field Ising model with $g=1,~h=0$, and the chaotic system has $g=-1.05,~h=0.5$ \cite{banuls2011strong}. In our numerics, we will take $n=7$ when studying subsystems of unequal sizes and $n=6$ when studying subsystems of equal sizes. For these systems, we will also compute the velocities $v_E$ and $v_B$.

The second model we will study is a $0$-dimensional fast scrambler, the Majorana fermionic system of Kitaev \cite{kitaev2} (see also \cite{sachdev1993gapless,Sachdev:2015efa}) with Hamiltonian
\be
H_{K} = \sum_{j<k<\ell <m} J_{jk\ell m} \, \chi_j \chi_k \chi_\ell \chi_m, \qquad \overline{J_{jk\ell m}^2} = \frac{3!}{(N-3)(N-2)(N-1)} J^2, \label{eq:kitaev-model}
\ee
where the $\chi_j$ are spinless Majorana fermions, and $j,k,\ell,m=1,\dots N$, $J_{jk\ell m}$ are random couplings with mean zero and variance $\overline{J_{jk\ell m}^2} $.\footnote{This is slightly different than in \cite{kitaev2}, 
because we are studying the system in the limit $\beta J = 0$, while the system in \cite{kitaev2} is considered in the limit $\beta J = \infty$. Thus, we don't expect the Lyapunov exponent to saturate the chaos bound \eqref{bound-on-chaos}, and instead we expect it to be proportional to the coupling $J$.} We can study Majorana fermions using spin chain variables by a nonlocal change of basis known as the Jordan-Wigner transformation
\be
\chi_{2i-1} = \frac{1}{\sqrt{2}}\, X_1 X_2 \dots X_{i-1} Z_i, \qquad \chi_{2i} = \frac{1}{\sqrt{2}}\, X_1 X_2 \dots X_{i-1} Y_i,
\ee
such that $\{\chi_j, \chi_k\} = \delta_{jk}$. With this representation, $N$ Majorana fermions require $N/2$ qubits. In our numerics, we will take $N=14$ and $J=1$.

Finally, we will consider scrambling with a Haar-random unitary as a baseline for a scrambled system. For scrambling channels, at late times the quantities we use to diagnose scrambling often don't reach their extremal values, but rather asymptote to Haar-random values. In Appendix~\ref{appendix-haar}, we analytically compute the second R\'enyi entropies of the Haar-random unitary channel considered as a state. To study the mutual informations, we will numerically sample from the Haar ensemble. For the size of the channels we will study ($\dim U = 2^6 \times 2^6$ or  $\dim U = 2^7 \times 2^7$), these properties are self-averaging and only require one sample.

\subsection{Spin chains}

\subsubsection*{Unequal sized subsystems}

\begin{figure}[htp]
\begin{center}
\includegraphics[scale=.38]{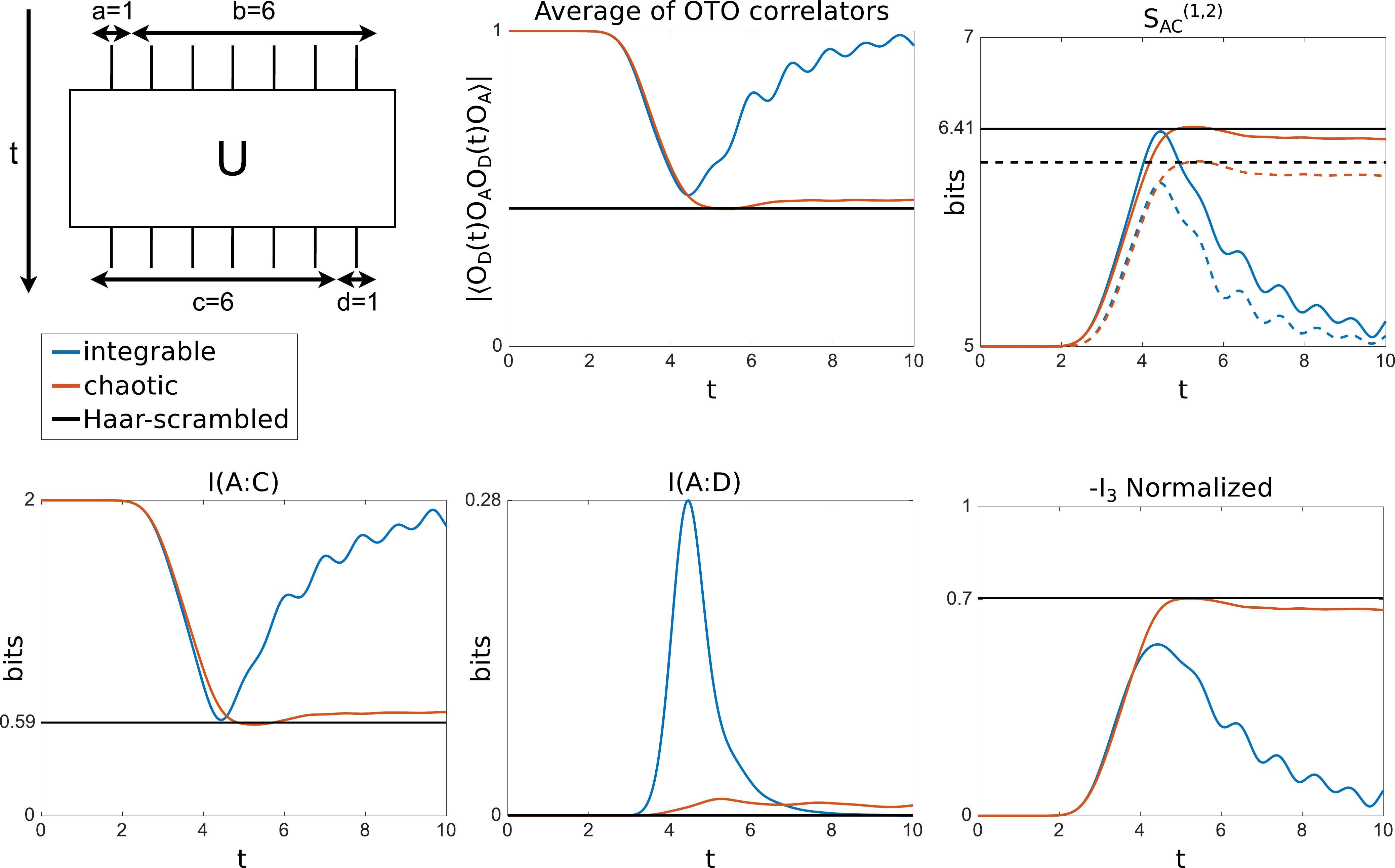}
\end{center}
\caption{Chaos and scrambling in the integrable (blue; $g=1, h=0$ Hamiltonian) and chaotic (orange; $g=-1.05, h=0.5$ Hamiltonian) spin chain unitary channels and the Haar-random channel (black). 
{\bf Top Left:} configuration of channel ($n=7$ spins; input subsystems of $a=1$, $b=6$ spins; output subsystems of $c=6$, $d=1$ spins). For the spin chains, the channel is the time evolution operator ($U=e^{-iHt}$). For the Haar-channel, $U$ is sampled from the Haar ensemble.  
{\bf Top Middle:} the average of OTO correlators shows the butterfly effect. At later times, the chaotic system asymptotes to the Haar-scrambled value, but the integrable system doesn't remain decorrelated. 
{\bf Top Right:} $S_{AC}$ (solid) and $S^{(2)}_{AC}$ (dotted) shows roughly the same behavior as the OTO average. 
{\bf Bottom Left:} $I(A:C)$, a trivial function of $S_{AC}$, shows that for the chaotic channel an initial $2$ bits of information between the subsystems gets delocalized so that at late times only a small amount ($0.59$ bits) remains. We can also read off the butterfly velocity, $v_B = 2.5$.
{\bf Bottom Middle:} the spike in $I(A:D)$ for the integrable channel shows that information is not delocalized by integrable time evolution. For the chaotic and Haar channels information is delocalized, and $I(A:D)$ is always near vanishing.
{\bf Bottom Right:} The negative of the tripartite information normalized by its maximum value ($2$ bits) is a simple diagnostic of scrambling. 
}
\label{fig:spin-chains}
\end{figure}

Our first setup for the spin chains is shown in the top-left corner of Fig.~\ref{fig:spin-chains}, which contains $7$ spins with open boundary condition. We study the $2^7 \times 2^7$-dimensional unitary operators as states
\be
| U \rangle = \frac{1}{2^7} \sum_{i=1}^{2^7}  \ket{i}_{AB} \otimes U \ket{i}_{CD},
\ee
where $AB$ is the input to the channel and $CD$ is the output. In the input, we take $A$ to be the first spin, and $B$ to be the final six spins. Similarly, we take $C$ to be the first six spins, and $D$ to be the last spin. This puts $A$ and $D$ at maximal spatial separation. For the spin chains, we take $U=e^{-iH_{sc}t}$, with $H_{sc}$ given by \eqref{eq:spin-chain-H}. As a comparison, we consider a Haar channel with $U$ a Haar random unitary operator. 
Since we are considering uniform input, the inverse temperature is always taken to be vanishing ($\beta=0$).

Our spin chain numerics for this unequal subsystem setup are shown in the rest of Fig.~\ref{fig:spin-chains}. In the top-middle panel, we show the average over OTO correlation functions $|\langle \OO_D(t)\,  \OO_A \,  \OO_D(t)\, \OO_A \rangle_{\beta=0}|$. Of course, from \eqref{corr=renyi} we know this is also equal to $2^{n-a-d-S_{AC}^{(2)}}$, which in this case can be verified explicitly. The chaotic spin chain asymptotes to just above the Haar-scrambled value, which can be computed from Appendix~\ref{appendix-haar} and is equal to $4^{-a} + 4^{-d} - 4^{-a-d} = 7/16$. This is also the value given by assuming all the correlation functions in the average where neither an operator in $A$ nor an operator in $D$ are the identity decay to zero. (There are $9$ such terms, with $16$ total correlation functions in the sum.) The Haar-scrambled value is far above the absolute minimum $4^{-a}=1/4$ of OTO correlator average, given by plugging the maximum possible value for the second R\'enyi $(\max S_{AC}^{(2)}=7)$ into \eqref{corr=renyi}. (To reach this value, some of the correlators would have to become negative.) Additionally, we see Lyapunov behavior of the OTO correlator decay beginning around $t=2$. At later times (around $t=4$), the integrable system does not asymptote to the Haar-scrambled value but instead has a recurrence and recorrelates.

We see the similar behavior in the other related quantities computed on subsystem $AC$. In the top-right panel of Fig.~\ref{fig:spin-chains}, we plot the entropies $S_{AC}$ and $S^{(2)}_{AC}$, and in the bottom-left panel we plot $I(A:C)$. The entropies $S^{(2)}_{AC}$ are directly related to the OTO average by \eqref{corr=renyi}, and $I(A:C)=a+c - S_{AC}$. (As a reminder $a = 1$ and $c  = 6$ are constant.) At early times we see roughly Lyapunov behavior of the second R\'enyi entropy as suggested from \eqref{eq:lyapunov-second-renyi}. In all of these curves, the quantity is roughly unchanged until a time of order $c$ and then begins to decay exponentially. As expected, $S^{(2)}_{AC}<S_{AC}$, and these entropy curves have roughly the same shape. Also as expected, localized information between $A$ and $C$ gets delocalized by chaos. For the chaotic channel, $I(A:C)$ begins at $2$ bits (representing the four different choices of input operators) and then decays to just above the Haar-scrambled value of roughly $0.6$ bits. Using the results from Appendix~\ref{appendix-haar} and  \eqref{Haar-random-IAC}, we can bound this late-time residual information to be less than $0.8$ bits
 \be 
 I(A:C)_{Haar} \le 1 + \log_2 (7/8) \approx 0.8 ~ \textrm{bits}.
 \ee
 For the integrable channel, the information doesn't get delocalized and $I(A:C)$ returns close to its initial value of $2$ bits. The decay in both cases is delayed until roughly $t=2$. The ratio of this delay to the distance between $A$ and $D$ lets us extract the butterfly velocity for these chains ($v_B = 2.5$).

 In the bottom-middle panel of Fig.~\ref{fig:spin-chains}, we plot $I(A:D)$. For the chaotic channel, this quantity can never become very large: under time evolution the information in $A$ gets spread across \emph{all} the degrees of freedom, so there can never be significant localized information about the input $A$ in the output subsystem $D$. From Appendix~\ref{appendix-haar}, we also note that the Haar-scrambled $I(A:D)$ is  exponentially small in the overall system size (Eq.~\eqref{Haar-random-IAD}). However, the integrable channel has a very sharp peak after a time of order $c$. In Appendix~\ref{sec:memory-effect}, we explain in the context of  CFT that this memory effect is the failure of integrable systems to efficiently delocalize  information---i.e. scramble---due to entanglement propagation by noninteracting quasi-particles. This peak ($0.28$ bits) is far from the maximum ($2$ bits), but corresponds precisely to the point in the bottom-left panel plot of $I(A:C)$ where the integrable channel is loosing information between $A$ and $C$. This supports our hypothesis of integrable channels moving around localized information rather than actually scrambling.

Finally, we plot the negative of the tripartite information in the bottom-right panel of Fig.~\ref{fig:spin-chains}, normalized by its maximal value ($-I_3(A:C:D)/2a = -I_3(A:C:D)/2$). As explained in \S\ref{sec:tripartite}, as a measure of multipartite entanglement this quantity is a simple diagnostic of scrambling; the chaotic channel asymptotes to the Haar-scrambled value of $I_3$ after a time $O(n)$. The integrable channel initially has an increase in $I_3$, but never reaches the Haar-scrambled value (due to the memory spike in $I(A:D)$) and instead has a recurrence beginning after a time of order $O(n)$. A long time average of this quantity would make it clear that the integrable channel doesn't scramble.

 As a final point, the fact that the Haar-scrambled value of the normalized negative $I_3$ is less than unity ($-I_3/2 \approx 0.7$) is not a relic of small $n$. This ``residual'' information can be bounded using the results from Appendix~\ref{appendix-haar} and depends only on the subsystem sizes
\be
\frac{-I_3(A:C:D)_{Haar}}{2a} \ge 1 - \frac{1 + \log_2(1-2^{-a-d-1})}{2a}.\label{eq:residual}
\ee
Even for large systems, the residual information vanishes like $1/2a$. The point is that  for $1 \ll a,d <b,c$, there's still always roughly one bit of information between $A$ and $C$ that doesn't delocalize across the entire output. 

\subsubsection*{Equal sized subsystems}

\begin{figure}[htp]
\begin{center}
\includegraphics[scale=.38]{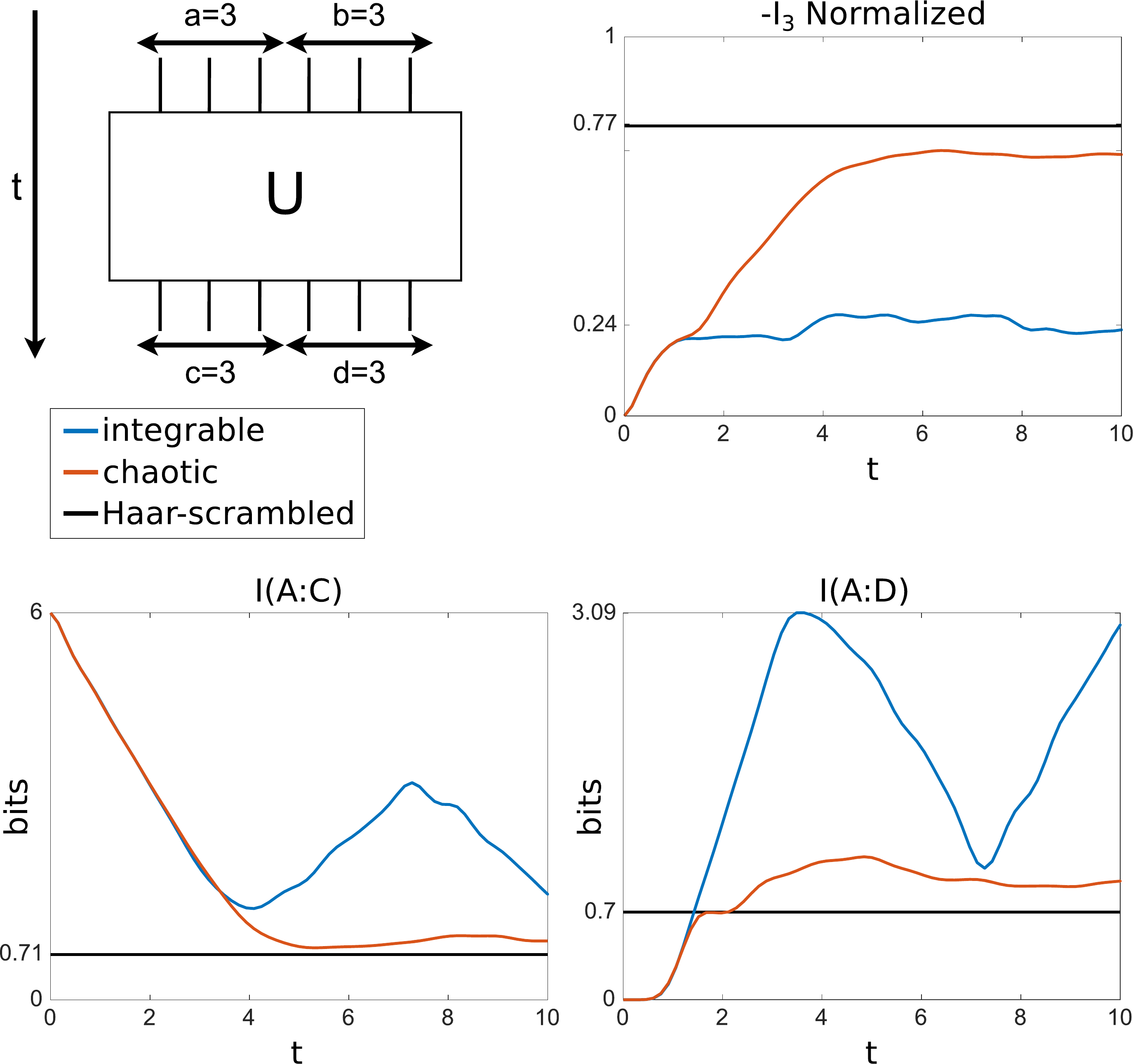}
\end{center}
\caption{Chaos and scrambling in the integrable (blue; $g=1, h=0$ Hamiltonian) and chaotic (orange; $g=-1.05, h=0.5$ Hamiltonian) spin chain channels and the Haar-random channel (black). 
{\bf Top Left:} configuration of channel ($n=6$ spins; input subsystems of $a=3$, $b=3$ spins; output subsystems of $c=3$, $d=3$ spins). For the spin chains, the channel is the time evolution operator ($U=e^{-iHt}$). For the Haar-channel, $U$ is sampled from the Haar ensemble.  
{\bf Top Right:} The negative of the tripartite information normalized by its maximum value ($6$ bits) is a simple diagnostic of scrambling.
{\bf Bottom Left:} shows the linear decrease of $I(A:C)$ in time with characteristic speed $v_E$ known as the entanglement or tsunami velocity.
{\bf Bottom Right:}  the spike of the unequal subsystem configuration (Fig.~\ref{fig:spin-chains}) is broadened to linear increase followed by decrease in $I(A:D)$ for the integrable channel.
}
\label{fig:spin-chain-unequal-numerics}
\end{figure}

For comparison, we also consider the same spin chains, but with $n=6$ and equal sized subsystems of $a=b=c=d=3$ spins.  $A$ and $D$ are still taken to be opposite ends of the chain, but their boundaries are no longer separated spatially. This setup is shown in the top-left panel of Fig.~\ref{fig:spin-chain-unequal-numerics}. As noted before, this is the setup often used when quenches are discussed. Here, we only plot the quantities that have interesting differences from the previous configuration.

In the top-right panel, we plot the normalized negative tripartite information $-I_3/6$. Interestingly, for the integrable channel the tripartite information does not exhibit a recurrence, but rather saturates at a very low value (roughly at $0.24$, compared to $0.7$ for the chaotic channel and $0.77$ for the Haar-random channel). Thus, in this configuration of subsystems it appears to be a robust measure of scrambling that doesn't require any long time average. 

In the lower-left panel, we see that $I(A:C)$ for the integrable and chaotic channels have near identical behavior until just before $t=4$. This is the expected linear growth decrease in mutual information with velocity $v_E$
\be
I(A:C) =6 -  v_E s t,
\ee
where we have $s=2$ for our spin chain channels, and we can read off $v_E = 0.625$. The relationship between $v_E$ and $v_B$ for these spin chains is consistent with results in holographic systems in the sense that $v_E \le v_B$, but it curious that these $1+1$-dimensional channels have $v_E \neq v_B$. Finally, unlike the previous subsystem configuration, the linear decrease of $I(A:C)$ begins immediately since the distance between subsystems $A$ and $D$ is $0$ spins.

In the lower-right panel, we see that $I(A:D)$ for both spin chains also lay on top of each other, but only  until just before $t=2$. At this time, the chaotic channel saturates just above the Haar-scrambled value, while the integrable channel begins a rough pattern of linear growth to $3$ bits followed by return to the Haar-scrambled value. This growth and collapse is the analog of the memory spike we discuss in Appendix~\ref{sec:memory-effect}, broadened by the fact that the spatial separation between $A$ and $D$ ($0$ spins) is no longer larger than the size of the regions ($3$ spins) as required to see the spike.

\subsection{Majorana fermion fast scrambler}
\subsubsection*{Unequal sized subsystems}

\begin{figure}[htp]
\begin{center}
\includegraphics[scale=.38]{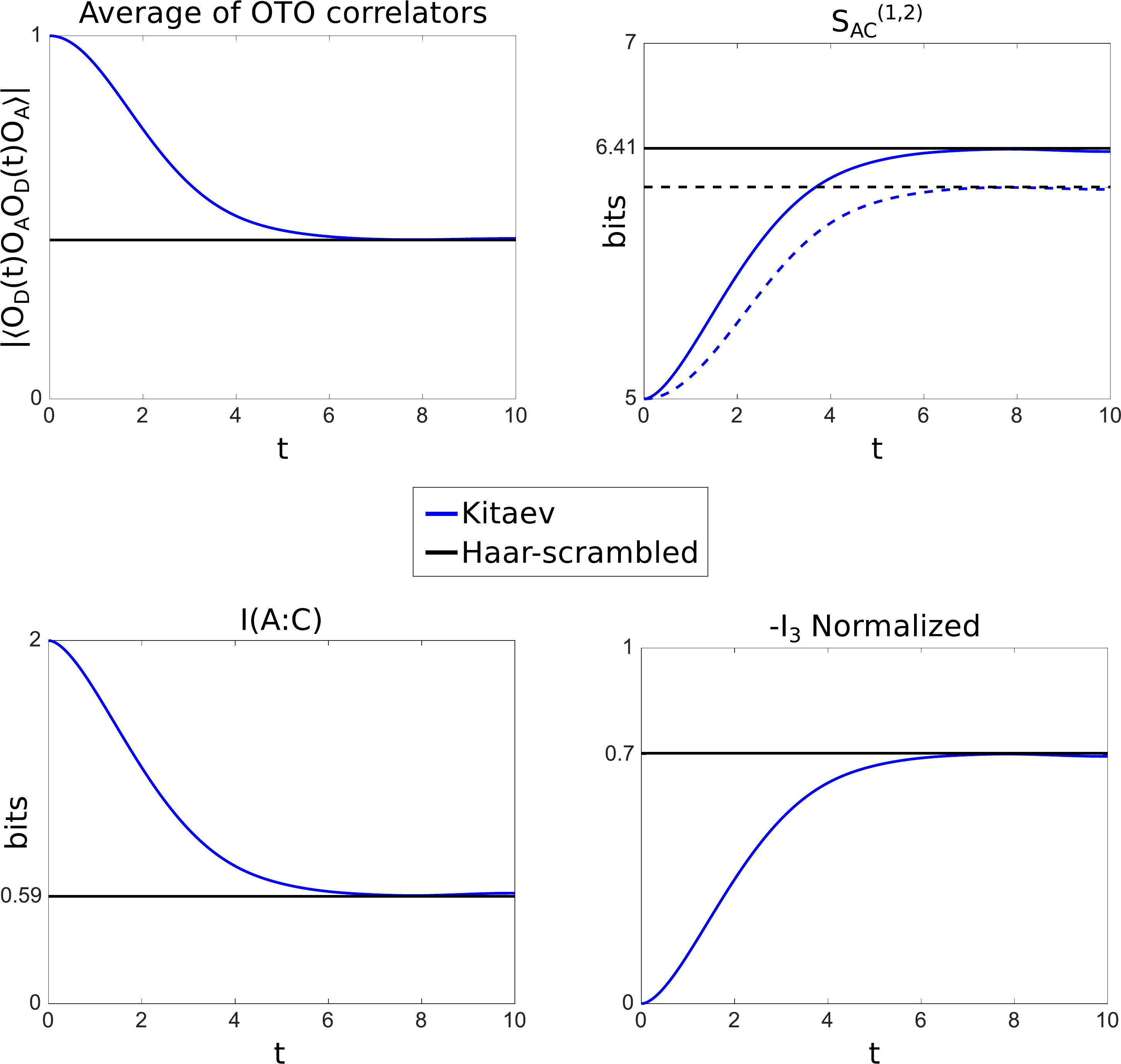}
\end{center}
\caption{Chaos and scrambling in the $0$-dimensional Majorana fermion fast scrambler unitary channel (blue) and Haar-channel (black) with $n=7$ spins; input subsystems of $a=1$, $b=6$ spins; output subsystems of $c=6$, $d=1$ spins.  
{\bf Top Left:} the average of OTO correlators decays immediately, showing the butterfly effect.
{\bf Top Right:} $S_{AC}$ (solid) and $S^{(2)}_{AC}$ (dotted) shows roughly the same behavior as the OTO average. 
{\bf Bottom Left:} $I(A:C)$, a trivial function of $S_{AC}$, show that for an initial $2$ bits of information between the subsystems in the fermion channel gets delocalized so that at late times only a small amount ($0.59$ bits) remains.  
{\bf Bottom Right:} The negative of the tripartite information normalized by its maximum value ($2$ bits) is a simple diagnostic of scrambling.
}
\label{fig:kitaev-numerics}
\end{figure}

Finally, we consider the $0$-dimensional Majorana fermion fast scrambler \eqref{eq:kitaev-model} with $N=14$ fermions represented with $n=7$ spins. In Fig.~\ref{fig:kitaev-numerics}, the inputs and outputs are divided unequally, with inputs subsytems of $a=1$, $b=6$ spins and outputs subsystems of $c=6$, $d=1$ spins. (There is no spatial arrangement for the $0$-dimensional system.) We plot the OTO average in the top-left panel, the entropies $S^{(1,2)}_{AC}$ in the top-right panel, $I(A:C)$ in the bottom-left panel, and the normalized tripartite information in bottom-left panel. (We do not plot $I(A:D)$ since for both the fermion channel and the Haar-channel it never becomes greater than $6 \times 10^{-4}$ bits.)

The main difference between the fermion channel and the spin chains in the unequal configuration is the relevant time scale of the butterfly effect. In the spin chain channels, the initially delay before the OTO correlator decay scales with $c$, the distance between $A$ and $D$. This is because operators in $A$ have to grow to encompass the entire spin chain so that the OTO correlators between operators in $A$ and $D$ can be affected. In the fermion channel, there is no notion of spatial locality so the correlators begin to decay immediately.

We don't plot the Majorana fermion channel for equal system sizes since it has essentially the same behavior: all quantities quickly asymptote to Haar-scrambled values similar to Fig.~\ref{fig:kitaev-numerics}. This is in slight contrast to chaotic spin chain (plotted in Fig.~\ref{fig:spin-chains}), in which the relevant quantities never quite reach the Haar-scrambled values. This suggests that the Majorana fermion system has stronger scrambling power than the chaotic spin chain. Nevertheless, since the late-time values of these quantities in both channels always asymptote to very near the Haar-scrambled values, local measurements cannot differentiate the time-evolved chaotic channels from the Haar-random channel. Thus, the butterfly effect implies scrambling.

\section{Perfect tensor model}\label{sec:perfect-tensor-model}
Now that we understand the relationship between strong chaos and the scrambling behavior of quantum channels, we will present a tensor network model of a scrambling channel with ballistic operator growth.\footnote{See also \cite{Casini:2015zua} for a similar recent tensor network model of ballistic entanglement propagation.} This model serves two purposes.

First, it is useful as a tractable model of ballistic scrambling. The network implements the expected entanglement structure of chaotic time evolution with a (discretized) time independent Hamiltonian. It also serves as a concrete toy model to study the growth of computational complexity in scrambling quantum channels consisting of local quantum circuits.\footnote{
In fact, there is a well-defined notion of the complexity of randomness, called unitary $t$-designs, and lower bounds on the complexity growth under random quantum circuits in this sense have been rigorously established \cite{Brandao2012}.
}

Second, it provides a model for the interior of the eternal AdS black hole \cite{Maldacena:2001kr}. In \cite{Hartman:2013qma}, it was proposed that the interior connecting the two asymptotic regions can be represented by a flat tensor network whose length is proportional to the total time evolution on the boundary. In \cite{Stanford:2014jda} and \cite{Roberts:2014isa}, this proposal was explored in a larger variety of black hole states that were perturbed by shock waves.

Here, we provide a concrete model of such a network (i.e. we specify the tensors). This is in the spirit of previous work on the AdS ground state: in \cite{swingle2012entanglement} it was suggested that the ground state of AdS can be represented by a hyperbolic tensor network (such as MERA \cite{PhysRevLett.99.220405}), and then an explicit tensor network model was proposed in \cite{Pastawski:2015qua} (see also \cite{Yang:2015uoa}).\footnote{This model has the additional nice property of implementing the holographic quantum error correction proposal of \cite{Almheiri:2014lwa}.}

\begin{figure}[htb!]
\centering
\includegraphics[width=0.60\linewidth]{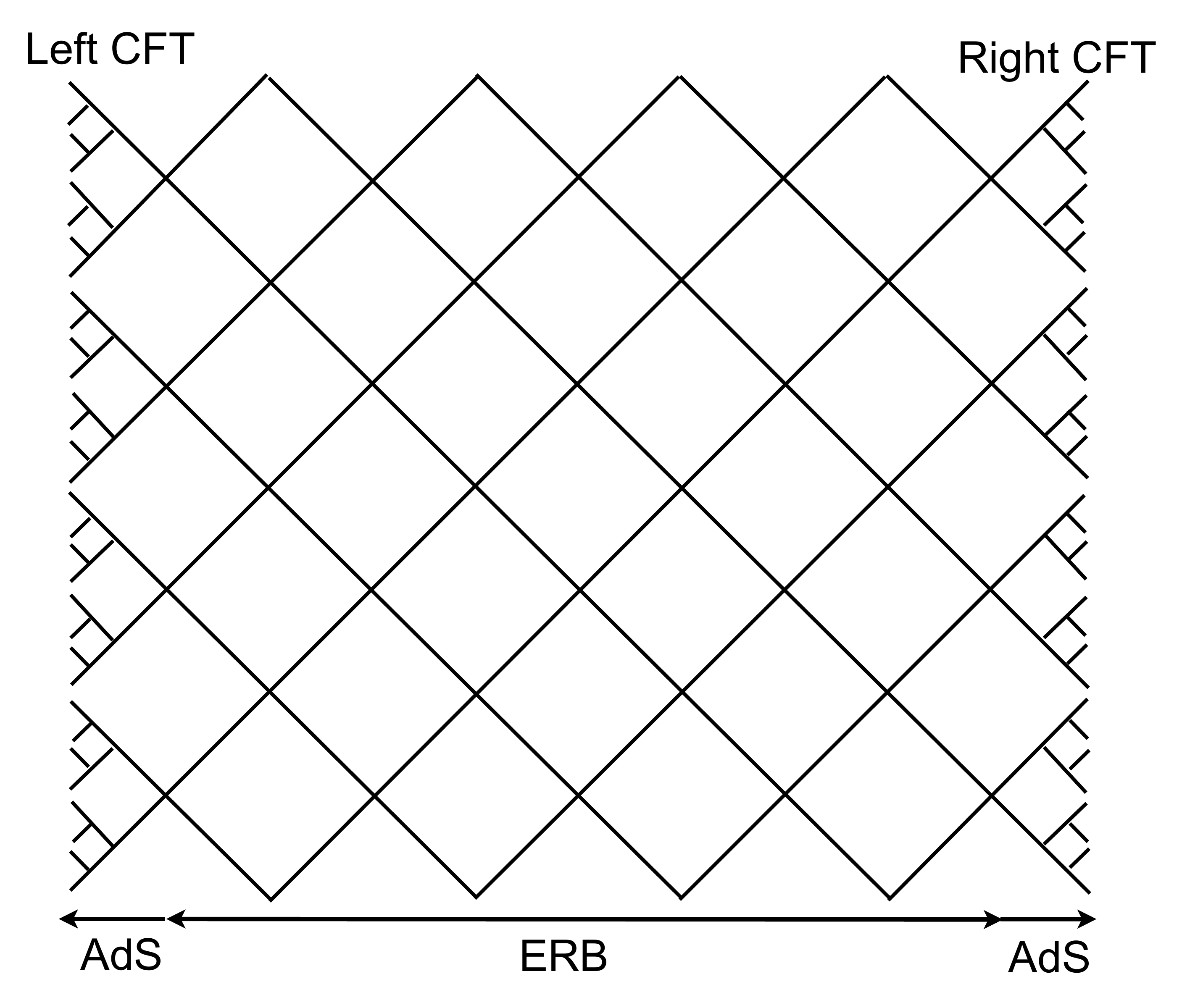}
\caption{Tensor network representation of the Einstein-Rosen bridge. A four-leg tensor lives at each node. We will consider a network of perfect tensors. 
} 
\label{fig_ER_bridge}
\end{figure}

Before we begin, let us review the proposal of \cite{Hartman:2013qma}. The tensor network representation of the thermofield double state is shown in Fig.~\ref{fig_ER_bridge}. At the left and right ends, we have a hyperbolic network, representing the two asymptotically AdS boundaries. This network extends infinitely from the UV into the IR thermal scale $\beta$ at the black hole horizon. Then, the middle is flat representing the black hole interior. The entire network grows as $t$ grows by adding more layers in the middle flat region. 

We would like to further elaborate on this proposal of tensor network representation of the black hole interior. We will study networks of perfect tensors and demonstrate chaotic dynamics by finding ballistic growth of local unitary operators and the linear growth of the tripartite information until the scrambling time. For the rest of discussion, we take the infinite temperature $\beta=0$ limit so we can ignore the hyperbolic part and focus in on the planar tiling of tensor networks representing the interior.

\subsection{Ballistic growth of operators and linear growth of $I_3$}

Let us review the definition of perfect tensors. Consider a tensor $\mathcal{T}$ with $2n$ legs and bond dimension $v$. A tensor can be represented as a pure state 
\begin{align}
|\Psi\rangle = \sum_{i_{1},\ldots, i_{2n}}\mathcal{T}_{i_{1},\ldots, i_{2n}}|i_{1},\ldots, i_{2n}\rangle,
\end{align}
with a proper normalization. We call a tensor $\mathcal{T}$ perfect if it is associated with a pure state $|\Psi\rangle$, called a perfect state, which is maximally entangled along any bipartition. Namely, 
\begin{align}
S_{A} = n, \qquad \forall A \ \ \mbox{s.t} \ \ |A|=n,
\end{align}
where for tensors of bond dimension $v$ we measure entropy in units of $\log v$.
The qutrit tensor Eq.~\eqref{eq:qutrit} is an example of a perfect tensor. There are known methods for constructing perfect tensors via the framework of quantum coding theory. Also, a Haar random tensor becomes a perfect tensor at the limit of $v \rightarrow \infty$. 

\subsubsection*{Growth of local operators}

Imagine a flat planar tiling of $4$-leg perfect tensors as shown in Fig.~\ref{fig_ER_bridge} which may be thought of as a discretized time-evolution by a strongly-interacting Hamiltonian. 
We can examine time evolution of a local unitary operator $V$ and observe linear growth of spatial profiles of operators $V(t)$ by using a basic property of perfect tensors. Let $|\Psi\rangle$ be a $4$-spin perfect state and denote $4$ legs by $a,b,c,d$. Consider a single-body unitary operator $U_{a}\not=I$ acting exclusively on $a$. Since $ab$ and $cd$ are maximally entangled, there always exists a corresponding operator $VU_{cd}\not=I$ acting exclusively on $cd$ such that 
\begin{align}
U_{a} |\Psi\rangle = U_{cd} |\Psi\rangle,
\end{align}
or in the tensor representation, we have 
\begin{align}
\includegraphics[height=1.0in]{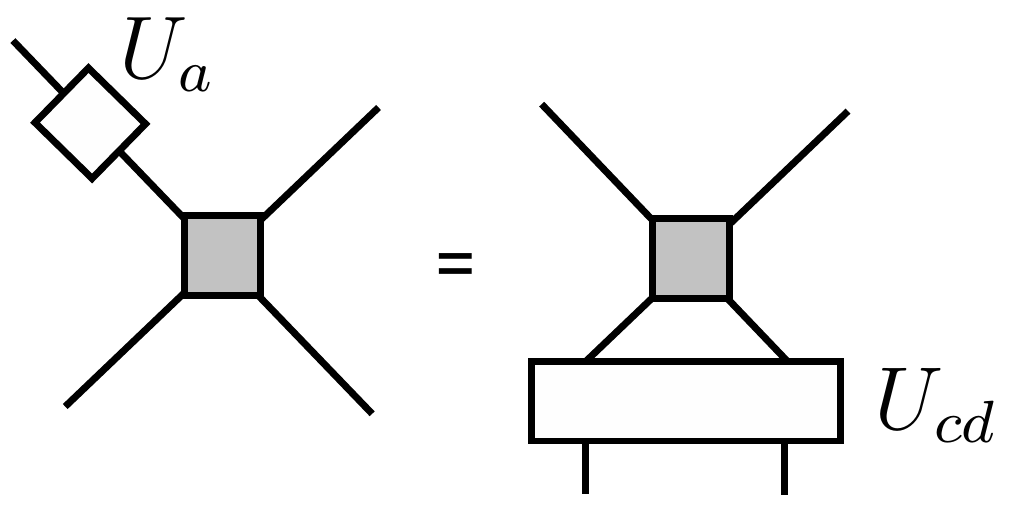},
\end{align}
where a gray square represents a four-leg perfect tensor. 
One can prove that $U_{cd}$ must act non-trivially both on $c$ and $d$. Namely, if $U_{cd}$ were a single-body operator acting only on $c$ (i.e. $U_{cd}=U_{c}$), then one would have $U_{a}U_{c}^{\dagger}|\psi\rangle=|\psi\rangle$. However this contradicts with the fact that $ac$ and $bd$ are maximally entangled. To see the contradiction, one can simply use $U_{a}U_{c}^{\dagger}|\psi\rangle\langle \psi|=|\psi\rangle\langle \psi|$ and take a partial trace over $b,d$ on both sides of the equation. If $a,c$ is maximally entangled with $b,d$ we obtain $U_aU_c^\dagger=I$ is the identity operator, which is not possible.
The conclusion is that, due to the perfectness of the tensors, each two-qudit unitary associated with perfect tensors always expands a single-body operator to a two-body operator. This observation is consistent with linear ballistic propagation of entanglement for single connected regions predicted for chaotic systems \cite{Kim:2013aa}.

\begin{figure}[htb!]
\centering
\includegraphics[width=0.70\linewidth]{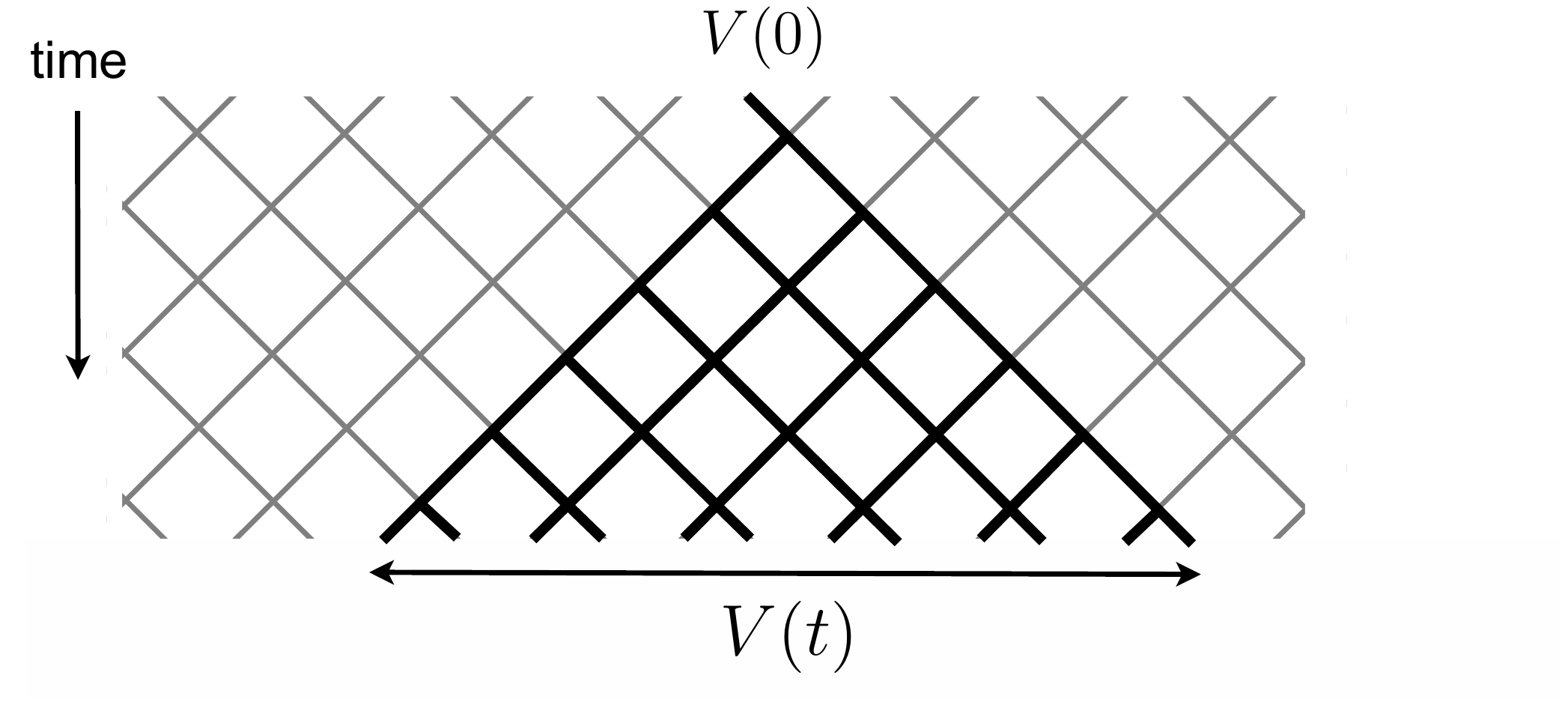}
\caption{Ballistic growth of local operators by perfect tensors. As a reminder, the horizontal radius of the operator $V$ grows with the butterfly velocity as $v_B t$, and the vertical depth of the circuit grows as $v_E t$.
} 
\label{fig_ballistic_propagation}
\end{figure}

The implication of this ballistic expansion of unitary operators under perfect tensors is quite interesting. The size of the region of nontrivial support for $V(t)$ increases linearly as shown in Fig.~\ref{fig_ballistic_propagation}. At $t=L/2$, for a lattice of linear size $L$, a local operator will evolve into a global operator supported over the entire lattice. The growth of OTO correlation functions originates from this linear growth of spatial profiles of local operators. Namely, for a local operator $W$ which is separated in space from $V(t=0)$, the commutator $[V(t),W]$ becomes non-negligible after $t=L/2$ indicative of the butterfly effect.\footnote{A qualitatively similar behavior occurs when Haar-random unitary operators are used instead of perfect tensors, which we checked numerically. For an analytical discussion a random tensor network in the context of a holographic state rather than a channel, see \cite{RandomModelHolgraphy}. (See also \cite{hastings2015random}.) }

\subsubsection*{Growth of tripartite information in time}
\begin{figure}[htb!]
\centering
\includegraphics[width=0.55\linewidth]{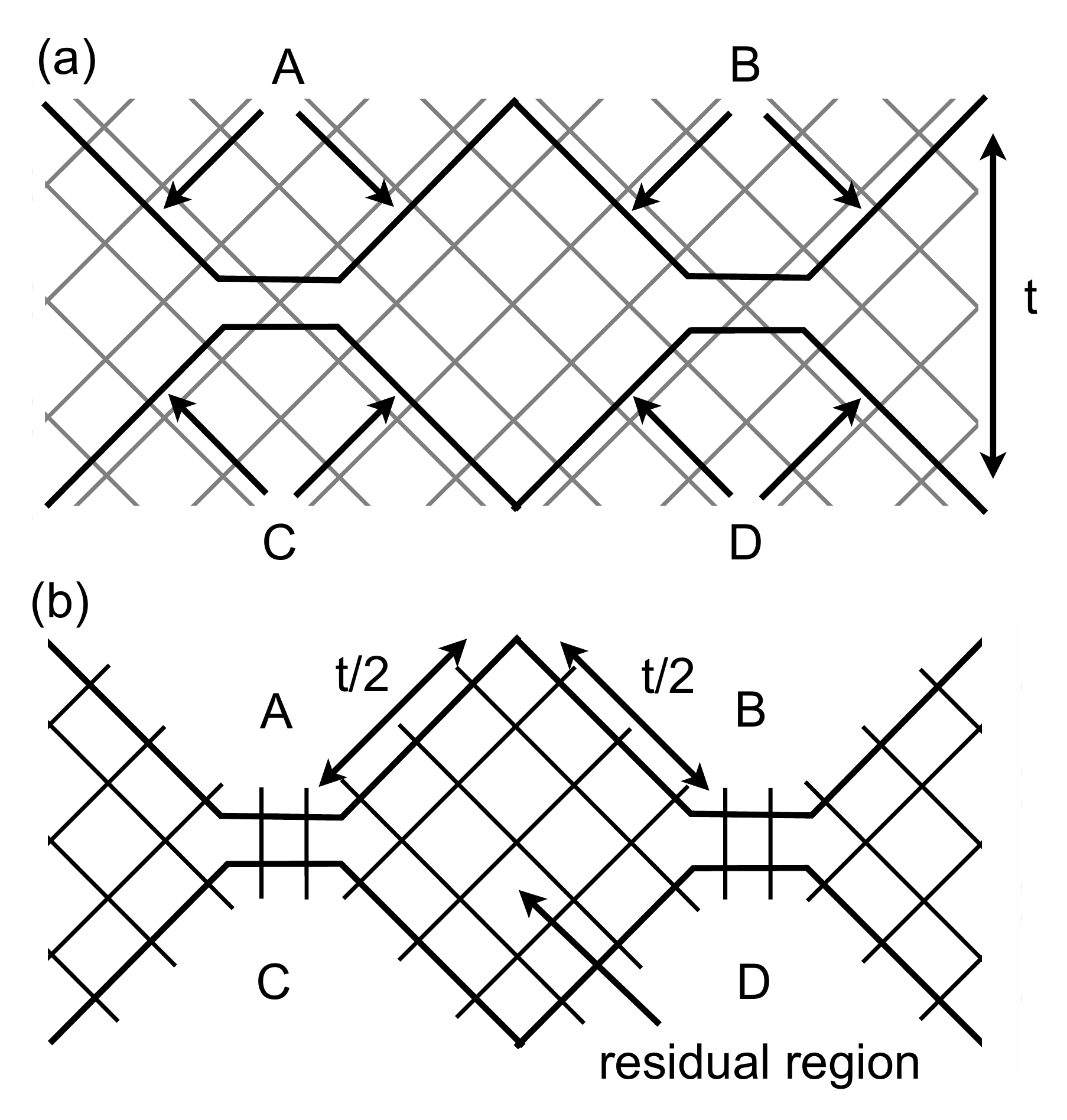}
\caption{(a) Computing $I_3$ for $A,B,C,D$ of equal size. Local moves are shown. (b) Residual tensor networks. Square-like tensor networks are responsible for the tripartite information. 
} 
\label{fig_I3}
\end{figure}

Let us then compute the tripartite information for a network of perfect tensors. The entire system is split into four regions $A,B,C,D$ of equal size as in Fig.~\ref{fig_I3}(a). The growth of entanglement entropy can be exactly calculated by using a method developed in \cite{Pastawski:2015qua}. Recall that, for a perfect state $|\Psi\rangle$ with four spins, there always exist a two-qubit unitary operator $D_{ab}$ such that 
\begin{align}
D_{ab} |\Psi\rangle = |EPR\rangle_{ac} \otimes |EPR\rangle_{bd}.
\end{align}
In other words, $D_{ab}$ disentangles a perfect state into two decoupled EPR pairs as graphically shown below
\begin{align}
\includegraphics[height=1.0in]{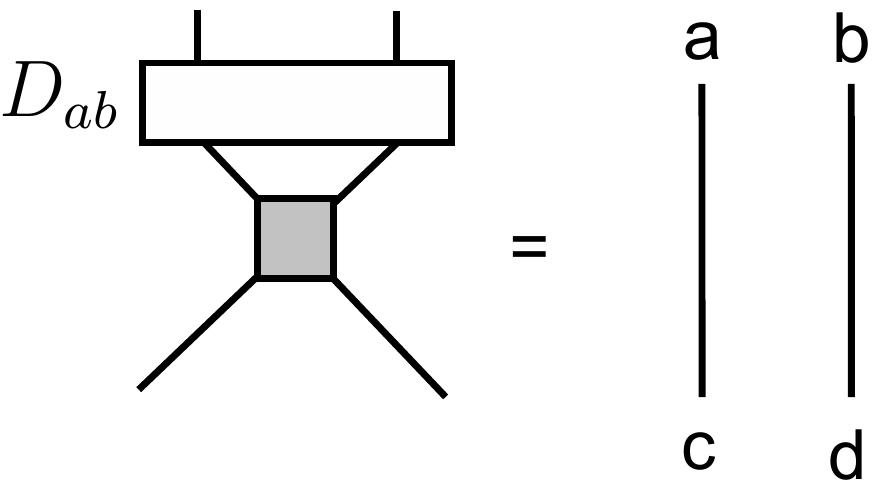}.
\end{align}
A key observation is that the process of finding a minimal surface by local updates can be viewed as entanglement distillation by applications of disentanglers. This led to the proof of the Ryu-Takayanagi formula for single intervals in networks of perfect tensors \cite{Pastawski:2015qua}. 

In general, calculation of entanglement entropies for disjoint regions is challenging even for networks of perfect tensors. Indeed, the verification of the Ryu-Takayanagi formula is given only for single intervals for a network of perfect tensors in~\cite{Pastawski:2015qua}. Here we assume that the planar tensor network is translationally invariant in both time and spatial directions. To be specific, we also assume that the network consists of the qutrit perfect tensor introduced in \eqref{eq:qutrit}. For such a perfect tensor tiling, an analytical calculation of the tripartite information is possible for time $t$ shorter than the scrambling time $t_*=L/2$. Namely, one can prove the following:
\begin{align}
I_3(A:B:C)= - 2t, \qquad 0 \leq t \leq L/2,  \label{eq:I3_tensor}
\end{align}
where $L$ is the linear length of the system and as a reminder for qutrits we measure entropy in units of $\log 3$. Below, we sketch the derivation.

For a network of perfect tensors, entanglement properties can be studied by applying local disentanglers which correspond to distillations of EPR pairs. The disentanglers map each region unitarily to the minimal surface bounding it, as is shown in Fig. \ref{fig_I3}. For time $t\leq L/2$, one observes that minimal surfaces for $A,C$ collide with each other, and similarly for $B,D$. Let us distill entanglement as shown in Fig.~\ref{fig_I3}(b) by applying some appropriate local unitary transformations on each region and remove decoupled spins. Regions $AC$ and $BD$ possess EPR-like entanglement along the collided surface of geodesic lines. In the middle of the network, we find square regions which are responsible for four-party entanglement among $A,B,C,D$. Such regions, which are not included inside causal wedges of boundary regions, are referred to as residual regions \cite{Pastawski:2015qua}. These become essential in understanding entanglement properties behind  the horizons of the multi-boundary black holes considered in \cite{Marolf:2015vma}. At the time step $t$, there will be a pair of square residual regions with linear length $t/2$ as shown in Fig.~\ref{fig_I3}(b). In Appendix~\ref{appendix-tensor}, we study multipartite entanglement for rectangular residual regions. Namely, we show that each residual region contributes to the tripartite information by $-t$. We thus obtain Eq. \eqref{eq:I3_tensor}. 

\subsection{Recurrence time}

We have shown that the network of perfect tensors, as shown in Fig.~\ref{fig_ER_bridge} and Fig.~\ref{fig_ballistic_propagation}, serves as a toy model of scrambling dynamics. 
A naturally arising question is whether such a system stays scrambled after the scrambling time $t_*=L/2$. 
In this section, we study the recurrence time of the planar network of perfect tensors. For concreteness, we will restrict our considerations to those with qutrit perfect tensors. 
We assume periodic boundary conditions in the spatial directions of the network.

Imagine that we inject some Pauli operators from the top of the tensor network and obtain output Pauli operators on the bottom. We are interested in the minimal time step necessary for a network to output the initial Pauli operators again. To find the recurrence time, we inject two-body Pauli $Z$ operators from the top left corner of the tensor network and compute the output Pauli operator on the bottom. We define the recurrence time $t_{rec}$ to be the minimal time step $t_{rec}$ necessary for the network to output the initial two-body Pauli $Z$ operators. Recall that the tensor network based on stabilizer tensors maps Pauli operators to Pauli operators. Since Pauli operators can be treated as classical variables, one can efficiently find the recurrence time via numerical methods. 

\begin{figure}[htb!]
\centering
\includegraphics[width=0.7\linewidth]{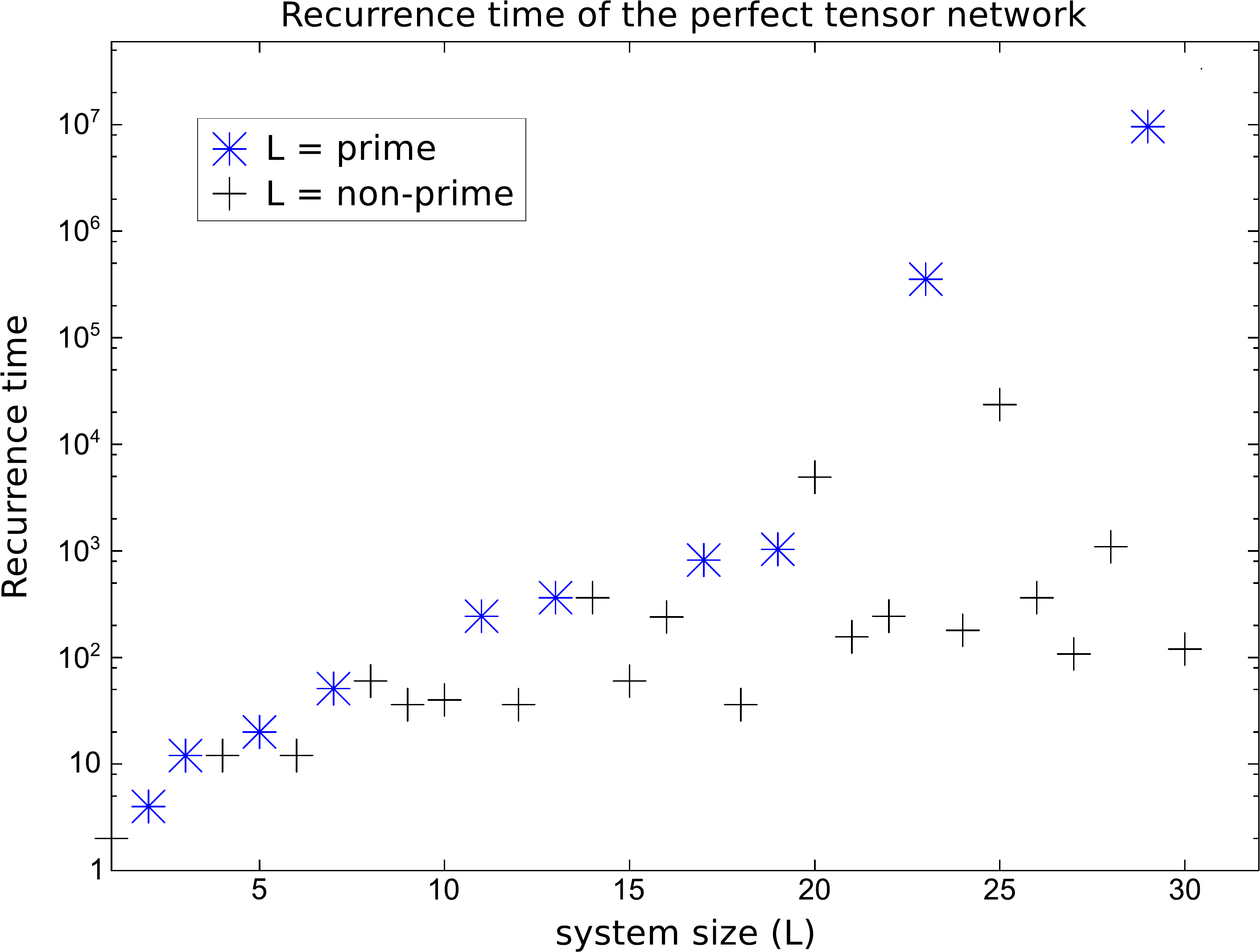}
\caption{The exponential growth of recurrence time for perfect tensor networks of prime size (blue) and non-prime size (black).
} 
\label{fig_recurrence_time}
\end{figure}

The recurrence time crucially depends on the system size $L$ as shown in the plot in Fig~\ref{fig_recurrence_time}. Note that the plot uses a logarithmic scale. When the system size is $L=3^{m}$, the recurrence time grows only linearly: $t_{rec}=4L$. This expression can be analytically obtained. The linear growth is due to the fact that the qutrit tensor can be viewed as a linear cellular automaton over $\mathbb{F}_{3}$ which has scale invariance under dilations by factor of $3$. For such special system sizes, the trajectories of time-evolution of Pauli operators form short periodic cycles. This is similar to the classical billiard problem where trajectories of a billiard ball are not ergodic for fine-tuned system sizes and fine-tuned angles. Yet, the billiard problem is ergodic for generic system sizes. 

Likewise, the perfect tensor network has longer recurrence time for generic values of system sizes. When $L$ is a prime number, the growth is rather fast, and seems exponential as shown in Fig.~\ref{fig_recurrence_time}. (We do not have an analytic proof of this statement.) Assuming the exponential growth of the recurrence time for prime $L$ ($t_{rec}\approx e^{kL}$), let us find out the growth for typical values of $L$. For typical values of $L$, we expect that $t_{rec}$ grows faster than any polynomial functions. This is because given a positive integer $n$, the probability for its largest prime factor to be larger than, say $\sqrt{n}$, is finite.\footnote{In general, the probability for the largest prime factor to be larger than $1/n^{u}$ is given by the Dickman function \cite{dickman1930frequency}.} 
Assuming that $L$ is not a prime number, let us decompose it as $L=L_{1}L_{2}$. Then, due to the translation invariance, one can show that $t_{rec}(L)\geq t_{rec}(L_{1}),t_{rec}(L_{2})$. As such, the recurrence time $t_{rec}(L)$ will be lower bounded by $t_{rec}(p)$ where $p$ is the largest prime factor of $L$. This argument implies an exponential growth of the recurrence time for typical values of $L$.

The recurrence time of the perfect tensor network is much longer than that of integrable systems, but is much shorter than that of chaotic systems. By construction, perfect state network based on the stabilizer formalism can have at most exponential recurrence time. This is essentially because unitary circuits implemented by stabilizer-type tensors belong to the so-called Clifford group which is a subgroup of unitary transformations that map Pauli operators into Pauli operators.\footnote{This means that under time evolution a simple Pauli operator $X$, $Y$, or $Z$ can only grow into a product of Pauli operators (rather than a sum of products of Pauli operators as would be generically expected by \eqref{BCH-expansion}).} 
Quantum circuits solely consisting of Clifford operators are classically simulable since transformations of Pauli operators can be efficiently characterized by pairs of classical bits. In this sense, the stabilizer perfect tensor network exhibits marginally chaotic behaviors. The classical simulability enables us to study chaos and scrambling behaviors in quantum channels at relatively early times in a computationally tractable manner.

This highlights an important point about perfect tensor networks as models of holography. In many cases they can exhibit key features expected of holographic systems (such as the error correcting and bulk reconstruction properties of the model presented in \cite{Pastawski:2015qua}). However, since the recurrence time of the perfect tensor network is exponential in the system size and not doubly exponential, it's clear that it fails to capture a very important feature: the possibility of exponential computational complexity. In particular, the (comparatively) quick recurrence means that the longest minimal perfect tensor networks are far less complex than the degree of complexity that generic holographic states are expected to reach. 

One possible resolution is to modify the stabilizer perfect tensor by applying some single qudit non-Clifford rotation, such as a rotation around the $Z$ axis by some angle $\theta$. An inclusion of a single non-Clifford operator to the full Clifford group enables us to efficiently approximate an arbitrary unitary operator, an important result known as the Solovay-Kitaev theorem~\cite{Nielsen_Chuang}. As such, we speculate that non-Clifford modification of perfect tensors would create a tensor network with doubly exponential recurrence time. This resolution is along the spirit of the billiard problem since the Clifford transformations are fine-tuned operations. 

Another possible resolution of this is that to reach the more complicated states (which are not at all understood holographically, see e.g. \cite{Almheiri:2012rt,Almheiri:2013hfa,Shenker:2013yza,Susskind:2015toa}), one needs to consider superpositions of such tensor networks which do not have a geometric description and thus would not be expected to have a semiclassical bulk interpretation.

Regardless, a network of perfect tensors is very capable of scrambling. This observation leads us to envision that a certain measure of complexity can be attached to each tensor in the network, in particular, $-I_{3}$ up to proper normalization. This would represent the complexity of forming the four-leg perfect tensor from a product state.
It would be interesting to see if some kind of upper bound on the gate complexity can be imposed by considering an integral of $-I_{3}$ over all the tensors in the network.

\section{Discussion}\label{sec:discussion}

In this paper, we have shown that the butterfly effect---as expressed by the decay of out-of-time-order (OTO) correlation functions---implies the information-theoretic notion of scrambling. The butterfly effect is manifested by the growth of simple operators under time evolution to complicated operators of high weight. These time-evolved operators will then have large commutators will all other operators in the system. If we think of the initial simple operator as an input to a unitary quantum time-evolution channel, then the output will be an operator spread over the entire system. All information associated with the input will be delocalized;  the output system is scrambled.

The method of characterizing scrambling/chaos via the framework of quantum channels may also find interesting applications in studying thermalization in many-body quantum systems. We have already demonstrated the usefulness of our approach by studying the tripartite information in several different examples: numerical results in integrable/non-integrable spin chains and the nonlocal interacting Majorana fermion model of Kitaev, and both analytical and numerical results in a perfect tensor network model of discretized time evolution. It would be interesting to study many-body/single-body localization and delocalization transitions in the setup of quantum channels. A closely related question may concern the information-theoretic formulation of the Eigenstate Thermalization Hypothesis (ETH). The state interpretation of the channel is able to consider a set of initial states as well as to probe off-diagonal elements in the Hamiltonian.

In order for quantum information to really be processed, it has to interact with the other information distributed across the system. Said another way, to process information the channel has to be capable of scrambling. This suggests that there is a strong connection between quantum chaos and computation. As a surrogate for a definition of computation, let's consider the computational complexity of the quantum circuit or channel. For tensor network models, this is simply the number of tensors in the minimal tensor network.

As a simple example, let's consider the quantum channel that only contains swap gates.  The channel doesn't scramble, and information can only be moved around. As discussed multiple times, the swap channel has a quick recurrence and can never get very complex. The only output states accessible are those related to permutations of the input, all the multipartite states cannot be accessed. For a system of $n$ qubits, the complexity can only ever reach $O(n)$ (the complexity of swapping localized information from one end of the system to the other using local swap operations). The maximal complexity of a state of $n$ qubits is $O(2^n)$; thus, for the simple swap channel most of the possible output states are entirely inaccessible. It is essentially only capable of classical computation.

Quantum computation requires interaction, and strong chaos is a signature of a strongly interacting system. 
Thus, in some sense, we speculate chaos must be the capacity for a system to do computation. This suggests that strongly chaotic-systems must be fast computers. In fact, in \cite{Brown:2015bva,CAFollowup} it was recently hypothesized that black holes are the fastest computers in nature. Given that black holes are already known to be nature's densest hard drives \cite{tHooft:1993gx,Susskind:1994vu} and most chaotic systems \cite{Maldacena:2015waa}, it seems reasonable to suspect that a system's computational power must be limited by its degree of chaos. It would be interesting to try and make this dependence on chaos for computation more direct.

\section*{Acknowledgments}
This work began at KITP, and the authors would like to acknowledge the KITP programs ``Entanglement in Strongly-Correlated Quantum Matter'' and ``Quantum Gravity Foundations: UV to IR.'' We would also like to thank Tarun Grover, Aram Harrow, Patrick Hayden, Matt Headrick, Isaac Kim, Alexei Kitaev, John Preskill, Steve Shenker and Douglas Stanford for discussions, and Douglas Stanford for comments on the draft. 

PH and XLQ are supported by the David and Lucile Packard foundation. XLQ is also partially supported by the Templeton foundation.  

DR is supported by the Fannie and John Hertz Foundation and is also very thankful for the hospitality of the Stanford Institute for Theoretical Physics during a stage or two of this work. DR also acknowledges the U.S. Department of Energy under cooperative research agreement Contract Number DE-SC0012567. This paper was brought to you by the butterfly effect.

BY is supported by the David and Ellen Lee Postdoctoral fellowship and the Government of Canada through Industry Canada and by the Province of Ontario through the Ministry of Research and Innovation.

\appendix

\section{Haar scrambling}\label{appendix-haar}
In this paper, we've generally considered scrambling by a one-parameter family of unitary operators $U(t)=e^{-iHt}$ and found that for chaotic systems, increasing time $t$ leads to more efficient scrambling. Instead, we will now take $U$ to be a Haar random unitary operator. This is useful as a baseline for scrambling. We expect that the late-time values of entropies and informations computed from scrambling operators $U(t)$ will asymptote to Haar random values. Additionally, we will see that Haar-random values of quantities such as $I(A:C)$ and $I_3(A:C:D)$ are not necessarily maximal, but exhibit ``residual'' information regardless of system size.

At its fastest, Page scrambling has a complexity of $n \log n$ gates, while Haar scrambling is nearly maximally complex requiring $\sim e^n$ gates. In \cite{Page:1993df}, it was proven that Haar scrambling implies Page scrambling. As we will now show, Haar scrambling also implies the tripartite scrambling. However, the implication does not work in the other direction: since the late-time values of $I(A:C)$, $I(A:D)$, and $I_3(A:C:D)$ approach Haar-scrambled values, entanglement is not enough \cite{Susskind:2014moa} to diagnose typicality (in the Haar-random sense). This possibly has a strong bearing on the paradoxes of \cite{Almheiri:2012rt,Almheiri:2013hfa} as discussed in \cite{Susskind:2015toa}.

To proceed with this analysis, we will consider an expectation over density matrices constructed from Haar-random states. These tools were used by Page to analyze the entropy of subsystems for random states \cite{Page:1993df}, and our approach will be similar to \cite{Hayden:2007cs} and Appendix~A of \cite{Shenker:2013pqa}. In fact, our calculation is very similar to that in \cite{Shenker:2013pqa}. However, beware that the results do not simply carry over; since we are pairing together input and output subsystems of possibly different size  (in our notation, the fact that $a\neq c$), we will find a very different result.

Our setup will be the usual division into subsystems $ABCD$, with the state given by \eqref{circuit-state}, and $U$ a random $2^n \times 2^n$ unitary matrix taken from the Haar ensemble. The Haar average lets us consider expectations over a number of unitary matrices and is non-zero only when the number of $U$s equals the number of $U^\dagger$s. For instance, with two $U$s and two $U^\dagger$s, the formula for the average is
\begin{align}
\int dU \ U_{i_1 j_1}U_{i_2 j_2}U^*_{i_1' j_1'} U^*_{i_2' j_2'} &= \frac{1}{2^{2n}-1}\Big(\delta_{i_1 i_1'}\delta_{i_2 i_2'}\delta_{j_1j_1'}\delta_{j_2j_2'} + \delta_{i_1 i_2'}\delta_{i_2 i_1'}\delta_{j_1j_2'}\delta_{j_2j_1'} \Big) \label{haar-average-formula} \\ &\hspace{40pt} -\frac{1}{2^n(2^{2n}-1)}\Big(\delta_{i_1i_1'}\delta_{i_2 i_2'}\delta_{j_1j_2'}\delta_{j_2j_1'} + \delta_{i_1 i_2'}\delta_{i_2 i_1'}\delta_{j_1j_1'}\delta_{j_2j_2'}  \Big).\notag
\end{align}
This formula will let us compute the average over the trace of the square of the density matrix $\rho_{AC}$
\be
\int dU \, \tr \, \{ \rho_{AC}^2\} = \frac{1}{2^{2n}}\int dU ~ U_{k \ell m o}\, U^*_{k' \ell m' o} \,   U_{k' \ell' m' o'}\, U^*_{k \ell' m o'}, \label{U-average-rho-ac}
\ee
where as in Appendix~\ref{appendix-proof}, $k=1\dots 2^a$ are $A$ indices, $\ell=1\dots 2^b$ are $B$ indices, $m=1\dots 2^c$ are $C$ indices, and $o=1\dots 2^d$ are $D$ indices. In applying the average \eqref{haar-average-formula} to \eqref{U-average-rho-ac}, note that both $k$ and $\ell$ in \eqref{U-average-rho-ac} are ``$i$''-type indices in \eqref{haar-average-formula}, and similarly $m$ and $o$ are ``$j$''-type indices. After a quick game of delta functions, we find
\be
\int dU \, \tr \, \{ \rho_{AC}^2\} = 2^{-a-c} + 2^{-b-d} - 2^{-a-d-n} - 2^{-b-c-n}. \label{haar-trace-rhoac-raw}
\ee
As of now, we have been completely general about the size of the subsystems. For simplicity, let's now focus on the case of $a+c=n$ and $b+d=n$ as was often considered in our numerics. Without loss of generality, let's also take $a\le b$ and $d \le c$.\footnote{Ref.~\cite{Shenker:2013pqa} neglects the bottom line of \eqref{haar-average-formula} as subleading. For the subsystems we consider, the first term on the second line of \eqref{haar-average-formula}  is actually the same order (in $n$) as the terms in the first line and cannot be neglected.}  Simplifying and throwing away exponentially subleading pieces, we get
\be
\int dU \, \tr \, \{ \rho_{AC}^2\} \simeq 2^{1-n}\, \big(1-2^{-a-d-1}\big).\label{eq:Haar-average-rho-squared-AC}
\ee
Using this with an appropriate caveat,\footnote{To use this result to compute the Haar average of the R\'enyi entropy 
\be
\int dU \,S_{AC}^{(2)} = -\int dU \,\log_2 \tr \, \{ \rho_{AC}^2\}, \notag
\ee
we need to commute the Haar average with the $\log$. This can be checked numerically and holds as long as $n$ is sufficiently large (which in this case ``large'' means about $n=4$). For large $n$, the system self-averages so that any single sample is extremely likely to be at the mean value. This lets the average commute with the $\log$. 
} 
we can compute the Haar average of the R\'enyi entropy 
\be
\big(S^{(2)}_{AC}\big)_{Haar} = n - 1 - \log_2(1-2^{-a-d-1}). \label{Haar-random-SAC}
\ee
This is rather interesting: the maximal value for $S^{(2)}_{AC}$ is $n$. Therefore, the Haar-scrambled state never reaches this maximal value. On the other hand, this is not unexpected. The corrections to Page's entropy of a subsystem formula for the divisions we are considering are expected to be $O(1)$ \cite{Page:1993df}. 

Next, let us use \eqref{Haar-random-SAC} to put a bound on the mutual information $I(A:C)$. We can do this using the fact that $S_{AC} > S_{AC}^{(2)}$, and we find
\be
I(A:C) \le 1 + \log_2(1-2^{-a-d-1}),\label{Haar-random-IAC}
\ee
indicating the possibility of residual information between $A$ and $C$ that is independent of $n$, even in the Haar-scrambled limit.\footnote{One might have thought that it would be possible to make a better bound with the Fannes-Audenaert inequality \cite{fannes1973continuity,audenaert2006sharp} and using the $2$-norm to bound the $1$-normal as in \cite{Hayden:2007cs} and \cite{Shenker:2013pqa}. However, such an approach actually leads to a bound that's actually much looser than the simple one given by \eqref{Haar-random-IAC}.} By considering more equal partitions of the input ($1\ll a\le b$), there will be more residual information between $A$ and $C$, though the fraction of residual information $I(A:C)/2a$ decreases. 

Let's complete our discussion by trying to bound $I(A:D)$. Following the approach outlined above, we find
\be
\int dU \, \tr \, \{ \rho_{AD}^2\} = 2^{-a-d} + 2^{-b-c} - 2^{-a-c-n} - 2^{-b-d-n},
\ee
which is the same as \eqref{haar-trace-rhoac-raw} with $c \Leftrightarrow d$. Taking our simplifying assumption $a+c=n$ and $b+d=n$, with $a\le b$ and $d \le c$, we see that the three latter terms are exponentially smaller in $n$ than the first term. We find
\be
\big(S^{(2)}_{AD}\big)_{Haar} = a + d + O(2^{-2n+a+d}),
\ee
and  we can bound the mutual information as
\be
I(A:D) \le 0 + O(2^{-2n+a+d}).\label{Haar-random-IAD}
\ee

Finally, we note that in the general case where $a+c\neq n$ and $b+d\neq n$, the mutual information \eqref{Haar-random-SAC} is different. With $a\le c$, we would instead find
\be
I(A:C) \le \log_2(1+4^{a-d} - 4^{-d})\label{Haar-random-IAC-general}.
\ee
Here, if $a=d$ we still have a residual bit of information, but with $a<d$ we do not.\footnote{We thank Alexei Kitaev for pointing out that our first result \eqref{Haar-random-SAC} is not general.}

\section{Entanglement propagation in CFT}\label{sec:memory-effect}

In Fig.~\ref{fig_swap}(b), we pointed out a strong resemblance to Feynman diagrams of a $2\to2$ scattering process. The swap gate resembles a diagram that contribute to a noninteracting theory: the only allowed operation is that the particles can swap locations between the inputs and the outputs (or they can do nothing). On the other hand, the perfect tensor resembles a Feynman diagram that contributes to a scattering process in an interacting theory. This is not a coincidence; the strength of chaos should be related to the strength of the coupling, see e.g. \cite{Maldacena:2015waa}.

With this point of view, let us consider entanglement propagation in CFT. The general setup considered in CFT is a global quench; the system is preprepared in a groundstate of a Hamiltonian $H_0$ and then the Hamiltonian is suddenly changed to a different Hamiltonian $H$ such that the system is now in a finite energy configuration. The system is then evolved with the new Hamiltonian and certain entanglement entropies saturate at their thermal values. This often referred to as thermalization. For two-dimensional CFT, the entanglement entropy of a single connected region after a global quench was shown to grow linearly in time, saturating at its thermal value at a time of order half the length of the region \cite{Calabrese:2005in,Calabrese:2007rg,Calabrese:2009qy}. To explain this, \cite{Calabrese:2005in} proposed that entanglement is carried by pairs of entangled noninteracting quasi-particles that travel ballistically in opposite directions.\footnote{See also \cite{Casini:2015zua} for a generalization of the non-interacting quasi-particle model to interacting systems and a discussion of the entanglement velocity.} The quasi-particles would travel at the entanglement velocity $v_E$, which is equal to unity in two-dimensional CFT.
Entanglement entropy increases as the entangle pairs are split between the region and its complement.

\begin{figure}
\begin{center}
\includegraphics[width=1\linewidth]{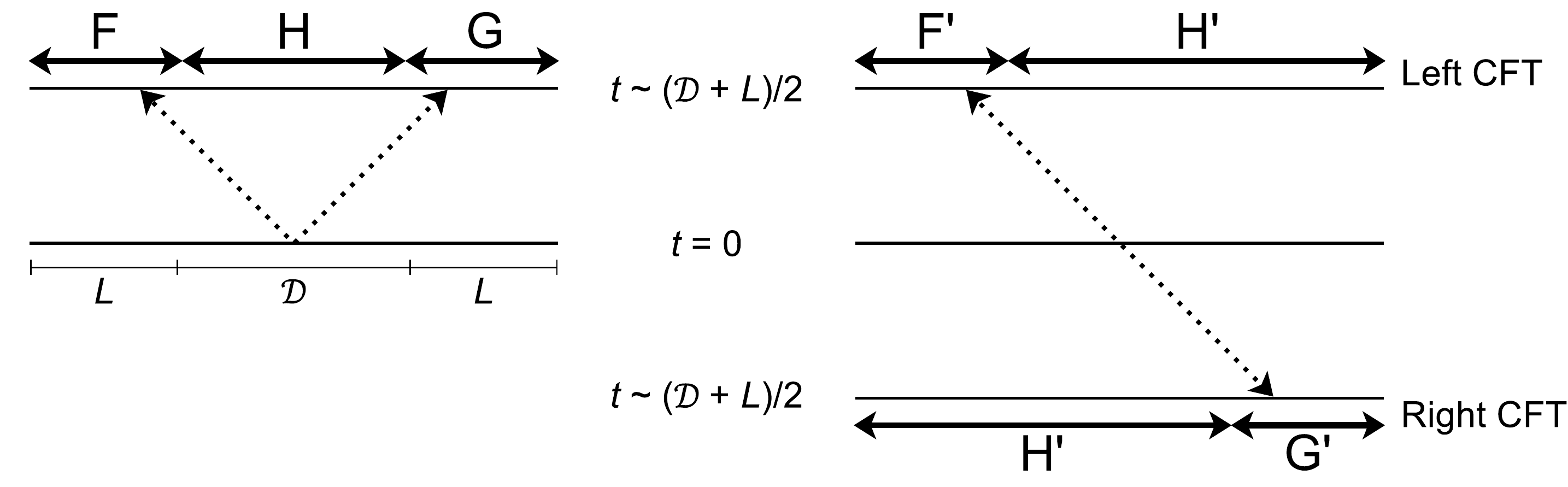}
\caption{Quench in CFT showing path of a quasi-particle EPR pair and two intervals of length $L$ separated by distance  $\mathcal{D}>L$.
{\bf Left:} for two intervals in the noninteracting quasi-particle model $S_{FG}$ has a dip at time $t\sim (\mathcal{D}+L)/2$. 
{\bf Right:} a simplified description in terms of two entangled CFT involves one partner of the EPR pair traveling in the left CFT and one partner traveling in the right CFT. This can be reinterpreted as a quantum channel. To make contact with the notation in the paper, we relabel as $F'\to A$, $G'\to D$, and $H'\to BC$.
}\label{fig:quench}
\end{center}
\end{figure}

This model of entanglement propagation is sufficient to explain the pattern of entanglement growth after a quench of a single interval (in fact, the result is universal \cite{Calabrese:2005in,Calabrese:2007rg,Calabrese:2009qy}), but gives puzzling results for the entanglement entropy of two separated disjoint intervals in interacting (e.g. holographic) systems \cite{Asplund:2014coa,Balasubramanian:2011at,Allais:2011ys,Leichenauer:2015xra,Asplund:2015eha}.\footnote{See also \cite{Hartman:2013qma,Liu:2013iza,Liu:2013qca} for holographic investigations of entanglement growth after a global quench.} Let us label the two intervals as $F$ and $G$, both of size $L$, and the rest of the system as $H$. We will take $F$ and $G$ to be separated by a distance $\mathcal{D}$, and all the scales are taken to be much greater than the thermal correlation length $L,\mathcal{D},\gg \beta$. Additionally, we require $\mathcal{D}>L$. This setup is depicted in the left-hand side of Fig.~\ref{fig:quench}.

In the quasi-particle model, after a quench $S_{FG}$ will grow linearly for a time $D$ and then saturate at its thermal value. However, at time of order $(\mathcal{D}+L)/2$ it will exhibit a dip. In the quasi-particle picture, entangled pairs created in the region between $F$ and $G$ are beginning to enter $F$ and $G$, respectively, causing the entanglement entropy of $FG$ with the rest of the system $H$ to dip. For holographic systems, this is known not to happen: after $S_{FG}$ saturates, it remains saturated \cite{Asplund:2014coa,Balasubramanian:2011at,Allais:2011ys,Leichenauer:2015xra,Asplund:2015eha}.

This puzzle was explored in depth in the context of two-dimensional CFT in \cite{Asplund:2015eha}. There, it is shown that the quasi-particle picture cannot be universal and must depend on the spectrum. Indeed, \cite{Asplund:2015eha} concludes that for interacting CFTs ``entanglement scrambles''---meaning there's no memory effect or dip in $S_{FG}$. Here, we would like to put these results in the context of unitary quantum channels. We will show that ``entanglement scrambling'' is precisely scrambling as diagnosed by the tripartite information. Furthermore, we will argue that the cause of such entanglement scrambling is chaotic dynamics. Strong chaos implies a picture of strongly interacting quasi-particles.

To make the connection, one simply has to realize that a global quench can be simply understood as the time evolution of the thermofield double state \cite{Hartman:2013qma}. That is, a global quench is the channel $|TFD(t)\rangle$ given by \eqref{eq:time-evolved-TFD}. 
This was also pointed out in \cite{Asplund:2015eha}, where the thermofield double state was used to simplify the setup of the two-interval calculation while retaining the basic puzzle.\footnote{In fact, in \cite{Asplund:2015eha} the memory effect was diagnosed by considering properties of the second R\'enyi entropy for the intervals in question.
} In this new setup, the puzzle is a memory effect between an interval on the left CFT $F'$ and interval on the right CFT $G'$, where the spatial separation $\mathcal{D}$ and interval sizes $L$ are large. After a time of order $(\mathcal{D}+L)/2$, the quasi-particle model predicts a dip in the entanglement between $F'G'$ and the rest of the system $H'$. This new setup is shown in the right-hand side of Fig.~\ref{fig:quench}.

Now, let us relabel the subsystems: $F'\to A$, $G'\to D$, and $H'\to BC$. With the perspective of the unitary channel setup (Fig.~\ref{circuit}), the memory effect is simply a question of whether $I(A:D)$ has a spike. Integrable systems will have a spike and can be modeled by entanglement carrying quasi-particles (the swap gate in Fig.~\ref{fig_swap}(b) provides an explicit cartoon of such noninteracting quasi-particles). Chaotic systems are strongly interacting, and the quasi-particle picture breaks down (the perfect tensor in Fig.~\ref{fig_swap}(b) provides the cartoon for the interacting system). This memory effect was shown explicitly in the bottom-middle panel of Fig.~\ref{fig:spin-chains} for the integrable spin chains we studied in section \S\ref{sec:numerics}.

This connection to the work of \cite{Asplund:2015eha} allows us to probe scrambling and chaos in particular CFTs. For instance, the results of section~$4.4$ in \cite{Asplund:2015eha} suggest that the $D1$-$D5$ CFT at the ``orbifold point'' does not scramble (in the sense of tripartite information) as expected for a free theory. It would be interesting to make additional connections between scrambling/chaos and CFT results.

\section{Proof of Eq.~\eqref{corr=renyi}} \label{appendix-proof}
\begin{figure}[htb!]
\centering
\includegraphics[width=0.75\linewidth]{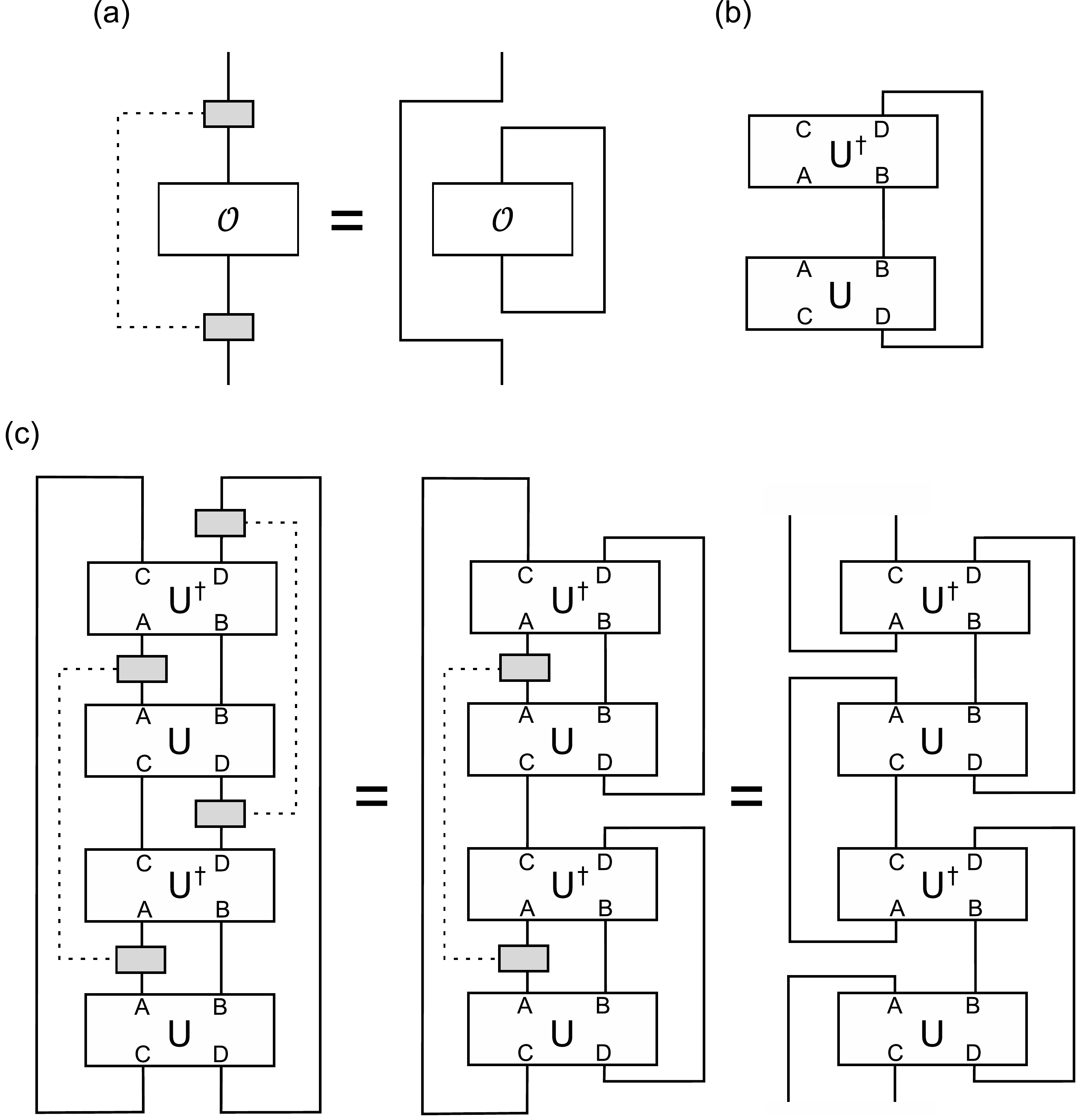}
\caption{Diagrammatic proof of the OTO average $|\langle   \OO_D(t) \,  \OO_A\, \OO_D(t)\, \OO_A\rangle_{\beta=0}|$. Dotted lines indicate summation over operators (gray rectangles), and in all diagrams the normalization is suppressed.
{\bf (a)} The operator identity Eq.~\ref{operator-identity}.  
{\bf (b)} $\rho_{AC}$ for the unitary channel.
{\bf (c)} Calculation of the average. 
In the final panel, the lines at the top and the bottom are appropriately connected.
} 
\label{diagram_OTO}
\end{figure}

The proof of the relation
\be
|\langle   \OO_D(t) \,  \OO_A\, \OO_D(t)\, \OO_A\rangle_{\beta=0}| = 2^{ n-a-d-S_{AC}^{(2)} },
\ee
is probably most easily understood diagrammatically as shown in Fig.~\ref{diagram_OTO}. For simplicity of discussion, we assume a system consisting of qubits while our discussion straightforwardly generalizes to a system consisting of qudits by considering generalized Pauli operators. Thus, this proof applies to lattice systems with a finite-dimensional Hilbert space at each site.\footnote{In a continuum limit, we would need some notion of the operator identity \eqref{operator-identity}, which is the completeness condition for a basis of operators. Naively, due to the infinite Hilbert space dimension, \eqref{corr=renyi} is trivially true; the R\'enyi entropy is UV-divergent and the correlation function average is vanishing due to normalizing by the total number of operators. However, the connection between our results and entanglement propagation in CFT (see Appendix~\ref{sec:memory-effect}) suggests that perhaps a recasting of the relation \eqref{corr=renyi} in terms of mutual information might lead to a sensible continuum limit.}

To proceed with the proof, we need to make use of an operator identity. Consider a partition of a system $AB$ with $A_j$ a complete basis of operators in $A$. Then, for any operator $\mathcal{O}$ on the entire system $AB$, we have 
\be
\sum_{j} A_j\, \mathcal{O}\, A_j = |A| \, I_A \otimes \tr_A \, \{\mathcal{O}\}, \label{operator-identity}
\ee
where the sum $i$ runs over the entire basis, and $|A|$ is the size of the Hilbert space of $A$. The set of all qubit Pauli operators supported in $A$ forms a complete basis of orthonormal operators, and \eqref{operator-identity} can be easily verified by decomposing $\mathcal{O}$ into that basis. A diagrammatic depiction of this identity is shown in Fig.~\ref{diagram_OTO}(a).

We would like to use this to evaluate the averaged correlator
\be
2^{-2a-2d-n}\sum_{ij}  \tr\,  \{D_i(t) A_j D_i(t) A_j \} = 2^{-2a-2d-n}\sum_{ij}  \tr\,  \{ U^\dagger D_i U A_j U^\dagger D_i U A_j \}.
\ee
Here the prefactor $2^{-2a-2d}$ is the inverse of number of operators in $A$ and $D$, and $2^{-n}$ is the normalization factor such that the quantity equals to $1$ if all operators are identity operators.  
Let's apply \eqref{operator-identity} to $D_i U A_j U^\dagger D_i$ to do the sum over $i$. This gives
\be
2^{-2a-d-n}\sum_{j}  \tr\,  \{U A_j U^\dagger \,  \tr_D \, \{U A_j U^\dagger\}\otimes I_D  \} \label{proof-mid-step},
\ee
where note that we have made use of the cyclicity of the trace.
 At this point, it's useful to adopt indices. We will use $k=1\dots 2^a$ for $A$ indices, $\ell=1\dots 2^b$ for $B$ indices, $m=1\dots 2^c$ for $C$ indices, and $o=1\dots 2^d$ for $D$ indices. This lets us rewrite \eqref{proof-mid-step} as
\be
2^{-2a-d-n}\sum_{j}  U_{k_1 \ell m o} (A_j)_{k_1 k_1'} U^*_{k_1' \ell m' o}    U_{k_2 \ell' m' o'} (A_j)_{k_2 k_2'} U^*_{k_2' \ell' m o'} ,
\ee
where repeated indices imply summation. Now, we apply \eqref{operator-identity} again, specifically to $(A_j)_{k_1 k_1'} U^*_{k_1' \ell m' o}    U_{k_2 \ell' m' o'} (A_j)_{k_2 k_2'}$. This sets $k_1'=k_2$ and $k_2' = k_1$ (and multiplies by $2^a$) to give
\be
2^{-a-d-n} U_{k_1 \ell m o} U^*_{k_2 \ell m' o}    U_{k_2 \ell' m' o'} U^*_{k_1 \ell' m o'}.\label{proof-near-final-step} 
\ee
Now, we remember how to express the density matrix $\rho$ of our channel (see Fig.~\ref{diagram_OTO}(b))
\be
\rho = 2^{-n}\, U_{k \ell m o} \, U^*_{k' \ell' m' o'}. 
\ee
Applying this to \eqref{proof-near-final-step} and then using the definition of the second R\'enyi entropy \eqref{def-of-renyi} 
gives our desired result \eqref{corr=renyi}. This whole proof, up to factors of normalization, is shown in Fig.~\ref{diagram_OTO}(c).

\subsubsection*{Finite temperature}

It is easy to generalize this formula for finite temperature $\beta>0$. Define
\be
Z(\beta):= \tr (e^{-\beta H}), \qquad \ket{\Psi(\beta)} := |TFD(\beta,t)\rangle.\label{eq:state-for-finite-temp-renyi}
\ee
To get an expression in terms of an entropy, we need to distribute the operators around the thermal circle
\be
|\langle   \OO_D(t) \,  \OO_A\, \OO_D(t) \, \OO_A\rangle_{\beta}|~\approx~Z(\beta)^{-1}|\tr\,  \{   \OO_D(t-i\beta/4) \,  \OO_A\, \OO_D(t-i\beta/4) \, \OO_A\}|.\label{eq:finite-beta-OTO}
\ee
Here, rather taking a thermal expectation value we are evolving the operators in $D$ in Euclidean time (and then renormalizing by $Z(\beta)$). The trace of these Euclidean-evolved correlators is expected to be related to the thermal expectation of the original OTO correlators as long as the temperature is high enough. Following our proof Fig.~\ref{diagram_OTO} but with the time argument for the unitary operators as $U(t-i\beta/4)$, we find
\be
= \frac{Z(\beta/2)^{2}}{Z(\beta)}2^{-a-d-S_{AC}^{(2)}(\beta/2) }.
\ee
where $S_{AC}^{(2)}(\beta/2)$ is evaluated for the state $\ket{\Psi(\beta/2)}$ defined in \eqref{eq:state-for-finite-temp-renyi}.

\subsubsection*{Higher order OTO correlators}

Finally, we will briefly comment on another possible generalization. The OTO correlation functions we studied here are observables for the chaotic dynamics of a thermal system perturbed by a single operator. In  \cite{Shenker:2013yza}, chaos is studied in holographic thermal systems that are perturbed by multiple operators. For two perturbations, the relevant observable is a six-point OTO correlation function of the form
\be
\langle W(t_1) \, V(t_2) \, Q \, V(t_2) \, W(t_1) \, Q \rangle_\beta,
\ee
where $W, V, Q$ are all simple Hermitian operators. (For simplicity, we will consider the case where $\beta=0$.) This observable is related to the effect of simple perturbations $W, V$ made at times $t_1, t_2$ on measurements of $Q$ at $t=0$. This correlation can be simplified by summing over a basis of operators in three regions associated with the $W, V, Q$ as we did for four-point functions earlier in this appendix. 

However, it's easy to see that one cannot get something as simple as a R\'enyi entropy: since there's two explicit times $t_1, t_2$, we can form density matrices $\rho(t_1)$, $\rho(t_2)$, and $\rho(t_1 -t_2)$, where $\rho(t)=|U(t)\rangle \langle U(t) |$ is the density matrix associated with the quantum channel of time evolution by $U(t) = e^{-iHt}$. The averaged six-point function will be related to contraction of these density matrices with a complicated permutation. This may be considered as a more generic entanglement property of the system, which is beyond R\'enyi entropies.
In the case of finite temperature, for the four-point functions we were able to evolve the operators in Euclidean time in order to symmetrize the time arguments of the unitary operators, as is discussed around  \eqref{eq:finite-beta-OTO}. However, cannot do that for the six-point functions since we need all three operators to be separated from each other in Lorentzian time.

\section{Tensor calculus} \label{appendix-tensor}

\begin{figure}[htb!]
\centering
\includegraphics[width=0.50\linewidth]{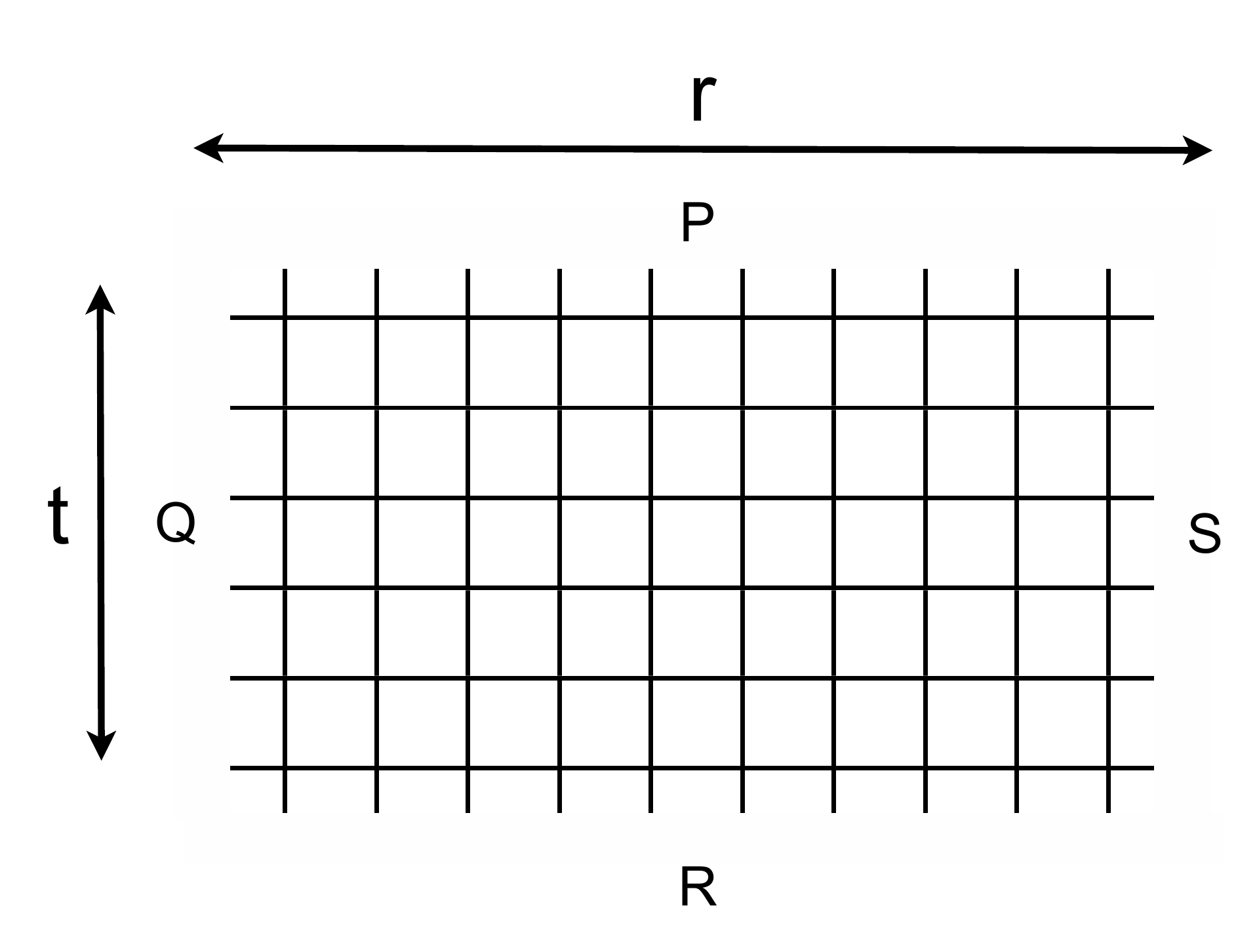}
\caption{A residual region in a planar tensor network. 
} 
\label{fig_rectangular}
\end{figure}

In this appendix, we provide more details about the perfect tensor calculation sketched in Fig. \ref{fig_I3}. 
We have seen that the two rectangular residual regions, which are not contained in any of causal wedges, are responsible for multipartite entanglement arising in a network of perfect tensors. To calculate $I_{3}$, we typically need to consider rectangular residual regions. In this appendix, we study multipartite entanglement in a rectangular network of perfect tensors as shown in Fig.~\ref{fig_rectangular} where tensor legs are split into four subsets $P,Q,R,S$. We assume that $P,R$ contain $r$ legs and $Q,S$ contain $t$ legs. 

In \cite{Pastawski:2015qua}, it was shown that, for any planar network of perfect tensors with non-positive curvature, the Ryu-Takayanagi formula for single intervals holds exactly. Keeping this in mind, let us summarize properties of entanglement in an arbitrary rectangular tiling of perfect tensors: 
\begin{equation}
\begin{split}
&S_{P}=r,\quad S_{Q}=t,\quad S_{R}=r,\quad S_{S}=t,\\ &S_{P Q}=r+t, \quad S_{Q R}=r+t, \quad S_{R S}=r+t, \quad S_{ P S}=r+t,
\end{split}
\end{equation}
where as a reminder, for tensors of bond dimension $v$, we measure their entropy in units of $\log v$.
Thus, the tripartite information $I_{3}$ is given by 
\begin{align}
I_{3}=- S_{PR}.
\end{align}
Note that the above statement holds for any perfect tensors and is not restricted to qutrit perfect tensors. But the value of $S_{PR}$ is non-universal for networks of perfect tensors since $PR$ consists of two spatially disjoint intervals. Below, we will prove that 
\begin{align}
I_{3}=- 2 \min(r,t),
\end{align}
for a network of four-leg qutrit perfect tensors.

The qutrit tensor network discussed in \S\ref{sec:perfect-tensor-model} can be described by the stabilizer formalism~\cite{Nielsen_Chuang}, and analytical calculations of entanglement entropies are possible. Let us recall a useful formula for entropy calculations for stabilizer states. Consider an $n$-qutrit pure state $|\psi\rangle$ specified by a set of $n$ independent Pauli stabilizer generators $g_{j}$ such that $g_{j}|\psi\rangle=|\psi\rangle$ for $j=1,\ldots,n$ with $[g_{i},g_{j}]=0$. The stabilizer group $\mathcal{S}^{(stab)}$ consists of all stabilizers $\mathcal{S}^{(stab)}=\langle \{ g_{j}\}^{\forall j} \rangle$. Therefore we have $g |\psi\rangle=|\psi\rangle$ for all $g\in \mathcal{S}^{(stab)}$. We are interested in entanglement entropy of $|\psi\rangle$ for some subset $A$ of qutrits. A useful formula to compute $S_{A}$ is the following \cite{Fattal04, Beni10}
\begin{align}
S_{A}= |A|- \log_{3} |\mathcal{S}^{(stab)}_{A}|, \label{eq:entropy_formula}
\end{align}
where $|A|$ is the number of qutrits in $A$ and $\mathcal{S}_{A}^{(stab)}$ is the restriction of $\mathcal{S}$ onto $A$ (i.e. a group of stabilizer operators which are supported exclusively on $A$). Note, $\log_{3} |\mathcal{S}^{(stab)}_{A}|$ can be understood as the number of independent stabilizers supported on $A$.

The stabilizer generators for the qutrit tensor are given by
\begin{equation}
\begin{split}
&Z \otimes Z \otimes Z \otimes I, \qquad
Z \otimes Z^{\dagger} \otimes I \otimes Z, \\
&X \otimes X \otimes X \otimes I, \qquad
X \otimes X^{\dagger} \otimes I \otimes X, 
\end{split}
\end{equation}
where $X^{\dagger}=X^{2}$ and $Z^{\dagger}=Z^{2}$. Graphically, stabilizer generators are given by
\begin{align}
\includegraphics[height=1.0in]{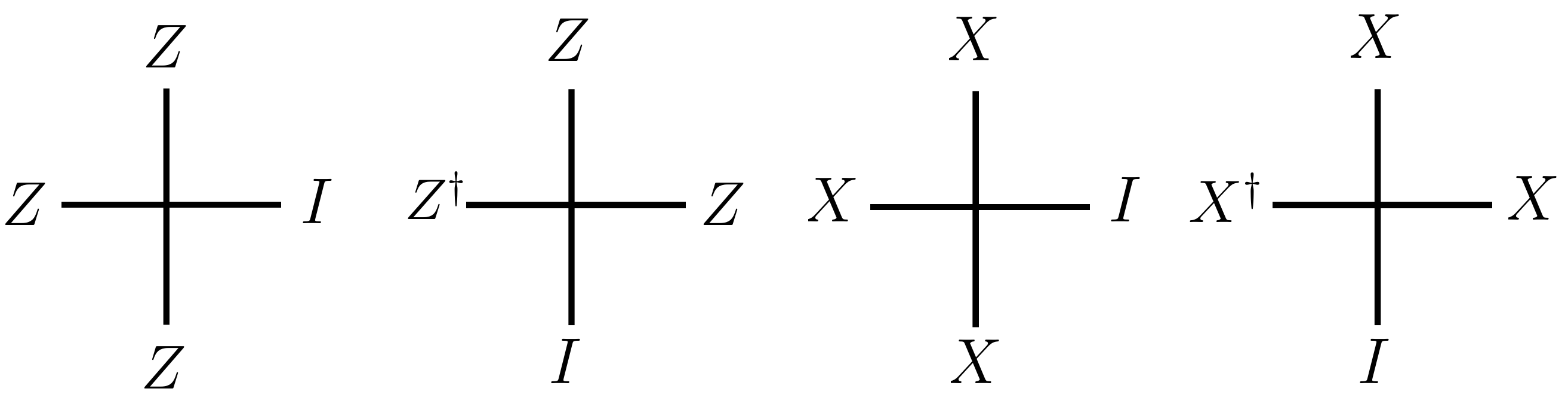},
\end{align}
where $Z|j\rangle = \omega^{j}|j\rangle$ and $X|j\rangle=|j+1\rangle$ with $\omega=e^{\frac{i2 \pi}{3}}$. Observe that stabilizer generators commute with each other. Also observe that there is no two-body stabilizer generator. This implies, from Eq ~\eqref{eq:entropy_formula}, that entanglement entropies for any subsets of two qutrits are two, and thus this stabilizer state is a four-leg perfect state.

We need to find the number of stabilizer generators which can be exclusively supported on $PR$. Let us first consider a contraction of two perfect tensors (i.e. $r=2$ and $t=1$). There are stabilizer generators supported only on upper and lower tensor legs
\begin{align}
\includegraphics[height=1.0in]{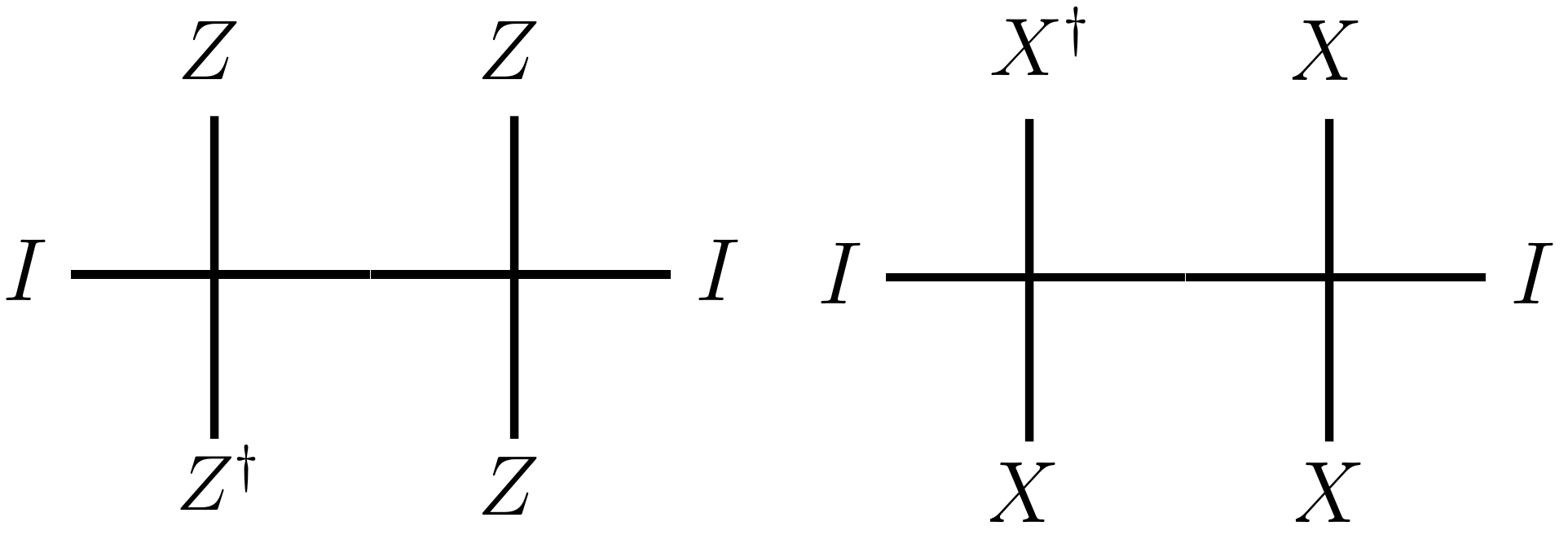}. \label{eq:stabilizer_polynomial}
\end{align}
So, $S_{PR}=|P|+|R|- 2=2$. Next, let us consider the case where $t=1$ and $r>2$. Since $X$-type and $Z$-type stabilizers are separable, one can treat them separately. We want to find all the stabilizer operators that are supported on $PR$. Here we consider an input Pauli operator $X(f)$ supported on $P$ where $f$ is a degree $r-1$ polynomial over $\mathbb{F}_{3}$. That is, for a polynomial
\begin{align}
f = c_{0} + c_{1}x + c_{2}x^{2} + \ldots + c_{r-1}x^{r-1},\qquad c_{j}\in \mathbb{F}_{3},
\end{align}
we define the Pauli-$X$ operator as
\begin{align}
X(f)= X^{c_{0}}\otimes X^{c_{1}}\otimes \ldots \otimes X^{c_{r-1}},
\end{align}
where $X_{j}$ acts on the $j$-th leg on $P$ for $j=0,\ldots, r-1$. Given an arbitrary degree $r-1$ polynomial $f_{0}$ over $\mathbb{F}_{3}$, one can write $f_{0}$ as follows
\begin{align}
f_{0} = (2+x) g_{0} + h_{0},
\end{align}
where $g_{0}$ is some degree $r-2$ polynomial while $h_{0}$ is some degree $0$ polynomial (in other words, a constant). Note that $X(2+x)$ is the Pauli $X$ operator on $P$ in \eqref{eq:stabilizer_polynomial}, whose ``output'' on $R$ is given by $X(1+x)$. Let us then look for a stabilizer operator whose action on the upper leg is given by $X(f_{0})$. When $t=1$, the output Pauli operator is supported exclusively on $R$ if and only if $h_{0}=0$. Namely, the output operator can be written as
\begin{align}
X(f_{1}) \qquad f_{1} = (1+x) g_{0}. 
\end{align}
Similar analysis holds for $Z$-type stabilizers. Therefore, there are in total $2r-2$ independent stabilizer generators supported on $PR$. Thus, $S_{PR}=r+r-(2r-2)=2$. Finally, let us consider the cases where $t>1$. For this purpose, we think of decomposing $f_{j}$ recursively as follows
\begin{align}
f_{j}=(2+x)g_{j} + h_{j}, \qquad f_{j+1}=(1+x)g_{j}.
\end{align}
The output has supports exclusively on $R$ if and only if $h_{j}=0$ for $j=0,\ldots,t-1$. This implies that there are in total $2(r-t)$ stabilizer generators supported on $AC$ for $t\leq r$, and there is no such stabilizer generator for $t>r$. Thus, one has $I_{3}=- 2 \min(r,t)$. 

In fact, the aforementioned result applies to a larger class of perfect tensors. Notice that stabilizer generators of the qutrit tensor can be written as tensor products of Pauli $Z$ or $X$ operators only. Such a stabilizer state is often referred to as a CSS (Calderbank-Shor-Steane) state, and a number of interesting quantum error-correcting codes belongs to this class. Let us assume that four-leg perfect tensors are based on CSS stabilizer states. Then, one is able to prove that $I_{3}$ is always given by $I_{3}=-2 \min(r,t)$ as long as the bond dimension $v$ is a prime number. We reached this conclusion by finding all the possible CSS-type four-leg perfect tensors with prime bond dimensions. However, this conclusion does not necessarily hold when bond dimension is not a prime number. We skip the derivation as it is similar to the one for the qutrit perfect tensor.

\mciteSetMidEndSepPunct{}{\ifmciteBstWouldAddEndPunct.\else\fi}{\relax}
\bibliographystyle{utphys}
\bibliography{channels}{}

\ifx\mcitethebibliography\mciteundefinedmacro
\PackageError{utphys.bst}{mciteplus.sty has not been loaded}
{This bibstyle requires the use of the mciteplus package.}\fi
\providecommand{\href}[2]{#2}\begingroup\raggedright\begin{mcitethebibliography}{10}

\bibitem{Hayden:2007cs}
P.~Hayden and J.~Preskill, ``{Black holes as mirrors: Quantum information in
  random subsystems},''
  \href{http://dx.doi.org/10.1088/1126-6708/2007/09/120}{{\em JHEP} {\bfseries
  0709} (2007) 120},
\href{http://arxiv.org/abs/0708.4025}{{\ttfamily arXiv:0708.4025 [hep-th]}}.

\bibitem{Sekino:2008he}
Y.~Sekino and L.~Susskind, ``{Fast Scramblers},''
  \href{http://dx.doi.org/10.1088/1126-6708/2008/10/065}{{\em JHEP} {\bfseries
  0810} (2008) 065},
\href{http://arxiv.org/abs/0808.2096}{{\ttfamily arXiv:0808.2096 [hep-th]}}.

\bibitem{Lashkari:2011yi}
N.~Lashkari, D.~Stanford, M.~Hastings, T.~Osborne, and P.~Hayden, ``{Towards
  the Fast Scrambling Conjecture},''
  \href{http://dx.doi.org/10.1007/JHEP04(2013)022}{{\em JHEP} {\bfseries 1304}
  (2013) 022},
\href{http://arxiv.org/abs/1111.6580}{{\ttfamily arXiv:1111.6580 [hep-th]}}.

\bibitem{Page:1993df}
D.~N. Page, ``{Average entropy of a subsystem},''
  \href{http://dx.doi.org/10.1103/PhysRevLett.71.1291}{{\em Phys. Rev. Lett.}
  {\bfseries 71} (1993) 1291--1294},
\href{http://arxiv.org/abs/gr-qc/9305007}{{\ttfamily arXiv:gr-qc/9305007
  [gr-qc]}}.

\bibitem{Roberts:2014isa}
D.~A. Roberts, D.~Stanford, and L.~Susskind, ``{Localized shocks},''
\href{http://arxiv.org/abs/1409.8180}{{\ttfamily arXiv:1409.8180 [hep-th]}}.

\bibitem{Shenker:2013pqa}
S.~H. Shenker and D.~Stanford, ``{Black holes and the butterfly effect},''
\href{http://arxiv.org/abs/1306.0622}{{\ttfamily arXiv:1306.0622 [hep-th]}}.

\bibitem{Shenker:2013yza}
S.~H. Shenker and D.~Stanford, ``{Multiple Shocks},''
\href{http://arxiv.org/abs/1312.3296}{{\ttfamily arXiv:1312.3296 [hep-th]}}.

\bibitem{Roberts:2014ifa}
D.~A. Roberts and D.~Stanford, ``{Two-dimensional conformal field theory and
  the butterfly effect},''
  \href{http://dx.doi.org/10.1103/PhysRevLett.115.131603}{{\em Phys. Rev.
  Lett.} {\bfseries 115} no.~13, (2015) 131603},
\href{http://arxiv.org/abs/1412.5123}{{\ttfamily arXiv:1412.5123 [hep-th]}}.

\bibitem{Shenker:2014cwa}
S.~H. Shenker and D.~Stanford, ``{Stringy effects in scrambling},''
  \href{http://dx.doi.org/10.1007/JHEP05(2015)132}{{\em JHEP} {\bfseries 05}
  (2015) 132},
\href{http://arxiv.org/abs/1412.6087}{{\ttfamily arXiv:1412.6087 [hep-th]}}.

\bibitem{Maldacena:2015waa}
J.~Maldacena, S.~H. Shenker, and D.~Stanford, ``{A bound on chaos},''
\href{http://arxiv.org/abs/1503.01409}{{\ttfamily arXiv:1503.01409 [hep-th]}}.

\bibitem{kitaev2}
A.~Kitaev, ``A simple model of quantum holography.''. Talks at KITP, April 7,
  2015 and May 27, 2015.

\bibitem{Hartman:2013qma}
T.~Hartman and J.~Maldacena, ``{Time Evolution of Entanglement Entropy from
  Black Hole Interiors},''
  \href{http://dx.doi.org/10.1007/JHEP05(2013)014}{{\em JHEP} {\bfseries 1305}
  (2013) 014},
\href{http://arxiv.org/abs/1303.1080}{{\ttfamily arXiv:1303.1080 [hep-th]}}.

\bibitem{Liu:2013iza}
H.~Liu and S.~J. Suh, ``{Entanglement Tsunami: Universal Scaling in Holographic
  Thermalization},''
  \href{http://dx.doi.org/10.1103/PhysRevLett.112.011601}{{\em Phys. Rev.
  Lett.} {\bfseries 112} (2014) 011601},
\href{http://arxiv.org/abs/1305.7244}{{\ttfamily arXiv:1305.7244 [hep-th]}}.

\bibitem{Liu:2013qca}
H.~Liu and S.~J. Suh, ``{Entanglement growth during thermalization in
  holographic systems},''
  \href{http://dx.doi.org/10.1103/PhysRevD.89.066012}{{\em Phys. Rev.}
  {\bfseries D89} no.~6, (2014) 066012},
\href{http://arxiv.org/abs/1311.1200}{{\ttfamily arXiv:1311.1200 [hep-th]}}.

\bibitem{Pastawski:2015qua}
F.~Pastawski, B.~Yoshida, D.~Harlow, and J.~Preskill, ``{Holographic quantum
  error-correcting codes: Toy models for the bulk/boundary correspondence},''
  \href{http://dx.doi.org/10.1007/JHEP06(2015)149}{{\em JHEP} {\bfseries 06}
  (2015) 149},
\href{http://arxiv.org/abs/1503.06237}{{\ttfamily arXiv:1503.06237 [hep-th]}}.

\bibitem{Nielsen_Chuang}
M.~A. {Nielsen} and I.~L. {Chuang}, {\em Quantum Computation and Quantum
  Information}.
\newblock Cambridge University Press, Cambridge, 2000.

\bibitem{Kitaev06}
A.~Kitaev and J.~Preskill, ``Topological entanglement entropy,'' {\em Phys.
  Rev. Lett.} {\bfseries 96} (2006) 110404.

\bibitem{Levin06}
M.~Levin and X.-G. Wen, ``Detecting topological order in a ground state wave
  function,'' {\em Phys. Rev. Lett.} {\bfseries 96} (2006) 110405.

\bibitem{Hayden:2011ag}
P.~Hayden, M.~Headrick, and A.~Maloney, ``{Holographic Mutual Information is
  Monogamous},'' \href{http://dx.doi.org/10.1103/PhysRevD.87.046003}{{\em Phys.
  Rev.} {\bfseries D87} no.~4, (2013) 046003},
\href{http://arxiv.org/abs/1107.2940}{{\ttfamily arXiv:1107.2940 [hep-th]}}.

\bibitem{Gharibyan:2013aha}
H.~Gharibyan and R.~F. Penna, ``{Are entangled particles connected by
  wormholes? Evidence for the ER=EPR conjecture from entropy inequalities},''
  \href{http://dx.doi.org/10.1103/PhysRevD.89.066001}{{\em Phys. Rev.}
  {\bfseries D89} no.~6, (2014) 066001},
\href{http://arxiv.org/abs/1308.0289}{{\ttfamily arXiv:1308.0289 [hep-th]}}.

\bibitem{Rangamani:2015qwa}
M.~Rangamani and M.~Rota, ``{Entanglement structures in qubit systems},''
  \href{http://dx.doi.org/10.1088/1751-8113/48/38/385301}{{\em J. Phys.}
  {\bfseries A48} no.~38, (2015) 385301},
\href{http://arxiv.org/abs/1505.03696}{{\ttfamily arXiv:1505.03696 [hep-th]}}.

\bibitem{Almheiri:2012rt}
A.~Almheiri, D.~Marolf, J.~Polchinski, and J.~Sully, ``{Black Holes:
  Complementarity or Firewalls?},''
  \href{http://dx.doi.org/10.1007/JHEP02(2013)062}{{\em JHEP} {\bfseries 02}
  (2013) 062},
\href{http://arxiv.org/abs/1207.3123}{{\ttfamily arXiv:1207.3123 [hep-th]}}.

\bibitem{Braunstein:2009my}
 cf.~S.~L. Braunstein, ``{Black hole entropy as entropy of entanglement, or
  it's curtains for the equivalence principle},''
  \href{http://arxiv.org/abs/0907.1190v1}{{\ttfamily arXiv:0907.1190v1
  [quant-ph]}}, for a similar prediction from different assumptions.

\bibitem{Almheiri:2013hfa}
A.~Almheiri, D.~Marolf, J.~Polchinski, D.~Stanford, and J.~Sully, ``{An
  Apologia for Firewalls},''
  \href{http://dx.doi.org/10.1007/JHEP09(2013)018}{{\em JHEP} {\bfseries 09}
  (2013) 018},
\href{http://arxiv.org/abs/1304.6483}{{\ttfamily arXiv:1304.6483 [hep-th]}}.

\bibitem{Maldacena:2001kr}
J.~M. Maldacena, ``{Eternal black holes in anti-de Sitter},''
  \href{http://dx.doi.org/10.1088/1126-6708/2003/04/021}{{\em JHEP} {\bfseries
  0304} (2003) 021},
\href{http://arxiv.org/abs/hep-th/0106112}{{\ttfamily arXiv:hep-th/0106112
  [hep-th]}}.

\bibitem{kitaev}
A.~Kitaev, ``Hidden correlations in the hawking radiation and thermal noise.''.
  Talk given at the Fundamental Physics Prize Symposium, Nov. 10, 2014.

\bibitem{sachdev1993gapless}
S.~Sachdev and J.~Ye, ``Gapless spin-fluid ground state in a random quantum
  heisenberg magnet,'' {\em Phys. Rev. Lett.} {\bfseries 70} no.~21, (1993)
  3339.

\bibitem{Sachdev:2015efa}
S.~Sachdev, ``{Bekenstein-Hawking Entropy and Strange Metals},''
\href{http://arxiv.org/abs/1506.05111}{{\ttfamily arXiv:1506.05111 [hep-th]}}.

\bibitem{RobertsSwingleVBUpcoming}
D.~A. Roberts and B.~Swingle. To appear.

\bibitem{Lieb:1972wy}
E.~Lieb and D.~Robinson, ``{The finite group velocity of quantum spin
  systems},''
\href{http://dx.doi.org/10.1007/BF01645779}{{\em Commun.Math.Phys.} {\bfseries
  28} (1972) 251--257}.

\bibitem{Hastings:2005pr}
M.~B. Hastings and T.~Koma, ``{Spectral gap and exponential decay of
  correlations},'' \href{http://dx.doi.org/10.1007/s00220-006-0030-4}{{\em
  Commun.Math.Phys.} {\bfseries 265} (2006) 781--804},
\href{http://arxiv.org/abs/math-ph/0507008}{{\ttfamily arXiv:math-ph/0507008
  [math-ph]}}.

\bibitem{hastings2010locality}
M.~B. Hastings, ``Locality in quantum systems,''
  \href{http://arxiv.org/abs/1008.5137}{{\ttfamily arXiv:1008.5137 [math-ph]}}.

\bibitem{Leichenauer:2015xra}
S.~Leichenauer and M.~Moosa, ``{Entanglement Tsunami in (1+1)-Dimensions},''
\href{http://arxiv.org/abs/1505.04225}{{\ttfamily arXiv:1505.04225 [hep-th]}}.

\bibitem{Casini:2015zua}
H.~Casini, H.~Liu, and M.~Mezei, ``{Spread of entanglement and causality},''
\href{http://arxiv.org/abs/1509.05044}{{\ttfamily arXiv:1509.05044 [hep-th]}}.

\bibitem{scramblingEntanglementWedge}
D.~Stanford, ``{Scrambling and the entanglement wedge},''. Unpublished.

\bibitem{banuls2011strong}
M.~C. Ba{\~n}uls, J.~I. Cirac, and M.~B. Hastings, ``Strong and weak
  thermalization of infinite nonintegrable quantum systems,'' {\em Phys. Rev.
  Lett.} {\bfseries 106} no.~5, (2011) 050405.

\bibitem{Brandao2012}
F.~G. Brandao, A.~W. Harrow, and M.~Horodecki, ``{Local random quantum circuits
  are approximate polynomial-designs},''
  \href{http://arxiv.org/abs/1208.0692}{{\ttfamily arXiv:1208.0692
  [quant-ph]}}.

\bibitem{Stanford:2014jda}
D.~Stanford and L.~Susskind, ``{Complexity and Shock Wave Geometries},''
  \href{http://dx.doi.org/10.1103/PhysRevD.90.126007}{{\em Phys.Rev.}
  {\bfseries D90} (2014) 126007},
\href{http://arxiv.org/abs/1406.2678}{{\ttfamily arXiv:1406.2678 [hep-th]}}.

\bibitem{swingle2012entanglement}
B.~Swingle, ``Entanglement renormalization and holography,'' {\em Phys.Rev.}
  {\bfseries D86} no.~6, (2012) 065007.

\bibitem{PhysRevLett.99.220405}
G.~Vidal, ``Entanglement renormalization,''
  \href{http://dx.doi.org/10.1103/PhysRevLett.99.220405}{{\em Phys. Rev. Lett.}
  {\bfseries 99} (Nov, 2007) 220405}.

\bibitem{Yang:2015uoa}
Z.~Yang, P.~Hayden, and X.-L. Qi, ``{Bidirectional holographic codes and
  sub-AdS locality},''
\href{http://arxiv.org/abs/1510.03784}{{\ttfamily arXiv:1510.03784 [hep-th]}}.

\bibitem{Almheiri:2014lwa}
A.~Almheiri, X.~Dong, and D.~Harlow, ``{Bulk Locality and Quantum Error
  Correction in AdS/CFT},''
  \href{http://dx.doi.org/10.1007/JHEP04(2015)163}{{\em JHEP} {\bfseries 04}
  (2015) 163},
\href{http://arxiv.org/abs/1411.7041}{{\ttfamily arXiv:1411.7041 [hep-th]}}.

\bibitem{Kim:2013aa}
H.~Kim and D.~A. Huse, ``Ballistic spreading of entanglement in a diffusive
  nonintegrable system,'' {\em Phys. Rev. Lett.} {\bfseries 111} no.~12, (09,
  2013) 127205--.

\bibitem{RandomModelHolgraphy}
P.~Hayden, S.~Nezami, X.-L. Qi, N.~Thomas, M.~Walter, and Z.~Yang,
  ``{Holographic duality from random tensor networks},''
\href{http://arxiv.org/abs/1601.01694}{{\ttfamily arXiv:1601.01694 [hep-th]}}.

\bibitem{hastings2015random}
M.~Hastings, ``Random mera states and the tightness of the brandao-horodecki
  entropy bound,'' \href{http://arxiv.org/abs/1505.06468}{{\ttfamily
  arXiv:1505.06468 [quant-ph]}}.

\bibitem{Marolf:2015vma}
D.~Marolf, H.~Maxfield, A.~Peach, and S.~F. Ross, ``{Hot multiboundary
  wormholes from bipartite entanglement},''
  \href{http://dx.doi.org/10.1088/0264-9381/32/21/215006}{{\em Class. Quant.
  Grav.} {\bfseries 32} no.~21, (2015) 215006},
\href{http://arxiv.org/abs/1506.04128}{{\ttfamily arXiv:1506.04128 [hep-th]}}.

\bibitem{dickman1930frequency}
K.~Dickman, {\em On the frequency of numbers containing prime factors of a
  certain relative magnitude}.
\newblock Almqvist \& Wiksell, 1930.

\bibitem{Susskind:2015toa}
L.~Susskind, ``{The Typical-State Paradox: Diagnosing Horizons with
  Complexity},''
\href{http://arxiv.org/abs/1507.02287}{{\ttfamily arXiv:1507.02287 [hep-th]}}.

\bibitem{Brown:2015bva}
A.~R. Brown, D.~A. Roberts, L.~Susskind, B.~Swingle, and Y.~Zhao, ``{Complexity
  Equals Action},''
\href{http://arxiv.org/abs/1509.07876}{{\ttfamily arXiv:1509.07876 [hep-th]}}.

\bibitem{CAFollowup}
A.~Brown, D.~A. Roberts, L.~Susskind, B.~Swingle, and Y.~Zhao, ``{Complexity,
  Action, and Black Holes},''
\href{http://arxiv.org/abs/1512.04993}{{\ttfamily arXiv:1512.04993 [hep-th]}}.

\bibitem{tHooft:1993gx}
G.~'t~Hooft, ``{Dimensional reduction in quantum gravity},'' in {\em {Salamfest
  1993:0284-296}}, pp.~0284--296.
\newblock 1993.
\newblock
\href{http://arxiv.org/abs/gr-qc/9310026}{{\ttfamily arXiv:gr-qc/9310026
  [gr-qc]}}.
\newblock

\bibitem{Susskind:1994vu}
L.~Susskind, ``{The World as a hologram},''
  \href{http://dx.doi.org/10.1063/1.531249}{{\em J. Math. Phys.} {\bfseries 36}
  (1995) 6377--6396},
\href{http://arxiv.org/abs/hep-th/9409089}{{\ttfamily arXiv:hep-th/9409089
  [hep-th]}}.

\bibitem{Susskind:2014moa}
L.~Susskind, ``{Entanglement is not Enough},''
\href{http://arxiv.org/abs/1411.0690}{{\ttfamily arXiv:1411.0690 [hep-th]}}.

\bibitem{fannes1973continuity}
M.~Fannes, ``A continuity property of the entropy density for spin lattice
  systems,'' \href{http://dx.doi.org/10.1007/BF01646490}{{\em Communications in
  Mathematical Physics} {\bfseries 31} no.~4, (1973) 291--294}.

\bibitem{audenaert2006sharp}
K.~M.~R. Audenaert, ``A sharp fannes-type inequality for the von neumann
  entropy,'' \href{http://dx.doi.org/10.1088/1751-8113/40/28/S18}{{\em Journal
  of Physics A: Mathematical and Theoretical} {\bfseries 40} no.~28, (2007)
  8127}, \href{http://arxiv.org/abs/quant-ph/0610146}{{\ttfamily
  arXiv:quant-ph/0610146 [quant-ph]}}.

\bibitem{Calabrese:2005in}
P.~Calabrese and J.~L. Cardy, ``{Evolution of entanglement entropy in
  one-dimensional systems},''
  \href{http://dx.doi.org/10.1088/1742-5468/2005/04/P04010}{{\em J. Stat.
  Mech.} {\bfseries 0504} (2005) P04010},
\href{http://arxiv.org/abs/cond-mat/0503393}{{\ttfamily arXiv:cond-mat/0503393
  [cond-mat]}}.

\bibitem{Calabrese:2007rg}
P.~Calabrese and J.~Cardy, ``{Quantum Quenches in Extended Systems},''
  \href{http://dx.doi.org/10.1088/1742-5468/2007/06/P06008}{{\em J. Stat.
  Mech.} {\bfseries 0706} (2007) P06008},
\href{http://arxiv.org/abs/0704.1880}{{\ttfamily arXiv:0704.1880
  [cond-mat.stat-mech]}}.

\bibitem{Calabrese:2009qy}
P.~Calabrese and J.~Cardy, ``{Entanglement entropy and conformal field
  theory},'' \href{http://dx.doi.org/10.1088/1751-8113/42/50/504005}{{\em J.
  Phys.} {\bfseries A42} (2009) 504005},
\href{http://arxiv.org/abs/0905.4013}{{\ttfamily arXiv:0905.4013
  [cond-mat.stat-mech]}}.

\bibitem{Asplund:2014coa}
C.~T. Asplund, A.~Bernamonti, F.~Galli, and T.~Hartman, ``{Holographic
  Entanglement Entropy from 2d CFT: Heavy States and Local Quenches},''
\href{http://arxiv.org/abs/1410.1392}{{\ttfamily arXiv:1410.1392 [hep-th]}}.

\bibitem{Balasubramanian:2011at}
V.~Balasubramanian, A.~Bernamonti, N.~Copland, B.~Craps, and F.~Galli,
  ``{Thermalization of mutual and tripartite information in strongly coupled
  two dimensional conformal field theories},''
  \href{http://dx.doi.org/10.1103/PhysRevD.84.105017}{{\em Phys. Rev.}
  {\bfseries D84} (2011) 105017},
\href{http://arxiv.org/abs/1110.0488}{{\ttfamily arXiv:1110.0488 [hep-th]}}.

\bibitem{Allais:2011ys}
A.~Allais and E.~Tonni, ``{Holographic evolution of the mutual information},''
  \href{http://dx.doi.org/10.1007/JHEP01(2012)102}{{\em JHEP} {\bfseries 01}
  (2012) 102},
\href{http://arxiv.org/abs/1110.1607}{{\ttfamily arXiv:1110.1607 [hep-th]}}.

\bibitem{Asplund:2015eha}
C.~T. Asplund, A.~Bernamonti, F.~Galli, and T.~Hartman, ``{Entanglement
  Scrambling in 2d Conformal Field Theory},''
\href{http://arxiv.org/abs/1506.03772}{{\ttfamily arXiv:1506.03772 [hep-th]}}.

\bibitem{Fattal04}
D.~Fattal, T.~S. Cubitt, Y.~Yamamoto, S.~Bravyi, and I.~L. Chuang,
  ``Entanglement in the stabilizer formalism,''
  \href{http://arxiv.org/abs/quant-ph/0406168}{{\ttfamily
  arXiv:quant-ph/0406168 [quant-ph]}}.

\bibitem{Beni10}
B.~Yoshida and I.~L. Chuang, ``Framework for classifying logical operators in
  stabilizer codes,'' {\em Phys. Rev. A} {\bfseries 81} (2010) 052302.

\end{mcitethebibliography}\endgroup

\end{document}